\documentclass[10pt]{article}

\usepackage{xcolor}
\usepackage{xspace}
\usepackage{subcaption}
\usepackage{caption}
\usepackage{pifont}
\usepackage{times}
\usepackage{graphbox}
\usepackage{graphicx}
\usepackage{epstopdf}
\usepackage{tocloft}
\usepackage[htt]{hyphenat}
\usepackage[hidelinks]{hyperref}
\usepackage{fancyhdr}
\usepackage{lastpage}
\usepackage{setspace}
\usepackage{relsize}
\usepackage{textcomp}
\usepackage{circledsteps}
\usepackage{url}
\usepackage[htt]{hyphenat}
\usepackage{lmodern}
\usepackage{geometry}
\usepackage{dblfloatfix}
\usepackage{titlesec}
\usepackage{blindtext}
\usepackage[super]{nth}
\usepackage{comment}
\usepackage{transparent}
\usepackage{wrapfig}
\usepackage{rotating}

\usepackage{amssymb}
\usepackage[framemethod=tikz]{mdframed}
\usepackage[numbers,sort&compress]{natbib}
\tikzset{
    vertical align/.style={
        baseline=-.5*(height("$+$")-depth("$+$"))
    }
}

\usepackage{calc}
\usepackage{amsmath,algorithm,algpseudocode}
\usepackage{cases}
\makeatletter
\renewcommand{\Function}[2]{%
  \csname ALG@cmd@\ALG@L @Function\endcsname{#1}{#2}%
  \def\jayden@currentfunction{#1}%
}
\newcommand{\funclabel}[1]{%
  \@bsphack
  \protected@write\@auxout{}{%
    \string\newlabel{#1}{{\jayden@currentfunction}{\thepage}}%
  }%
  \@esphack
}
\makeatother

\usepackage{pythonhighlight}
\definecolor{codegray}{rgb}{0.5,0.5,0.5}
\lstdefinestyle{pythonStyle}{
  basicstyle=\tiny\ttfamily\footnotesize,
  commentstyle=\color{codegray},
  frame=single,
  language=Python,
  stepnumber=1,
  numbers=left,
  numbersep=5pt,
  numberstyle=\tiny\color{codegray},
  tabsize=2,
  showspaces=false,
  showstringspaces=false,
  mathescape,
  moredelim=**[is][\color{red}]{~}{~},
  moredelim=**[is][\color{blue}]{<}{>},
  moredelim=**[is][\color{orange}]{@}{@},
  literate={\\~}{{\textasciitilde}}1
  {\\<}{{\unichar{"003C}}}1
  {\\>}{{\unichar{"003E}}}1
  {\\@}{{\unichar{"0040}}}1
}

\usepackage{CJK}
\usepackage[utf8]{inputenc}
\usepackage[T1]{fontenc}
\usepackage{soul}
\usepackage{kotex}
\usepackage{lipsum}
\usepackage[normalem]{ulem}
\makeatletter
\def\SOUL@hlpreamble{%
\setul{\dimexpr\dp\strutbox-2pt}{\dimexpr\ht\strutbox+\dp\strutbox-2pt\relax}
\let\SOUL@stcolor\SOUL@hlcolor
\SOUL@stpreamble
}
\makeatother

\newcommand\khlc[1][yellow]{
  \bgroup
  \markoverwith{\textcolor{#1}{\rule[-.5ex]{1pt}{2.5ex}}}
  \ULon
}
\newcommand{\todo}[1]{\bgroup\color{white}\textbf{\khlc[black]{TODO: [#1]}}\egroup\xspace}
\newcommand{\fixme}[1]{\bgroup\color{red}\textbf{\khlc{FIXME: [#1]}}\egroup\xspace}
\newcommand{\pointer}[1]{\bgroup\color{white}\textbf{\khlc[red]{POINTER: [#1 is working here]}}\egroup\xspace}

\newcommand{\reviewer}[1]{\bgroup\color{blue}#1\egroup\xspace}
\newcommand{\ranswer}[1]{\bgroup\color{red}#1\egroup\xspace}

\usepackage{tabularx}
\usepackage{array}
\usepackage{multirow}
\usepackage{booktabs}
\usepackage{colortbl}
\usepackage{hhline}

\makeatletter
\def\hlinewd#1{%
\noalign{\ifnum0=`}\fi\hrule \@height #1 %
\futurelet\reserved@a\@xhline}
\makeatother

\usepackage[framemethod=tikz]{mdframed}
\usepackage{marginnote}
\usepackage{pdfcomment}
\mdfsetup{
  hidealllines=true,
  innerleftmargin=2pt,
  innerrightmargin=2pt,
  innertopmargin=2pt,
  innerbottommargin=2pt,
  leftmargin=-2pt,
  rightmargin=-2pt,
  skipabove=-2pt,
  skipbelow=-2pt,
  backgroundcolor=blue!20}

\newlength{\markerHeight}
\newlength{\markerMargin}
\newlength{\linespace}
\newlength{\linedepth}
\sbox0{yf}
\definecolor{mylime}{RGB}{205, 220, 57}
\definecolor{mygreen}{RGB}{60, 200, 0}
\definecolor{myblue}{RGB}{0, 51, 204}

\colorlet{soulred}{red!20}
\colorlet{soulgreen}{green!20}
\colorlet{soulblue}{blue!20}

\usepackage{enumitem}
\setlistdepth{5}
\renewlist{itemize}{itemize}{5}
\setlist[itemize,1]{label=$\bullet$}
\setlist[itemize,2]{label=$\circ$}
\setlist[itemize,3]{label=$\ast$}
\setlist[itemize,4]{label=-}
\setlist[itemize,5]{label=$\cdot$}

\makeatletter
\def\SOUL@hlpreamble{%
\setul{\dimexpr\dp\strutbox-2pt}{\dimexpr\ht\strutbox+\dp\strutbox-2pt\relax}
\let\SOUL@stcolor\SOUL@hlcolor
\SOUL@stpreamble
}
\makeatother

\definecolor{themeColor}{RGB}{0, 51, 0}
\definecolor{titleColor}{RGB}{0, 51, 0}
\definecolor{sectionColor}{RGB}{0, 51, 0}
\definecolor{abstractBg}{RGB}{240, 240, 240}

\usepackage{lastpage}

\renewcommand{\headrulewidth}{0.4pt}
\renewcommand{\headrule}{
  \vskip1ex
  \hbox to\headwidth{
    \color{themeColor}\leaders\hrule height \headrulewidth\hfill
  }\vskip-1.2pt
}

\titleformat{\section}
  {\normalfont\bfseries\color{sectionColor}\Large}
  {\thesection.}{1em}{}

\titleformat{\subsection}
  {\normalfont\bfseries\color{sectionColor}\large}
  {\thesubsection.}{1em}{}

\titleformat{\subsubsection}
  {\normalfont\bfseries\color{sectionColor}}
  {\thesubsubsection.}{1em}{}

\newmdenv[
  backgroundcolor=abstractBg,
  linecolor=themeColor,
  innerlinewidth=1pt,
  roundcorner=5pt,
  skipabove=\topsep,
  skipbelow=\topsep,
  innertopmargin=10pt,
  innerbottommargin=10pt,
  innerleftmargin=10pt,
  innerrightmargin=10pt
]{abstractbox}

\renewenvironment{abstract}{
  \begin{center}
    \begin{abstractbox}
      \textbf{\textcolor{themeColor}{Abstract}}\\[4pt]
      \ignorespaces
}{
      \unskip
    \end{abstractbox}
  \end{center}
  \vspace{1em}
}

\newcommand{\papertitle}{
Compute Can't Handle the Truth:
 Why Communication Tax Prioritizes Memory and Interconnects in
 Modern AI Infrastructure}
\newcommand{\paperauthor}{Myoungsoo Jung}
\newcommand{\paperinfo}{
  \textit{Panmnesia, Inc.} \\
  \texttt{http://panmnesia.com} \\
  \texttt{mj@panmnesia.com}
}

\begin{document}

\begin{center}
    {\LARGE \textbf{\textcolor{titleColor}{\papertitle}} \par}
    \vspace{0.5em}
    {\large \paperauthor \par}
    \vspace{0.5em}
    {\paperinfo \par}
    \vspace{1em}
\end{center}

\begin{abstract}
Modern AI workloads, particularly large-scale language models (LLMs) and retrieval-augmented generation (RAG), impose stringent demands on memory resources, inter-device communication, and flexible resource allocation. Since traditional GPU-centric architectures face scalability bottlenecks, inter-GPU communication overhead often dominates runtime, severely limiting efficiency at large scales.

For better understanding of these challenges, this technical report first introduces fundamental AI concepts in an accessible manner, explaining how contemporary models represent and process complex, high-dimensional data. Specifically, we illustrate how Transformer architectures overcame previous modeling constraints and became foundational technologies underlying modern LLMs. We then analyze representative large-scale AI hardware configurations and data center architectures, detailing how the LLMs are executed within these infrastructures and identifying fundamental sources of scalability challenges in hierarchical deployments.

Based on the observations, we redesign a modular and composable data center architecture leveraging Compute Express Link (CXL), which can address the scalability issues of modern AI data centers. Specifically, the redesigned architecture can independently disaggregate and scale memory, compute, and accelerator resources, dynamically allocating them based on specific workload requirements. This report also explores various CXL topologies and hardware configurations to enable accelerator-centric architectures and facilitate efficient resource disaggregation through modular memory pool designs in data centers. Our empirical evaluations across diverse AI workloads demonstrate that this modular approach can improve scalability, memory efficiency, computational throughput, and operational flexibility.

On the other hand, to accommodate diverse accelerators and hardware scales, we explore and integrate dedicated accelerator-optimized interconnect technologies, collectively referred to as XLink, including Ultra Accelerator Link (UALink), NVIDIA's NVLink, and NVLink Fusion. XLink optimizes latency-sensitive intra-accelerator communication through high-throughput, direct connections, whereas CXL enables scalable, coherent inter-node memory sharing. Motivated by this insight, we introduce a hybrid interconnect approach, CXL-over-XLink, designed to minimize unnecessary long-distance data transfers across scale-out domains, improving overall scalability of scale-up architectures while ensuring memory coherence.

Upon establishing the CXL-over-XLink design, we further propose a hierarchical memory architecture that combines accelerator-local memory and flexible external memory pools to address diverse latency and capacity requirements. In this technical report, we also present optimization strategies for lightweight CXL implementations, high-bandwidth memory (HBM), and silicon photonics to efficiently scale these composable architectures. Finally, we evaluate key performance and scalability parameters critical for modern AI infrastructures.
\end{abstract}

\paragraph{\textcolor{themeColor}{Keywords}}: CXL, NVLink, NVLink Fusion, UALink, AI Infrastructure, Data Centers, Accelerators, GPUs, Machine Learning, Hardware Architecture.
\pagebreak

\tableofcontents
\pagebreak

\section{Introduction}
\label{section:1}
Artificial Intelligence (AI), particularly machine learning (ML), has experienced substantial growth over recent decades through cycles of incremental improvements and major breakthroughs \cite{Avoidingwinter,AIwinter,attentionall,ai1}. Historically, AI advancements primarily stemmed from increased computational capabilities. However, recent progress relies significantly on data availability and advancements in memory management techniques. Such improvements have enabled contemporary AI systems to effectively emulate complex cognitive functions, achieving human-level or superior performance in tasks like image interpretation, natural language understanding, conversational interactions, and creative content generation \cite{Imagenet,Bert,llama3herdmodels,Generative}.

Modern AI methodologies rely on transforming data into structured numerical representations, such as vectors and matrices, which enable computational models to learn complex patterns. Specifically, AI models iteratively adjust internal parameters during training using optimization techniques like gradient descent \cite{GD1,GD2,GD3,GD4,GD5}. The accuracy and effectiveness of these models depend on their capacity to represent data within high-dimensional numerical spaces \cite{Numerical,dimension,dimension1,dimension2,gpurpose}. As datasets grow in size and complexity, the memory and computational demands of AI models increase significantly, surpassing the practical limitations of traditional CPU-centric computing infrastructures.

To accommodate such extensive data and computational demands, specialized hardware accelerators such as graphics processing units (GPUs\footnote{In this technical report, the term `accelerators' includes GPUs, NPUs, and other specialized processing units. Although GPUs are primarily discussed, all accelerators mentioned can be collectively categorized as coprocessors and treated equivalently as data-processing acceleration hardware.}) have become widely adopted in industry and academia \cite{UseofGPU,UseofGPU1,Trillion,Trillion1,reduceperf,reduceperf1,efficientlarge,gpumem3,gpumem2,gpumem1,commop4,memdemand,Blackwell2,peakperf,c2c1,poseidon}. GPUs provide parallel processing capabilities and integrate substantial internal memory resources (e.g., high-bandwidth memory such as HBM3e \cite{hbm1,hbm2,hbm3,gpumem4,UseofGPU}), enhancing their suitability for contemporary AI workloads.

However, as AI models scale to billions or even trillions of parameters~\cite{Trillion,Trillion1,GB200200,swtransformer,efficientlarge,rag1,rag3,megatron,knowledgepack,parallel1,trillpara}, individual GPU memory capacities are insufficient by design to meet these demands. For instance, Llama 3 405B~\cite{llama3herdmodels}, with a context window of over a hundred thousand tokens, requires more than a hundred terabytes (TB) of total memory to accommodate embeddings, activations, and optimizer states. This demand exceeds the hundred gigabyte (GB) capacity available in current state-of-the-art GPUs such as NVIDIA's GB200 and GB300~\cite{GB200,GB200200,GB300,GB300300}. Given the variety of models concurrently deployed, modern AI infrastructures utilize thousands to tens of thousands of GPUs collaboratively, which leads to substantial inter-GPU communication overhead. Industry analyses indicate that such communication accounts for 35\%--70\% of total training time in large-scale AI deployments, severely limiting efficiency and scalability \cite{Overhead,commoverhead,chang2024flux,commop1,commop4,jiang2024lancet,10.1145/3695053.3731105}. Given that outcomes from AI research conducted across various disciplines have historically been challenging to directly translate into practical daily applications, it is noteworthy that significant real-world impacts, such as those exemplified by OpenAI’s ChatGPT, have emerged prominently only within the past two years. This recent shift highlights the unprecedented data exchanges, information volumes, and memory demands associated with recent large-scale AI workloads compared to previous periods.

Therefore, the central challenge in contemporary AI infrastructures is \textit{no longer purely computational but involves managing massive data transfer volumes, extensive memory resources, and intensive communication demands} characteristic of modern AI workloads. Traditional GPU architectures limit memory scalability due to tightly integrated memory controllers, restricting flexible memory expansion. External memory access via PCIe or storage introduces significant latency, typically ranging from hundreds of nanoseconds ($n$s) to tens of microseconds ($\mu$s), substantially reducing GPU utilization and overall performance \cite{reduceperf,reduceperf1,externalmem,pcieperf,pcieperf1}. To address these limitations, \textit{Compute Express Link} (CXL \cite{cxl1.0,cxl2.0,cxl3.2}) has emerged as a transformative memory and interconnect technology \cite{memwall,cxlintro,externalmem,hello,reduceperf,pond,breakwall,directcxl}. CXL can fundamentally redefine AI infrastructure by decoupling memory controllers from computational units, enabling independent and dynamic memory management. By externalizing memory controllers and aggregating memory into composable pools accessible by multiple computing nodes, CXL can significantly expand available memory capacity and reduce data communication latency. We believe that this approach can effectively resolve traditional scalability bottlenecks associated with large-scale AI deployments.

To address the challenges and opportunities in redesigning AI infrastructures, it is essential to thoroughly understand the complete system stack, ranging from high-level theoretical AI models and architectures to the low-level hardware configurations deployed in data centers. In this technical report\footnote{This technical document, excluding the section pertaining to XLink, is based entirely on the keynote speech delivered by Panmnesia, at the 2024 Summer Conference of the Institute of Semiconductor Engineers.}, we first provide a clear explanation of fundamental AI principles to comprehensively analyze the scalability challenges of AI. We then describe how recent models represent and process complex, high-dimensional data. We then systematically analyze key factors that have enabled the successful evolution of advanced AI models from the perspectives of data management, memory architectures, and optimized communication structures in AI infrastructures. To this end, we examine real-world AI deployments within large-scale data center architectures, focusing on how advanced GPU technologies are utilized and identifying architectural limitations that arise in these deployments.

Subsequently, considering the growing complexity and diverse performance requirements of modern AI workloads, we redesign a modular and composable data center architecture optimized for flexible and dynamic resource management through CXL. True composability requires effective disaggregation of memory, computational units, and accelerator resources. In such composable systems, resources can be independently scaled and precisely adapted to varying workload demands, being able to improve flexibility, scalability, and resource efficiency. Note that CXL technology is central to realizing this composable infrastructure. By enabling coherent memory sharing and scalable interconnect topologies, CXL supports dynamic allocation of memory and computational resources independently of CPU involvement. This architectural advancement substantially reduces latency, enhances memory utilization, and can address a wide range of AI workloads, from general training and inference tasks to specialized applications such as retrieval-augmented generation (RAG) \cite{RAG,rag1,rag2,rag3,ragwork1,ragwork2} and key-value (KV) caching \cite{KV,kvcache,kvcache1,gpumem1,gpumem3,memdemand1}.

On the other hand, we also propose a strategic integration of complementary interconnect solutions, specifically \emph{accelerator-centric interconnect link} (XLink), an umbrella term encompassing both \textit{Ultra Accelerator Link} (UALink \cite{ualink,ualink1,uascale}) and NVIDIA's \textit{NVLink/NVLink Fusion} \cite{nvlink1,nvlink2,nvlink5.0,fusion,fusion1}. XLink technologies optimize latency-sensitive intra-accelerator communication with high throughput and direct connections, but these technologies differ significantly in deployment scope and compatibility: NVLink exclusively targets NVIDIA GPU clusters, while UALink offers an open, Ethernet-based interconnect solution that may exclude NVIDIA GPUs. Meanwhile, both NVLink and UALink typically utilize single-hop Clos topologies, restricting their scalability to rack-level scale-up domains. To bridge this gap and integrate diverse accelerator clusters, we propose a hybrid architecture, \textit{CXL-over-XLink}, in which CXL serves as an inter-cluster fabric. By interconnecting accelerator clusters composed of either NVLink or UALink via CXL fabrics, this architecture enables coherent memory sharing and communication scalability beyond rack-level constraints. This approach can fundamentally reduce unnecessary long-distance communication, including \textit{remote direct memory access} (RDMA)-based data exchanges \cite{rdma3,rdma1,rdma2,rdma}, thereby providing scalable architectures tailored for modern AI infrastructures.

Upon establishing the CXL-over-XLink architecture, we further introduce a hierarchical memory structure to efficiently address diverse latency and capacity demands of contemporary AI workloads. Specifically, we propose integrating accelerator-local memory with lightweight, customized CXL interconnected together with XLink, providing rapid and cache-coherent access to frequently used, latency-sensitive datasets. In parallel, large-scale composable memory pools, connected via lightweight CXL interfaces, handle extensive datasets with relatively relaxed latency constraints. By employing this hybrid memory approach in CXL-over-XLink, our proposed architecture can effectively balance low-latency data access and scalable memory capacity, enhancing overall performance at reduced costs within scale-up domains.

Lastly, to enhance scalability and optimize deployment at larger infrastructure scales, we investigate various hardware-level strategies including HBM, silicon photonics, and cost-efficient, tiered CXL memory implementations. Note that recognizing structural similarities in communication and memory access patterns, we extend our analysis to scientific computing applications, evaluating message passing interface (MPI)-based high-performance computing (HPC) workloads \cite{Graph500,NASA.IS,LBM,tealeaf,heattransfer}. These evaluations moreover illustrate the broad applicability of CXL beyond strictly AI-focused tasks, highlighting its potential to enhance computational efficiency and performance across diverse computational domains.

\section{From RNNs to Transformers: Evolution in Sequence Modeling}
\label{section:2}
The purpose of this section is to explain why AI, previously confined largely to research domains, has recently transitioned into practical everyday applications. First of all, we introduce fundamental AI concepts in an accessible manner, highlighting how contemporary models represent and process complex, high-dimensional data, and describe how advancements in hardware acceleration have enabled this transition. To achieve this, we discuss the progression of sequence modeling techniques, beginning with early \textit{Sequence-to-Sequence} (Seq2Seq \cite{seq2seq1,seq2seq2,seq2seq3}) frameworks, advancing through attention mechanisms, and progressing to \textit{Transformer} architectures \cite{attentionall,Bert,radford2018improving}, ultimately leading to modern \textit{Large Language Models} (LLMs \cite{GPT4turbo,llamamodel,trillpara,trillpara1,trillpara2,trillpara3}).

\begin{figure}[t!]
    \centering
    \includegraphics[width=\linewidth]{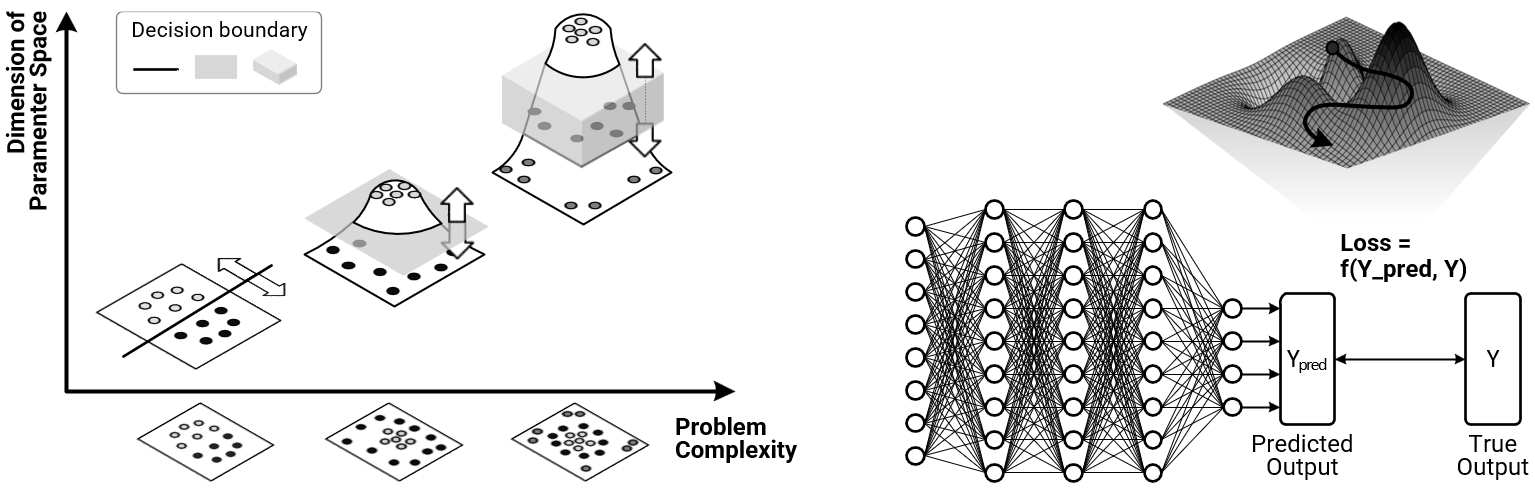}
    \begin{minipage}{0.56\linewidth}
        \vspace{-8pt}
        \begin{subfigure}{\linewidth}
            \begin{tabularx}{\textwidth}{
                p{\dimexpr.12\linewidth-2\tabcolsep-1.3333\arrayrulewidth}
                p{\dimexpr.24\linewidth-2\tabcolsep-1.3333\arrayrulewidth}
                p{\dimexpr.24\linewidth-2\tabcolsep-1.3333\arrayrulewidth}
                p{\dimexpr.24\linewidth-2\tabcolsep-1.3333\arrayrulewidth}
                p{\dimexpr.12\linewidth-2\tabcolsep-1.3333\arrayrulewidth}
                }
                \caption*{} &
                \caption{2D space.} \label{fig:db_2} &
                \caption{3D space.} \label{fig:db_3} &
                \caption{4D space.} \label{fig:db_4}
                \caption*{}
            \end{tabularx}
          \end{subfigure}
          \vspace{-22pt}
        \caption{Setting the decision boundary.}
        \label{fig:db}
    \end{minipage}
    \begin{minipage}{0.43\linewidth}
        \vspace{-10pt}
        \begin{subfigure}{\linewidth}
            \begin{tabularx}{\textwidth}{
                p{\dimexpr.53\linewidth-2\tabcolsep-1.3333\arrayrulewidth}
                p{\dimexpr.47\linewidth-2\tabcolsep-1.3333\arrayrulewidth}
                }
                \caption{Many parameters.} \label{fig:lf_1} &
                \caption{Gradient descent.} \label{fig:lf_2}
            \end{tabularx}
        \end{subfigure}
        \vspace{-6pt}
        \caption{Minimizing the loss function.}
        \label{fig:lf}
    \end{minipage}
\end{figure}

Specifically, in this section, we examine the practical importance and characteristics of time-series data, introducing foundational concepts underlying the Seq2Seq framework. We then address the strengths and limitations of early Seq2Seq implementations based on \textit{Recurrent Neural Network} (RNN \cite{rnn0,RNN,rnn1,rnn2}), describing how attention mechanisms resolved key shortcomings. Next, we explain how the Transformer architecture leveraged advances in ``parallel computing'' and ``hardware acceleration'' to overcome scalability limitations of earlier approaches, enhancing AI's applicability to real-world scenarios. Lastly, we outline how Transformers evolved into LLMs and emphasize their implications for hardware infrastructure and scalability, both critical factors in their widespread practical adoption.

\subsection{Understanding Time-Series Data and the Sequence-to-Sequence Framework}
\label{subsection:2_1}
Recent advances in AI technologies have achieved human-level accuracy across diverse real-world problems, attracting broad attention and driving innovation in both industry and academia. Before discussing sequence models, it is useful to briefly outline the foundational methods of AI model training and inference, particularly regarding the handling of time-series data.

\paragraph{Brief overview of AI training and inference methods.} To solve complex real-world problems, AI typically transforms input data into mathematical representations, such as numbers, vectors, or matrices mapped onto coordinate systems of specific dimensions. As depicted in Figure \ref{fig:db}, these transformed data representations allow the establishment of decision boundaries, which categorize input data into distinct groups based on identifiable patterns. The \emph{training} process involves the AI model navigating the parameter space to optimize these decision boundaries, enhancing their clarity and effectiveness. \emph{Inference}, conversely, refers to determining the category of new inputs based on established decision boundaries, with accuracy measured by the similarity between the model’s prediction and the actual outcome.

Determining a decision boundary during AI model training involves measuring how the predicted results differ from true values, through a metric known as a \emph{loss function}, which will be explained in detail shortly. The optimal decision boundary minimizes this loss, resulting in high inference accuracy and effective resolution of real-world problems. However, as real-world problems grow more complex, clearly separating data groups in low-dimensional spaces becomes challenging. For instance, as illustrated in Figure \ref{fig:db_2}, a two-dimensional plane (2D) may distinguish two groups easily using a simple linear boundary in ideal cases, but such simple scenarios are uncommon in practice. To address this issue, AI models expand their parameter spaces into higher dimensions, such as 3D and 4D (Figures \ref{fig:db_3} and \ref{fig:db_4}), allowing clearer and more precise representation of complex data. Although the illustrated examples focus on a limited number of dimensions, modern large-scale AI models use parameter spaces consisting of billions to trillions of dimensions \cite{fewshot,efficientlarge,rae2021scaling,chowdhery2023palm,le2023bloom,llama3herdmodels}. These extensive parameter spaces greatly enhance the models' ability to solve previously intractable problems but also significantly increase memory and communication requirements.

Defining effective decision boundaries also requires knowledge of the characteristics and roles of loss functions. Commonly used loss functions include mean squared error loss \cite{rumelhart1986learning,kim2021comparing,kato2021mse,ren2022balanced} and cross-entropy loss \cite{kullback1951information,zhang2018generalized,ho2019real,gordon2020uses,mao2023cross}, as mentioned earlier. As illustrated in Figure \ref{fig:lf}, the goal of AI model training is to configure model parameters such that the loss function achieves its minimum value. However, even when the dimensions of data representation could be optimally determined, the sheer number of parameters in AI models often makes it impractical to analytically find an exact minimum. Therefore, approximate algorithms such as \emph{gradient descent} \cite{fletcher1963rapidly,rumelhart1986learning,lecun1998gradient,andrychowicz2016learning} are widely employed. Gradient descent starts from arbitrary initial parameter values and iteratively updates these parameters based on gradient information, moving toward lower loss values. The algorithm terminates upon reaching a point where the gradient equals zero, indicating the minimum of the loss function.

Despite its effectiveness, gradient descent exhibits certain limitations. The step size, or the magnitude of parameter adjustments, is important: excessively large steps can overshoot minima, while excessively small steps prolong convergence. Moreover, initiating the search from arbitrary points may lead to convergence at local minima rather than the global minimum. To address these limitations, various optimization techniques, such as stochastic gradient descent (SGD \cite{rumelhart1986learning,gardner1984learning,amari1993backpropagation,eon1998online,krizhevsky2012imagenet,hardt2016train,li2019convergence,tian2023recent}) or adaptive moment estimation (Adam \cite{kingma2014adam,zhang2018improved,bock2018improvement,mehta2019cnn,yi2020effective}), have been developed, dynamically adjusting step sizes and search directions. Note that these optimizations, while varied, ultimately share the common goal of adjusting the model parameters so that the loss function reaches its minimum, which aligns with the aforementioned overall objective of AI training.

\begin{figure}[t!]
    \centering
    \includegraphics[width=0.9\linewidth]{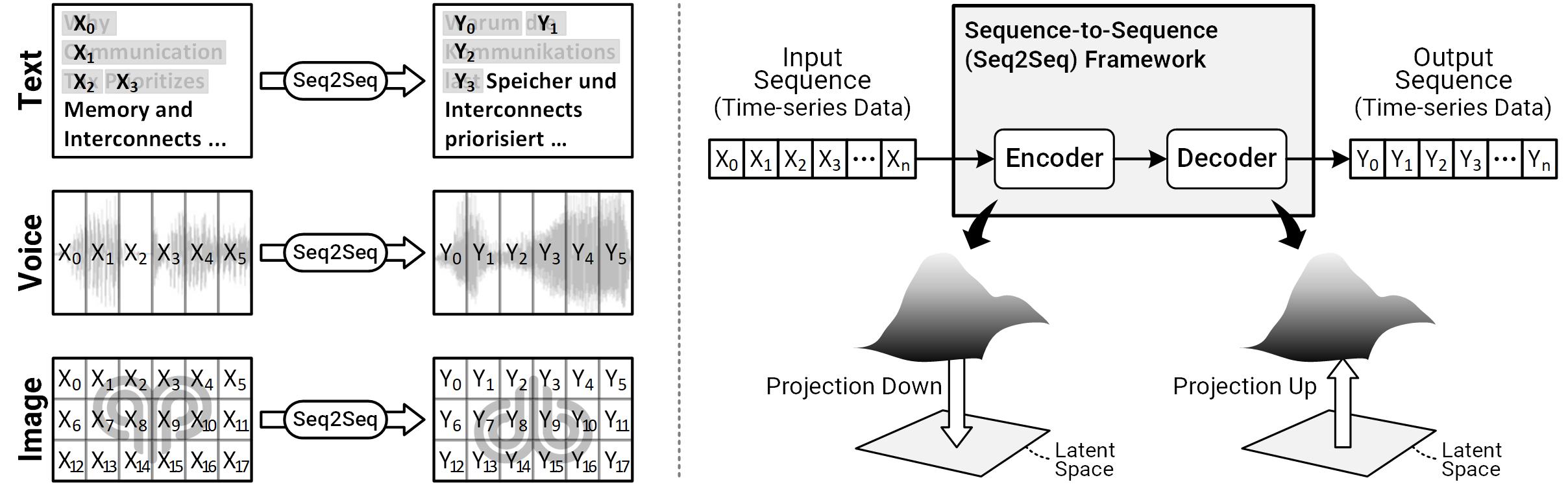}
    \begin{subfigure}{\linewidth}
        \begin{tabularx}{\textwidth}{
            p{\dimexpr.40\linewidth-2\tabcolsep-1.3333\arrayrulewidth}
            p{\dimexpr.60\linewidth-2\tabcolsep-1.3333\arrayrulewidth}
            }
            \caption{Time-series data sequences.} \label{fig:seq2seq1} &
            \caption{Fundamental concept of Seq2Seq model.} \label{fig:seq2seq2}
        \end{tabularx}
    \end{subfigure}
    \caption{Sequence-to-sequence (Seq2Seq) framework.}
    \label{fig:seq2seq}
\end{figure}

\paragraph{Early implementations and limitations: Recurrent neural networks.}

In practical AI applications, input data encompass diverse modalities such as images, audio, video, and text. Despite these variations, many real-world datasets possess temporal or sequential characteristics, which allow them to be generalized and analyzed as time-series data. As shown in Figure \ref{fig:seq2seq1}, transforming different types of input data into structured numerical sequences enables AI models to capture temporal dependencies essential for accurate predictions and sophisticated decision-making. Sequence modeling techniques have emerged specifically to handle these structured sequences.

One influential model designed to address this requirement is the Sequence-to-Sequence (Seq2Seq) framework. Introduced in 2014~\cite{seq2seq1}, Seq2Seq remains fundamentally significant today as the foundation for many advanced sequence modeling approaches. This section briefly describes its core structure. Specifically, the concepts of Encoding, Ordering, and Decoding -- essential for understanding Seq2Seq -- are schematically illustrated in Figure \ref{fig:seq2seq2} and discussed below:

\begin{enumerate}
  \item \textbf{Encoding:} The encoder (``projecting down'') compresses input sequences into concise numerical representations, known as latent \textit{vectors}. These vectors can capture essential context and temporal relationships within data, facilitating efficient sequence analysis.

  \item \textbf{Ordering:} Maintaining sequential order within internal processing (``sequential processing'') ensures each step incorporates context from preceding elements. This ordering preserves coherence and accuracy throughout the generated outputs.

  \item \textbf{Decoding:} The decoder (``projecting up'') reconstructs these internal representations into understandable output sequences. It leverages context-rich latent vectors to generate coherent outputs, such as translations, summaries, or predictive forecasts.
\end{enumerate}

The encoder-decoder framework in Seq2Seq models handles sequences of varying lengths and complexities. By compressing the entire input sequence into concise internal representations, the encoder ensures that contextual and temporal information is preserved. The decoder then leverages these representations to generate accurate and coherent output sequences, bridging the gap between complex data patterns and human interpretability. Thanks to this structured methodology, Seq2Seq models have become effective tools for various practical tasks, including language translation, summarization, speech recognition, anomaly detection, and predictive analytics.

\begin{figure}[t!]
    \centering
    \includegraphics[width=0.9\linewidth]{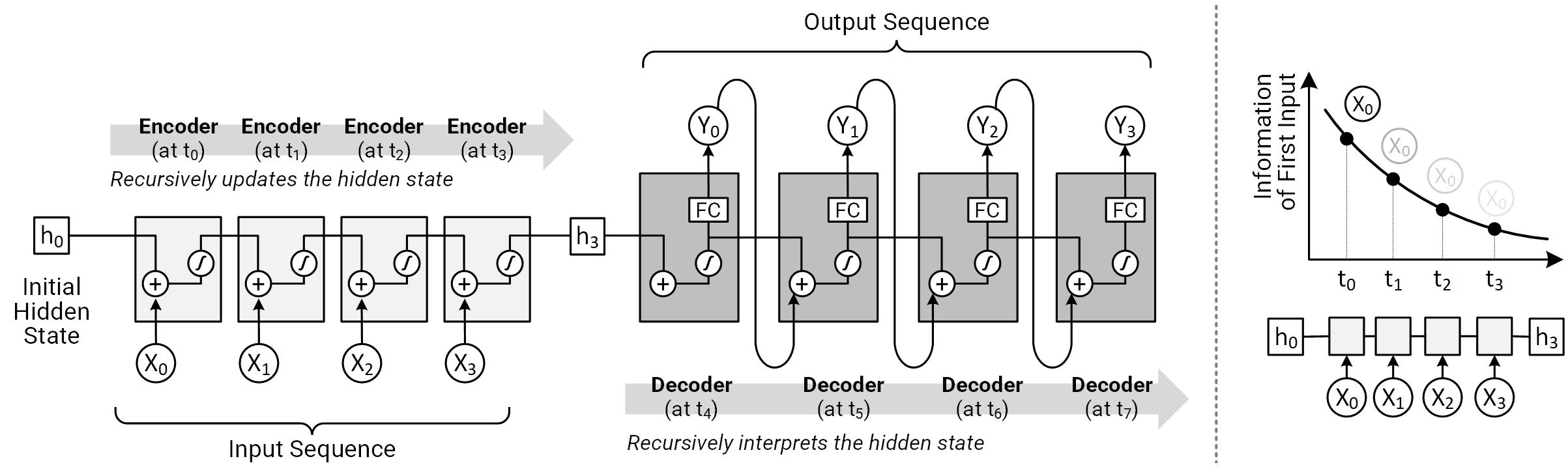}
    \begin{subfigure}{\linewidth}
        \begin{tabularx}{\textwidth}{
            p{\dimexpr.40\linewidth-2\tabcolsep-1.3333\arrayrulewidth}
            p{\dimexpr.35\linewidth-2\tabcolsep-1.3333\arrayrulewidth}
            p{\dimexpr.30\linewidth-2\tabcolsep-1.3333\arrayrulewidth}
            }
            \caption{Encoding process.} \label{fig:rnn_encoding} &
            \caption{Decoding process.} \label{fig:rnn_decoding} &
            \caption{Vanishing gradient.} \label{fig:vanishing_grad}
        \end{tabularx}
    \end{subfigure}
    \caption{RNN-based Seq2Seq architecture and limitation.}
    \label{fig:tray}
\end{figure}

Initial implementations of Seq2Seq models relied on RNNs, which naturally handle sequential data. As shown in Figure \ref{fig:rnn_encoding}, in RNN-based Seq2Seq architectures, the encoder's primary role is to read and understand the input sequence as explained previously. It achieves this by processing each data point sequentially and maintaining an internal memory, known as the \textit{hidden state}, which summarizes the essential context of previously seen data points. At each step, the encoder combines the current data point with the previous hidden state through mathematical operations involving non-linear functions (e.g., tanh and ReLU \cite{activation,relu}). These non-linear functions enable the network to capture complex and non-linear relationships within the data. Without such non-linear transformations, the network's ability to model intricate patterns and temporal dependencies in sequential data would be limited \cite{nonlin,nonlin1,nonlin2}. This sequential updating allows the encoder to preserve essential context and information from all preceding data points, gradually summarizing the entire input sequence into a condensed, meaningful internal representation.

As shown in Figure \ref{fig:rnn_decoding}, once the encoder completes this task of compressing the sequence, the decoder reconstructs the summarized representation into a comprehensible and structured output sequence. The decoder functions by generating output data points in a sequential manner, using not only the condensed hidden state representation but also its previously generated outputs as inputs for each subsequent step. To perform this reconstruction, the decoder employs \textit{fully connected} (FC) layers \cite{attentionall,fclayer}. These FC layers are specialized neural network layers designed to map abstract and compressed internal representations back into understandable, real-world outputs. Each layer transforms the internal information into a clearer and more concrete form, allowing the decoder to produce accurate and coherent sequences such as translations, summaries, or predictions.

Despite their initial success, RNN-based Seq2Seq models unfortunately faced serious limitations, which are illustrated by Figure \ref{fig:vanishing_grad}. A prominent challenge is the ``vanishing gradient problem \cite{vanishingprob,vanishingprob1},'' where information from early sequence elements gradually diminishes during training, degrading the model's ability to capture long-range dependencies. Simply put, this is analogous to the human phenomenon of forgetting older memories, or forgetfulness. In addition, \textit{due to their intrinsic sequential nature, RNNs cannot leverage parallel computing technologies}, limiting their scalability and computational efficiency. These constraints motivated the development of more advanced architectures designed to overcome the shortcomings of RNNs by enhancing long-range context retention and enabling parallel computation.

\subsection{A Paradigm Shift in Sequence Modeling}
\label{subsection:2_2}
\paragraph{Integration of attention mechanisms.} To overcome the limitations of conventional RNN architectures, especially the vanishing gradient problem and poor scalability, researchers introduced \textit{attention} mechanisms \cite{attentionbased,bahdanau2014neural,attentionall}, marking a significant evolution in sequence modeling. The fundamental idea behind attention is to allow the model to focus on the most relevant parts of the input sequence when generating each element of the output. Instead of compressing the entire input sequence into a single fixed-size hidden state (as done in traditional RNNs), attention mechanisms maintain access to all hidden states produced by the encoder and selectively assign weights to them based on their relevance to the current decoding step.

\begin{figure}[t!]
    \centering
    \includegraphics[width=0.9\linewidth]{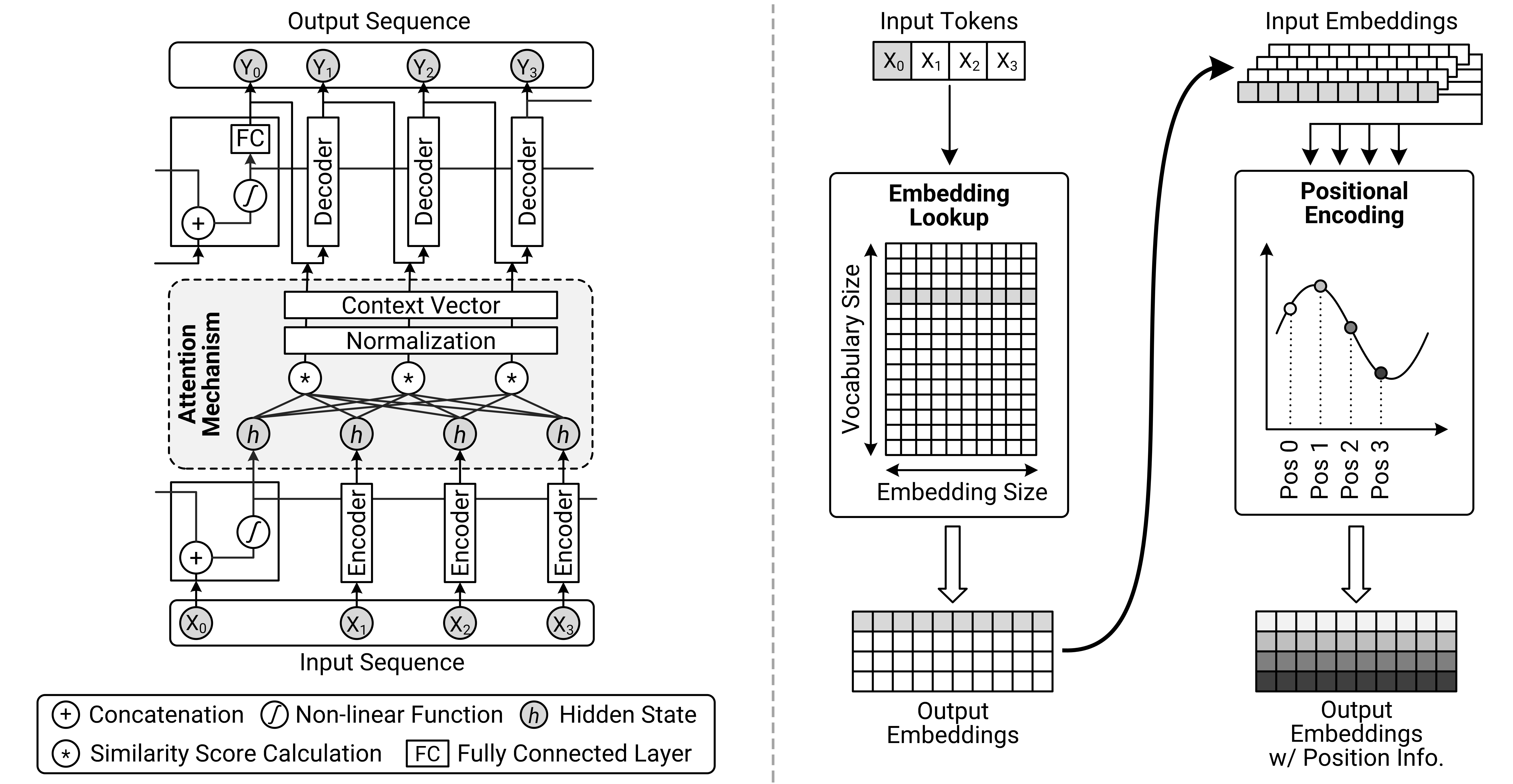}
    \begin{subfigure}{\linewidth}
        \begin{tabularx}{\textwidth}{
            p{\dimexpr.51\linewidth-2\tabcolsep-1.3333\arrayrulewidth}
            p{\dimexpr.23\linewidth-2\tabcolsep-1.3333\arrayrulewidth}
            p{\dimexpr.26\linewidth-2\tabcolsep-1.3333\arrayrulewidth}
            }
            \caption{Attention mechanism over RNNs.} \label{fig:attention_rnn} &
            \caption{Embedding layer.} \label{fig:embedding_trans} &
            \caption{Positional encoding.} \label{fig:position_trans}
        \end{tabularx}
    \end{subfigure}
    \caption{Attention-based RNNs and Transformer architecture.}
    \label{fig:attention_rnn_transformer}
\end{figure}

As shown in Figure \ref{fig:attention_rnn}, at each decoding time step, the attention mechanism computes a similarity score between the decoder's current hidden state and each encoder hidden state. These scores are then normalized to produce attention weights, typically using a softmax function \cite{softmax,softmax1,softmax2}. The decoder uses these weights to compute a weighted sum of the encoder hidden states, resulting in a context vector that captures the most relevant information from the input sequence. This vector is then used, along with the decoder's previous outputs, to generate the next token in the output sequence. Here, a \textit{token} refers to basic data units such as words, subwords, or characters.

This process provides two main advantages. First, attention mechanisms allow models to retain and access information from ``any part of the input sequence'', regardless of its length. By eliminating the need to compress all information into a fixed-size vector, this approach addresses the vanishing gradient problem and provides models with robust, content-based memory, enabling efficient referencing of relevant sequence elements during output generation \cite{lin2017structured,attentionbased,bahdanau2014neural,attentionall}. Second, attention enhances model interpretability. Attention weights indicate which parts of the input the model focuses on when generating outputs, offering intuitive insights into the reasoning and decision-making processes of complex models \cite{weight,weight1,weight2}.

Unfortunately, despite these improvements, models that used attention mechanisms within RNNs still inherited the sequential nature of computation from their underlying architecture (i.e., ordering). Each time step had to be computed in order, which limited the ability to leverage modern parallel processing hardware. This motivated the development of fully attention-based architectures, culminating in the Transformer model, which reshaped the landscape of sequence modeling.

\paragraph{From recurrence to parallelism: Transformer revolution.} A major breakthrough in sequence modeling occurred in 2017 with the introduction of the Transformer architecture \cite{attentionall}, marking a fundamental departure from the recurrent computational structure intrinsic to RNNs and conventional Seq2Seq models. Transformers entirely abandoned recurrence, adopting instead a fully attention-based design. In particular, the Transformer utilizes \emph{self-attention} \cite{selfattention,selfattention1,selfattention2,selfattention3,selfattention4}, which computes interactions among all positions within a sequence simultaneously. Unlike traditional attention mechanisms embedded within RNNs, self-attention independently calculates attention scores across all sequence positions in parallel, without relying on previously computed states or sequential dependencies between positions. This independence arises because each position's attention calculation is based directly on fixed input representations, known as \textit{embeddings}. By eliminating sequential processing constraints, Transformers exploit modern parallel processing hardware, boosting computational efficiency and scalability \cite{kitaev2020reformer,longertransformer,megatron}.

This parallel processing capability is enabled by the Transformer's embedding layers, which project discrete tokens into high-dimensional continuous vector spaces. As shown in Figure \ref{fig:embedding_trans}, these embeddings serve as input representations for the model, enabling simultaneous processing of all tokens in a sequence. Unlike RNN-based Seq2Seq models, which must process inputs sequentially due to their recurrent structure, Transformers operate directly on embeddings in parallel. Each token's embedding distinctly captures semantic meaning, allowing the self-attention mechanism to relate tokens across the entire sequence in parallel; the details of the self-attention will be explained, shortly. This simultaneous token-level interaction removes the step-by-step dependency inherent in recurrent models, enabling full parallelization of computations across all tokens.

However, the removal of explicit sequential structure introduces a new challenge: Transformers lack a sense of order in the input sequence, also referred to as ``ordering''. In RNNs, the order is preserved due to their sequential processing nature. To compensate for this, the Transformer introduces \textit{positional encoding}, which is a mechanism that injects information about the relative or absolute position of tokens directly into their embeddings (cf. Figure \ref{fig:position_trans}). These encodings are added to the input embeddings before they are processed by the self-attention layers, ensuring that the model can distinguish between tokens based on their positions in the sequence. Positional encoding can be implemented using fixed sinusoidal functions or learned embeddings \cite{positional,positional1,positional2}, both of which allow the model to capture sequence order while preserving the ability to compute in parallel.

This combination of embedding-based input representation and positional encoding enables Transformers to model long-range dependencies while fully leveraging parallel computation. As a result, the Transformer architecture achieves superior performance and scalability compared to RNN-based models. It has become the foundation for virtually all modern large-scale language models and sequence modeling tasks.

\begin{figure}[t!]
    \centering
    \includegraphics[width=\linewidth]{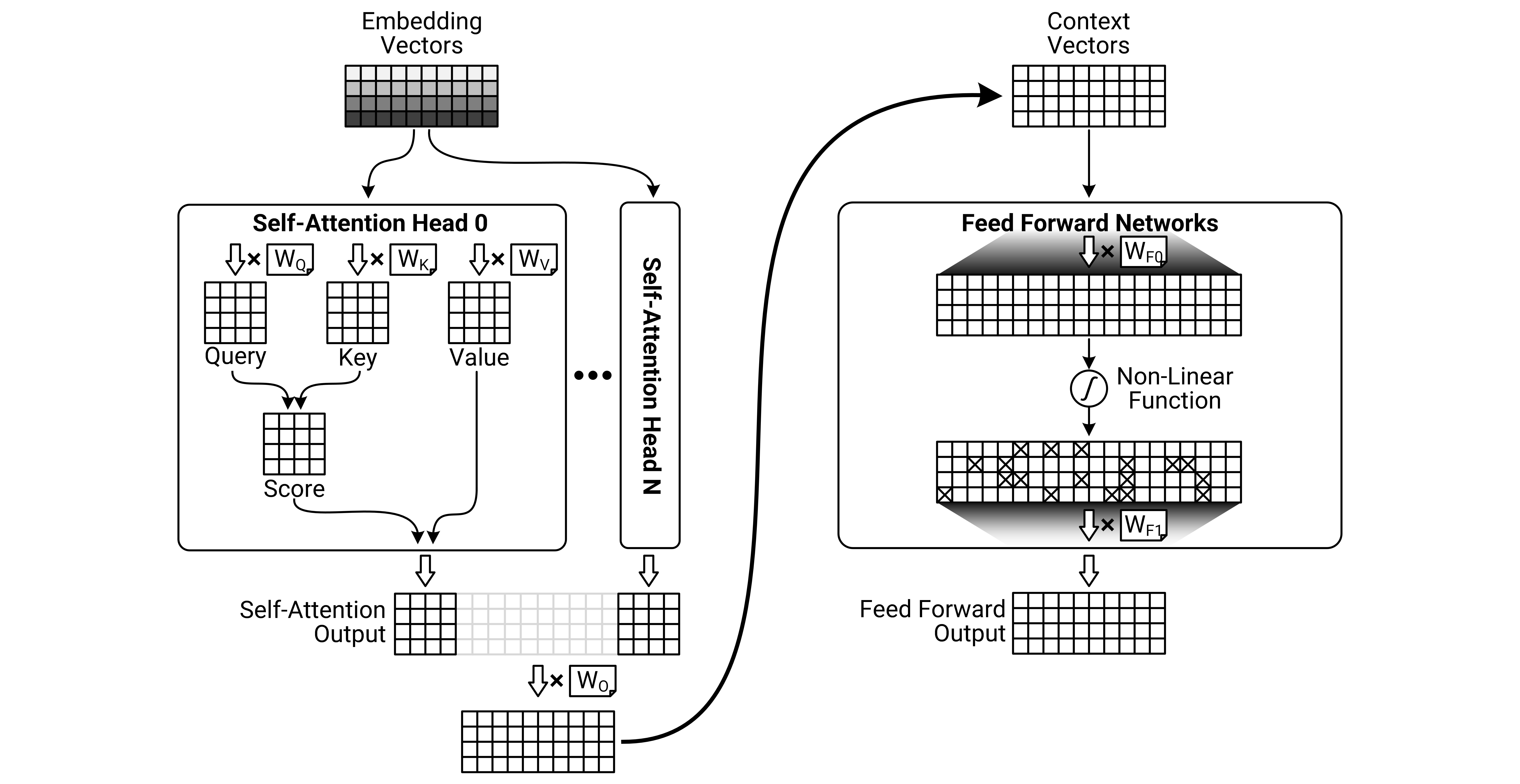}
    \begin{subfigure}{\linewidth}
        \begin{tabularx}{\textwidth}{
            p{\dimexpr.55\linewidth-2\tabcolsep-1.3333\arrayrulewidth}
            p{\dimexpr.32\linewidth-2\tabcolsep-1.3333\arrayrulewidth}
            }
            \caption{Multi-head self-attention mechanism.} \label{fig:self_attention} &
            \caption{Feed-Forward networks (FFNs).} \label{fig:ffn}
        \end{tabularx}
    \end{subfigure}
    \caption{Transformer layer operations: self-attention and FFNs.} \label{fig:transformer_operation}
\end{figure}

\paragraph{Self-attention and mixture of experts: Enhancing sequence understanding.} The self-attention mechanism is a cornerstone of the Transformer architecture, enabling the model to simultaneously relate different parts within a sequence. Unlike traditional attention mechanisms, which typically align separate input-output sequences, self-attention directly computes interactions among tokens within a single sequence. As depicted in Figure \ref{fig:self_attention}, this intra-sequence interaction is facilitated by three distinct vectors, \textit{query}, \textit{key}, and \textit{value}, each derived from individual token embeddings through separate linear transformations using learnable parameter matrices ($W^Q$, $W^K$, and $W^V$). Specifically, query vectors (Q) represent the information that each token seeks, key vectors (K) denote information each token provides, and value vectors (V) encapsulate the actual content shared by tokens. The self-attention process involves comparing each query vector with all key vectors using scaled dot-product operations, producing attention scores reflecting token similarities or relevancies. These scores are scaled by the square root of the key dimension, ensuring numerical stability during training, and subsequently normalized via a softmax function to yield attention weights. The context-aware representation for each token is then computed as a weighted sum of all value vectors, with weights determined by these attention scores.

Transformer architectures further enhance representational capability through \emph{multi-head attention} \cite{attentionall,li2021diversity,li2018multi} (or grouped query attention \cite{gqa1,gqa2,gqa3}), where several parallel self-attention computations (\emph{heads}) operate concurrently. Each attention head employs independent parameter matrices to generate distinct query, key, and value vectors, enabling the model to simultaneously capture diverse relationships and nuanced token interactions. This multi-head design broadens the model's ability to capture complex and long-range dependencies, thus enhancing overall model accuracy and contextual understanding without sequential processing constraints.

\begin{figure}[t!]
    \centering
    \includegraphics[width=\linewidth]{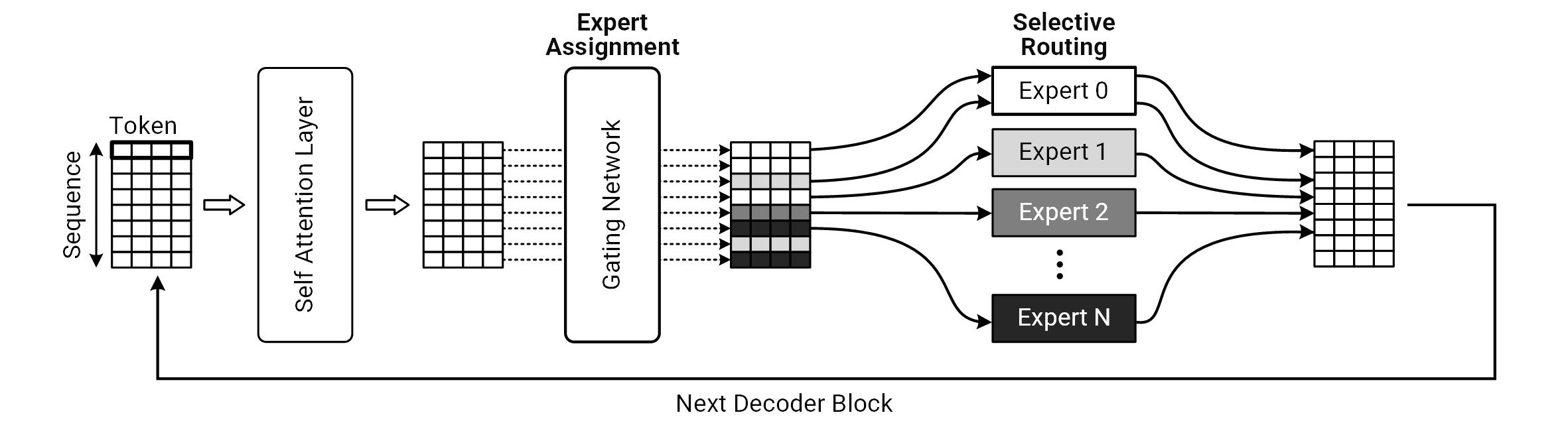}
    \caption{Mixture of experts (MoE) architecture.}
    \label{fig:moe_architecture}
\end{figure}

In addition, as shown in Figure \ref{fig:ffn}, Transformers incorporate specialized \emph{Feed-Forward Networks} (FFNs) within their architecture, distinct from the FC layers typically mentioned in classical neural networks. While self-attention can capture the contextual relationships among tokens, each token representation requires further refinement. To this end, FFNs apply separate, position-wise nonlinear transformations independently to each token embedding, complementing the self-attention mechanism. Specifically, each FFN comprises two linear transformations separated by a non-linear activation function (e.g., ReLU or GELU~\cite{relu,gelu}). The first linear transformation projects input embeddings into a ``higher'' dimensional representation space, enabling the model to capture complex, non-linear data patterns. The second linear layer subsequently projects these refined representations back to the ``original'' dimensional space. Consequently, FFNs enhance the context-aware embeddings produced by self-attention. As these FFN layers operate independently at each token position, they preserve the inherent parallel processing advantages characteristic of the Transformer architecture.

Even though FFNs refine token-level representations produced by self-attention, further improvements in handling diverse and complex data require increased model capacity and computational efficiency. To address this, Transformers incorporate advanced structures such as the \emph{Mixture of Experts} (MoE) architecture~\cite{outrageouslylarge,gshard,swtransformer}. Figure~\ref{fig:moe_architecture} illustrates an overview of MoE, designed to enhance model capacity and efficiency. An MoE model consists of multiple specialized neural networks, called \textit{experts}, each trained to address distinct data types or patterns. During inference, a gating network dynamically routes input tokens to selected experts according to learned patterns, activating only the necessary experts to optimize computational resources. This selective expert activation not only enhances computational efficiency but also improves model accuracy by enabling experts to specialize in particular subtasks. As a result, MoE-based Transformers achieve high scalability, robustness, and generalization, making them effective for large-scale and diverse datasets~\cite{swtransformer,gshard,moerobust,zoph2022designing,moe1}.

Note that MoE enables high computational efficiency through parallel processing by multiple independent experts, but aggregating the intermediate outputs of these experts necessitates significant inter-expert communication. Moreover, because the aggregated MoE outputs serve as inputs for subsequent layers such as self-attention, a sequential dependency exists; the self-attention can only begin processing once MoE aggregation is complete. As a result, efficient \textit{AI infrastructure design must include high-speed interconnects to facilitate rapid communication} and support this essential sequential processing order.

\subsection{From Transformers to Large Language Models}
\label{subsection:2_3}
The introduction of Transformers marked a pivotal shift in sequence modeling by replacing the recurrent computations in RNNs with parallelizable self-attention mechanism. This innovation enabled models to concurrently analyze relationships among all tokens in a sequence, accelerating computation and supporting the development of deeper, more complex neural architectures. This transition also served as a critical turning point for the widespread adoption of hardware accelerators such as GPUs, as the underlying computation patterns aligned well with massively parallel hardware.

\begin{figure}[t!]
    \centering
    \includegraphics[width=\linewidth]{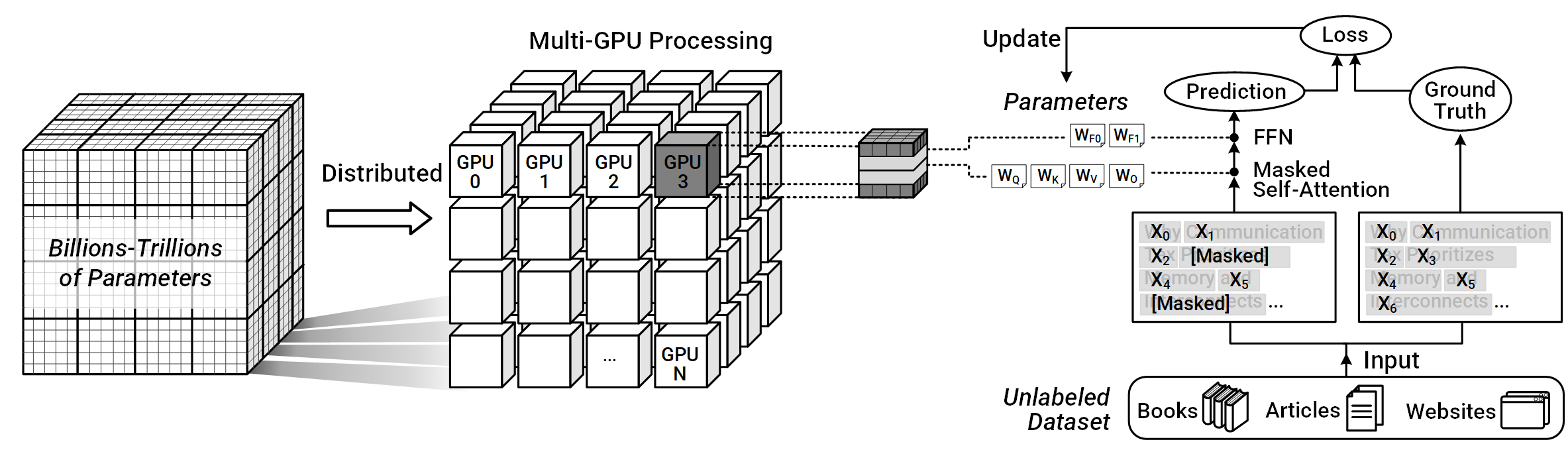}
    \caption{LLM pre-training through parallel processing.}
    \label{fig:llm_train}
\end{figure}

\paragraph{Massive-scale training and parameter optimization.} Building upon the scalability of Transformers, researchers developed LLMs, extending model capabilities by increasing their parameter counts and training them extensively on massive and diverse datasets. Modern LLMs, exemplified by models such as GPT-4 Turbo \cite{GPT4turbo,GPT4turbo1} and Google's Gemini \cite{Gemini2,geminiteam}, can contain billions or even trillions of parameters. These parameters, adjustable internal settings of neural network layers, are optimized during a comprehensive pre-training phase involving vast textual datasets, including books, scholarly articles, websites, and social media content \cite{Bert,roberta,xlnet}. This pre-training commonly employs self-supervised learning \cite{spanbert,albert}, an approach that enables models to derive meaningful representations from unlabeled data in an automatic manner. Typically, this involves predicting masked words, sentences, or segments from the given context. As shown in Figure \ref{fig:llm_train}, through extensive pre-training, LLMs encode linguistic knowledge, contextual comprehension, syntactic and semantic nuances, as well as general world knowledge directly into the model's internal parameters, specifically within the self-attention mechanism (query, key, and value vectors) and FFN layers (weights and biases).

Note that training LLMs requires intensive computation and significant parallel processing across many GPUs \cite{efficientlarge,gpipe}. These demands increase further with optimization techniques like mixed-precision arithmetic, distributed training across multiple nodes, and complex gradient synchronization. The parallel structure of Transformers addresses these issues by handling large parameter sets and extensive datasets across multiple GPU clusters in parallel. In contrast to the previous sequential models limited by recurrent operations, Transformers fully utilize modern parallel processing hardware, enabling scalable training. However, large-scale training using many GPUs typically takes weeks or even months of continuous operation \cite{longtrain,longtrain1,longtrain2}. \textit{This extended training time arises from large memory requirements and frequent data synchronization among GPUs}, highlighting key infrastructure challenges related to scalability, memory capacity, and interconnect design.

\paragraph{Ensuring coherence and generalization.} Many modern LLMs adopt an \textit{auto-regressive} approach during training and inference~\cite{xlnet,gpurpose,zhang2022opt}. The auto-regressive method sequentially predicts each token based solely on previously generated tokens, capturing the inherent sequential dependencies in linguistic data. Specifically, when generating a sentence, the model determines each word using only prior context, without incorporating information from subsequent tokens. This characteristic enables the auto-regressive approach to maintain logical coherence and contextual accuracy even in the absence of future context.

\begin{figure}[t!]
    \centering
    \includegraphics[width=0.9\linewidth]{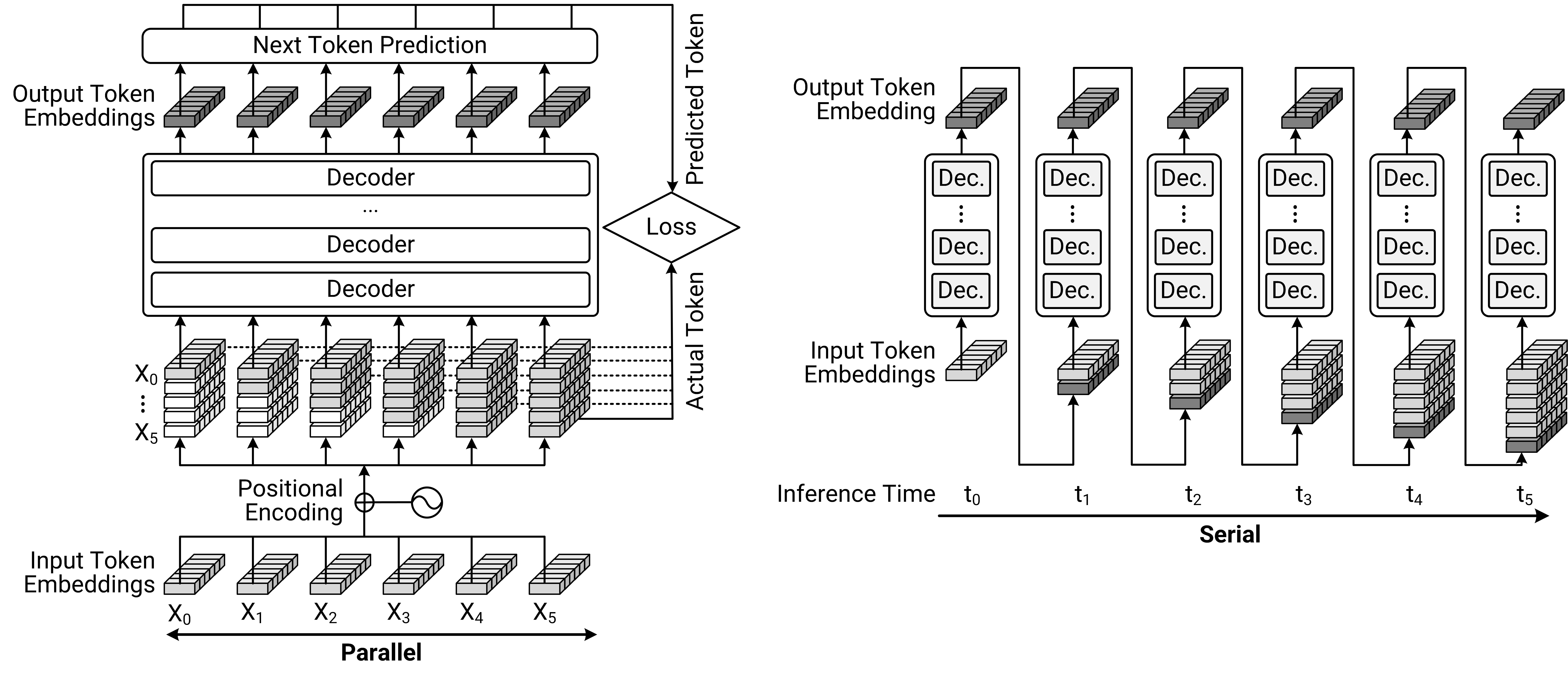}
    \begin{subfigure}{\linewidth}
        \begin{tabularx}{\textwidth}{
            p{\dimexpr.49\linewidth-2\tabcolsep-1.3333\arrayrulewidth}
            p{\dimexpr.49\linewidth-2\tabcolsep-1.3333\arrayrulewidth}
            }
            \caption{Training with ground-truth tokens.} \label{fig:autoreg_train} &
            \caption{Inference using previously generated tokens.} \label{fig:autoreg_infer}
        \end{tabularx}
    \end{subfigure}
    \caption{Auto-regressive model workflow.}
    \label{fig:autoreg}
\end{figure}

Figure \ref{fig:autoreg_train} illustrates the training process, in which the model learns to predict the next token based on preceding tokens within the provided sequences. This sequential approach helps the model learn patterns in grammar, syntax, and logical progression, modeling temporal dependencies and maintaining coherence within generated texts. During inference, as shown in Figure \ref{fig:autoreg_infer}, auto-regressive models generate tokens one at a time. Each new token then becomes part of the context for subsequent predictions. Although this sequential mechanism ensures coherent and contextually consistent outputs, it limits parallel computation, reducing inference speed relative to parallelized generation methods.

Despite these computational constraints, auto-regression remains widely used due to its proven capability to represent complex linguistic dependencies and produce accurate outputs, essential for high-quality language generation tasks. To mitigate inference limitations, modern LLMs also emphasize their extensive pre-training and strong generalization capabilities. By learning from large-scale and diverse textual datasets, these models form comprehensive internal language representations, enabling flexible application to various downstream tasks, often requiring minimal or no additional specialized training in a scenario known as zero-shot or few-shot learning \cite{fewshot,zeroshot,zeroshot1}. Such generalization expands the applicability of LLMs beyond traditional language tasks into multimodal areas, including image generation, video synthesis, audio processing, and interactive conversational systems.

\paragraph{Reducing redundancy and improving reliability in LLM inference.} As LLMs have become widely deployed across various applications, two critical techniques, \textit{Key-Value} (KV) caching \cite{attentionall,radford2019language,kvcache,kvcache1} and \textit{Retrieval-Augmented Generation} (RAG) \cite{rag1,rag2,rag3}, have been developed to overcome significant computational and accuracy-related challenges in LLM inference processes.

\begin{figure}[t!]
    \centering
    \includegraphics[width=\linewidth]{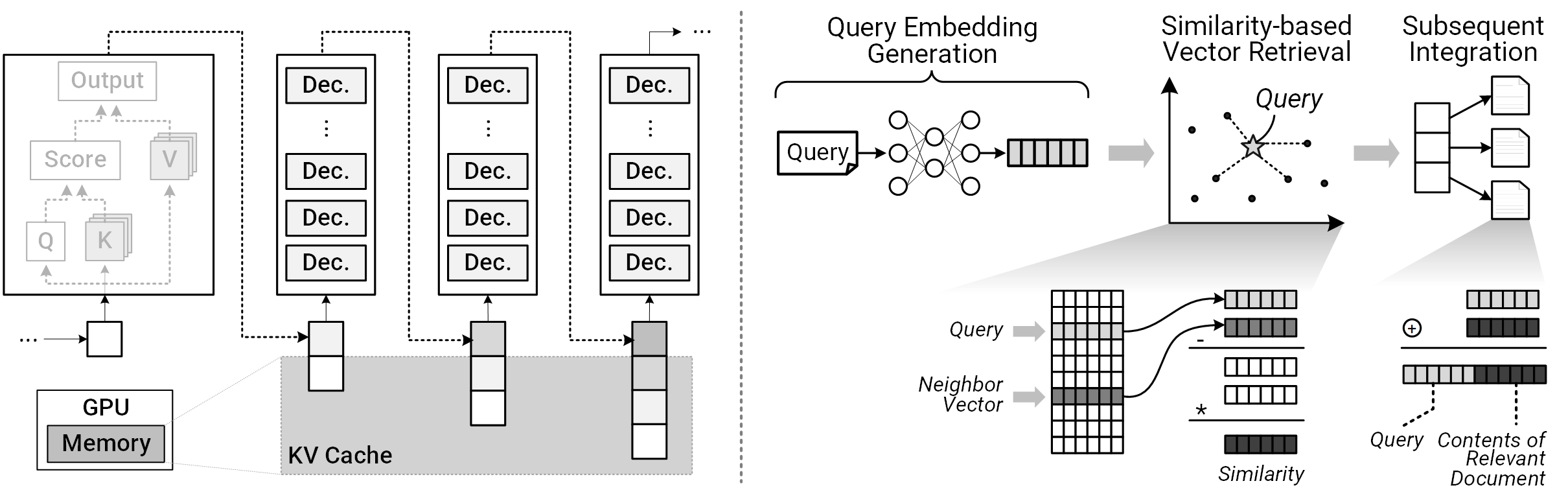}
    \begin{subfigure}{\linewidth}
        \begin{tabularx}{\textwidth}{
            p{\dimexpr.50\linewidth-2\tabcolsep-1.3333\arrayrulewidth}
            p{\dimexpr.50\linewidth-2\tabcolsep-1.3333\arrayrulewidth}
            }
            \caption{Caching key-value pairs in GPU memory.} \label{fig:optim_kv} &
            \caption{Searching external database.} \label{fig:optim_rag}
        \end{tabularx}
    \end{subfigure}
    \caption{Inference Optimization Techniques.}
    \label{fig:optim}
\end{figure}

KV caching addresses computational inefficiencies inherent in the self-attention mechanism of LLMs. Due to the auto-regressive inference approach, where each token generation depends on previously generated tokens, LLMs must repeatedly calculate self-attention scores involving all previously processed tokens at every generation step. Without optimization, this results in redundant and repeated calculations, slowing inference speed. As shown in Figure \ref{fig:optim_kv}, KV caching resolves this by storing the computed attention scores as key-value pairs directly in GPU memory after their initial calculation. Once stored, these cached results can be directly reused for subsequent inference steps without additional computation. This can reduce redundant computational overhead and accelerate the inference process, particularly for longer input sequences. However, this efficiency gain comes at the cost of increased memory demands. Depending on the model size, token length, and inference complexity, KV caching can occupy between 30\% and 85\% of the available GPU memory \cite{hooper2024kvquant,gpumem1,gpumem2,gpumem3}, considerably intensifying memory utilization and often surpassing the capacity of individual GPU modules.

On the other hand, RAG targets the inherent limitation of LLMs known as model hallucinations \cite{hallucination,hallucination1}, which are situations in which models produce plausible yet factually incorrect or contextually irrelevant outputs. Such inaccuracies arise because LLMs rely exclusively on internal knowledge learned during their training phase, lacking real-time or updated external context. RAG enhances model reliability by incorporating external knowledge retrieval directly into the inference workflow. When an input query is received, a RAG-equipped LLM first searches an external knowledge database, implemented as a specialized vector database \cite{vectordb,vectordb1,vectordb3} or retrieval system \cite{retrievalsys,retrievalsys1,retrievalsys2}, for contextually relevant information (cf. Figure \ref{fig:optim_rag}). This retrieved context is then combined with the original input query, providing the model with accurate, externally verified information from which to generate its final response. Although this can reduce hallucinations and improves factual correctness, it introduces additional computational steps, including query embedding generation, similarity-based vector retrieval, and subsequent integration of retrieved information into the inference process. Consequently, RAG imposes computational complexity and requires significant memory capacity for maintaining large-scale vector databases. Moreover, network latency and bandwidth become critical performance factors, as rapid and reliable retrieval from external sources impacts response accuracy and inference latency.

Note that KV caching and RAG are all essential to address the crucial bottlenecks within LLM inference. While KV caching optimizes inference speed by minimizing redundant computations within the self-attention mechanism, RAG enhances output reliability and accuracy by leveraging external knowledge sources. Despite of these advantages, \textit{deploying these inferencing techniques with LLM amplifies demands on GPU memory, computational resources, network bandwidth, and storage infrastructure}, further emphasizing the necessity for highly scalable and composable data center architectures capable of supporting diverse and intensive LLM workload requirements.

\section{Scaling LLMs: Multi-Accelerator and Data Center Deployments}
\label{section:3}
Modern data centers have evolved to handle diverse AI workloads, including recommendation systems \cite{recomm1,recomm2,recomm3}, ranking algorithms \cite{ranking1,ranking2}, and vision models \cite{vision,vision1}. However, LLM workloads uniquely stress infrastructure due to their intense memory and communication needs. In this section we first examine how fundamental LLM concepts map onto multi-accelerator systems. We then analyze architectural and modular strategies adopted by contemporary data centers utilizing thousands of GPUs or accelerators. Throughout this discussion, GPU and accelerator terms are used interchangeably.

At the end of this section, we discuss the limitations of tightly integrated CPU-GPU architectures, which restrict scalability, flexibility, and efficient resource utilization. Addressing these constraints to meet large-scale AI demands necessitates adopting modular, independently scalable designs for CPUs, GPUs, memory, and networking components.

\subsection{Mapping and Challenges of Deploying LLMs on Multi-Accelerator Systems}
\label{subsection:3_1}
The exponential scaling of modern LLMs, particularly those based on Transformer architectures, surpasses the memory and computational capabilities of individual GPUs \cite{megatron,knowledgepack,scaling}. This necessitates distributing large models across multiple GPUs. Each GPU within these multi-GPU setups manages specific subsets of parameters and computational tasks, enabling effective parallelization and distributed training.

\begin{figure}[t!]
    \centering
    \includegraphics[width=\linewidth]{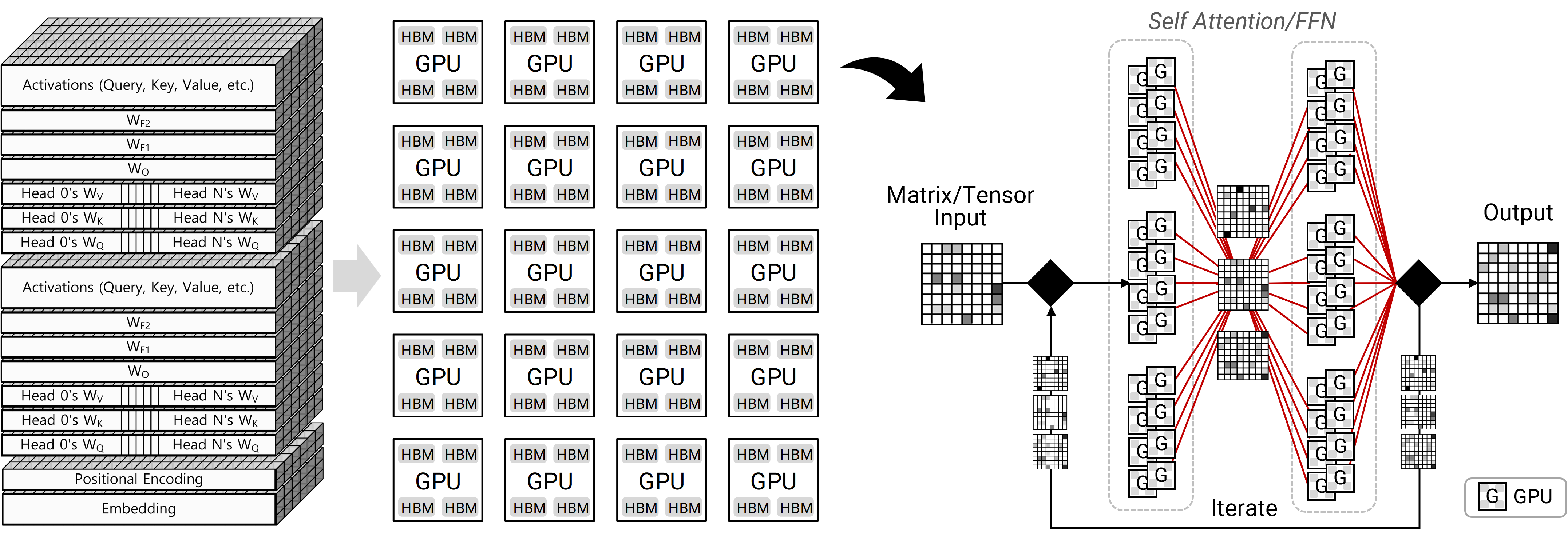}
    \caption{Transformer-based model partitioning and synchronization overhead.}
    \label{fig:model_partitioning}
\end{figure}

\paragraph{Multi-GPU LLM training: Model partitioning, parallelization, and overheads.} The primary challenge in multi-GPU LLM training is efficiently partitioning and synchronizing extensive model parameters, activations, and gradients across GPUs, ensuring coherent and effective distributed computation.

Figure~\ref{fig:model_partitioning} illustrates the partitioning strategy and tensor synchronization across GPUs in Transformer-based models, emphasizing the frequent exchange of outputs from both self-attention and FFN computations. The self-attention mechanism within Transformer architectures computes interactions across all tokens in a sequence. At first glance, it may appear that GPUs must generate and exchange partial query, key, and value vectors. However, each GPU can independently compute self-attention using only its assigned partial vectors, enabled by multi-head attention and grouped query attention, which partition one large attention layer into multiple parallel smaller layers. Nonetheless, periodic synchronization of these computed vectors and their gradients remains necessary to ensure global coherence and consistency across the distributed architecture \cite{synchro,synchro1,gpumem3}. This synchronization step is essential for accurate gradient computation and parameter updates, significantly increasing demands on inter-GPU communication bandwidth and memory resources.

In addition to self-attention, Transformer architectures include FFN layers, which facilitate independent token-level computations. Although FFN operations allow parallel execution, exchanging intermediate results and synchronizing gradients across GPUs remain required during forward and backward passes \cite{dean2012large,goyal2017accurate,grasync,grasync1,grasync2}. Such gradient synchronization necessitates frequent exchanges of intermediate gradient updates \cite{freqsync,freqsync1,freqsync2}, further elevating inter-GPU communication overhead.

\begin{figure}[b!]
    \centering
    \includegraphics[width=0.9\linewidth]{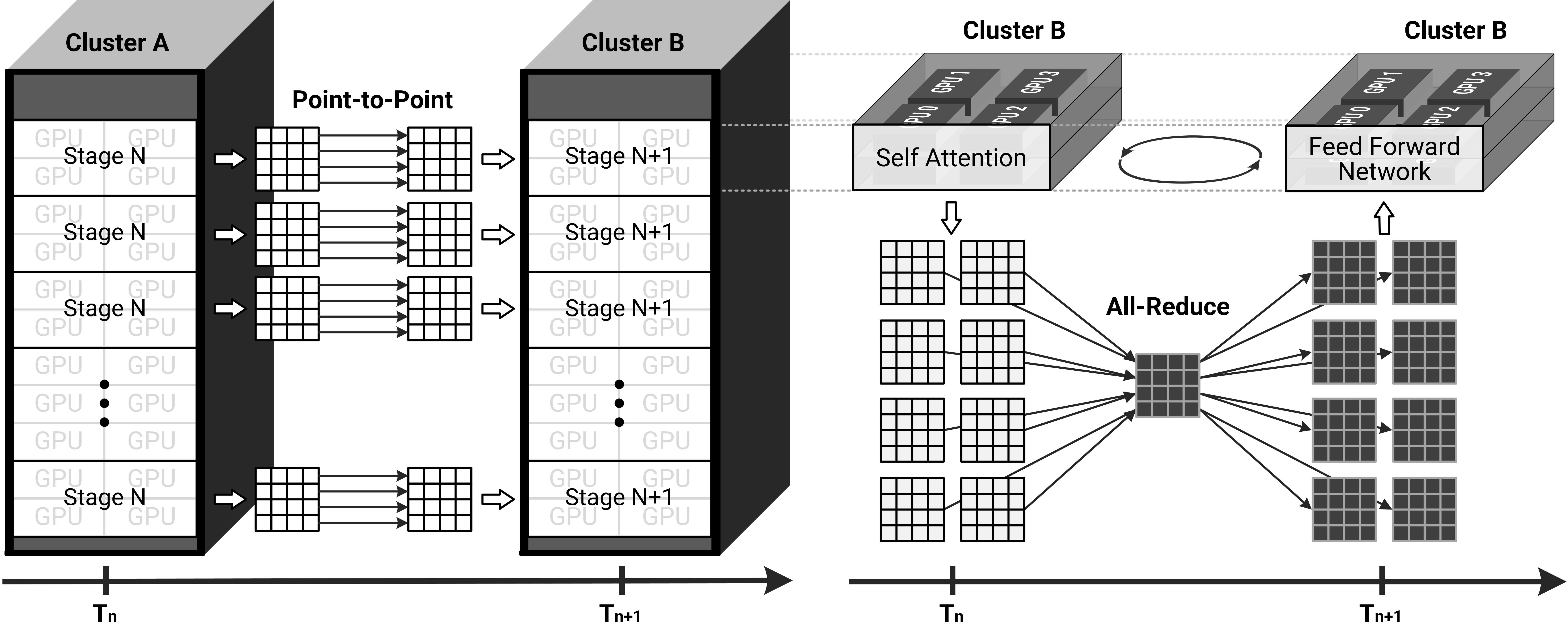}
    \begin{subfigure}{\linewidth}
        \begin{tabularx}{\textwidth}{
            p{\dimexpr.49\linewidth-2\tabcolsep-1.3333\arrayrulewidth}
            p{\dimexpr.49\linewidth-2\tabcolsep-1.3333\arrayrulewidth}
            }
            \caption{Pipeline parallelism (PP).} \label{fig:pipeline_parallelism} &
            \caption{Tensor parallelism (TP).} \label{fig:tensor_parallelism}
        \end{tabularx}
    \end{subfigure}
    \caption{Pipeline parallelism (PP) and tensor parallelism (TP) for transformer architecture.}
    \label{fig:pipeline_tensor_parallelism}
\end{figure}

Advanced parallelization techniques, such as \emph{pipeline parallelism} and \emph{tensor parallelism} \cite{gpipe,parallel1,parallel3,megatron,wang2022tesseract}, also play crucial roles in distributed LLM training. Figure \ref{fig:pipeline_parallelism} visualizes pipeline parallelism by illustrating the sequential ``stages'' of model execution, showing how each GPU cluster handles specific layers of the Transformer model to optimize resource utilization. Pipeline parallelism divides the Transformer model into sequential stages, with each stage processed on separate GPU clusters. Although this method increases available parallel computation, careful orchestration is required to minimize idle periods (pipeline bubbles \cite{pipefisher,gpipe,bubble2,parallel1,he2021pipetransformer}) caused by inter-stage data dependencies. Stages must synchronize their data handoffs to ensure smooth operation and maximum utilization of GPU resources. Pipeline parallelism is particularly effective for large models where the computation within each stage can fully utilize individual GPUs \cite{efficientlarge,gpipe,megatron}.

On the other hand, tensor parallelism complements pipeline parallelism by partitioning large tensor operations, such as matrix multiplications, across multiple GPUs. This approach enables simultaneous computation within layers, accelerating the processing of large tensor operations. However, tensor parallelism requires frequent synchronization of partial results across GPUs, typically using \textit{collective communication} operations such as All-Reduce, All-Gather, and Reduce-Scatter \cite{forum1994mpi,commop1,commop2,commop3,commop4,commop5}. These collective communication operations enable GPUs to exchange and aggregate intermediate computational results, maintaining consistency across parallel computations. Figure \ref{fig:tensor_parallelism} further illustrates tensor parallelism by depicting how large tensor computations are distributed across GPUs, highlighting the critical role of collective communication operations in synchronizing partial computations. Efficient implementation of tensor parallelism thus relies on optimized collective communication algorithms and high-performance interconnect infrastructures to maintain low communication latency and high throughput.

\begin{figure}[t!]
    \centering
    \includegraphics[width=0.9\linewidth]{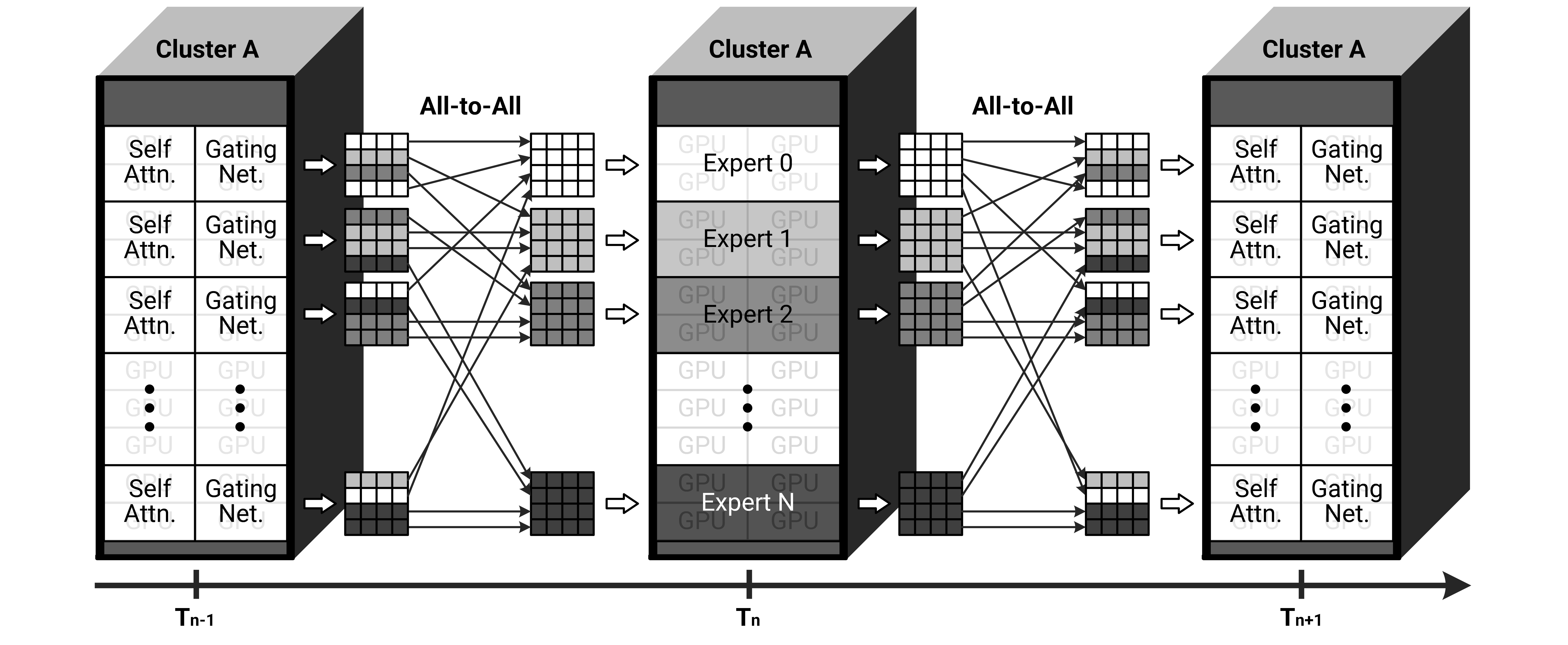}
    \caption{Expert parallelism (EP) for mixture of experts (MoE) architecture.}
    \label{fig:expert_parallelism}
\end{figure}

Furthermore, the adoption of dynamic computational strategies, exemplified by MoE architectures, adds complexity to training workflows \cite{moe,moe1}. Figure \ref{fig:expert_parallelism} shows the distribution of MoE expert modules across GPUs, emphasizing how each GPU independently manages distinct subsets of data computations. MoE models partition the network into multiple expert modules, each hosted on dedicated GPUs or GPU clusters. Each GPU acts as an independent expert performing distinct forward and backward computations for specific input data subsets \cite{gshard,gpumoe0,gpumoe,gpumoe1}. Input data sequences, represented tokens, are distributed across multiple GPUs based on predetermined criteria or routing strategies. Tokens or token segments representing parts of sentences or queries are allocated to GPUs according to the model's expert selection policy. After tokens are assigned, each GPU expert processes its designated computations individually. However, Transformer-based models require aggregating outputs from multiple experts to generate meaningful predictions, leading to ``frequent exchanges'' of intermediate results among GPUs. The figure illustrates the aggregation and synchronization process among GPUs in MoE architectures, showcasing the intensive exchange of intermediate computational results required for maintaining global model consistency. Regular aggregation and synchronization of gradients across experts are essential to maintain global model consistency. Consequently, \textit{MoE training escalates inter-GPU communication demands}, requiring sophisticated high-bandwidth, low-latency interconnect infrastructures.

\paragraph{Multi-GPU LLM inference: Optimization techniques and challenges.} In contrast to the training phase, inference is primarily known to emphasize computational speed and real-time responsiveness \cite{fastllm,fastllm1,fastllm2}. However, recent optimization techniques \cite{inferopt,inferopt1,inferopt2} for inference workloads have shifted part of the performance emphasis from pure computation toward increased memory capacity and inter-GPU communication bandwidth. These shifts result from advanced methods designed to reduce redundant calculations and enhance contextual accuracy, introducing substantial new system-level challenges.

\begin{figure}[t!]
    \centering
    \includegraphics[width=\linewidth]{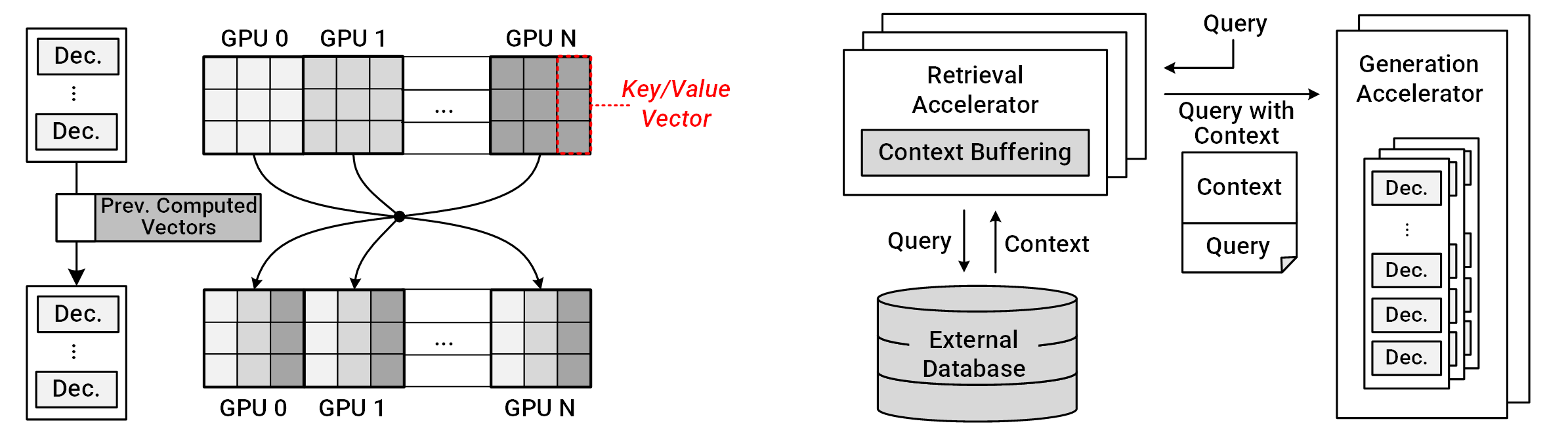}
        \begin{subfigure}{\linewidth}
        \begin{tabularx}{\textwidth}{
            p{\dimexpr.49\linewidth-2\tabcolsep-1.3333\arrayrulewidth}
            p{\dimexpr.49\linewidth-2\tabcolsep-1.3333\arrayrulewidth}
            }
            \caption{KV caching for token reuse across GPUs.} \label{fig:overhead_kv_caching} &
            \caption{RAG execution with large-scale retrieval integration.} \label{fig:overhead_rag}
        \end{tabularx}
    \end{subfigure}
    \caption{Communication and memory overhead in KV caching and RAG inference.}
    \label{fig:overhead_inference}
\end{figure}

Figures \ref{fig:overhead_kv_caching} and \ref{fig:overhead_rag} show inter-GPU communication and external data access patterns for inference optimizations such as KV caching and RAG, respectively; as shown in Figure \ref{fig:overhead_kv_caching}, KV caching exemplifies this shift by storing previously computed key and value vectors directly within GPU memory. While this technique reduces redundant computation, leading to significantly faster inference, it substantially elevates GPU memory demands \cite{memdemand,memdemand1}. As models and context windows scale, KV caches become massive, often requiring careful partitioning and frequent synchronization across GPUs \cite{kvcache1,sync1}. As a result, this places intensive pressure on memory management strategies and dramatically increases inter-GPU communication overhead to maintain cache coherence and efficiency.

On the other hand, as illustrated in Figure \ref{fig:overhead_rag}, RAG further intensifies demands on both memory and communication resources. By incorporating external knowledge bases into inference, RAG improves the accuracy and contextual relevance of outputs \cite{RAG,externalknowledge,externalknowledge1,externalknowledge2}. However, it requires GPUs to execute rapid, frequent queries to external databases, swiftly retrieve relevant data, and seamlessly integrate this external information into ongoing computations. These operations drastically increase memory usage to temporarily store the retrieved data, and place heightened demands on network bandwidth and low-latency communication infrastructure \cite{accrag,RAG,rag1}, complicating system design and performance optimization.

Note that auto-regressive inference methods also impose additional memory and interconnect-related constraints. Due to the sequential dependency inherent in token generation, each prediction explicitly relies on previously generated tokens, severely restricting parallel execution. Consequently, GPUs must exchange intermediate computational results as soon as possible and maintain synchronization across inference steps. This sequential dependency not only limits achievable parallelism but also escalates inter-GPU communication traffic, thereby intensifying demands for low-latency and high-throughput network connectivity and sophisticated memory management solutions to mitigate GPU idle times.

Considering all these factors, both training and inference workloads of modern LLMs increasingly emphasize memory capacity and inter-GPU communication infrastructure. Training requires sophisticated model partitioning, frequent synchronization, and advanced parallelization techniques due to parameter and activation sizes exceeding GPU memory capacities. Similarly, inference optimizations such as KV caching, RAG, and auto-regressive methods reduce redundant computations but further amplify memory usage and inter-GPU synchronization overhead. Therefore, modern AI infrastructures must adopt flexible architectures that efficiently manage memory and communication resources in addition to computational performance, addressing the comprehensive needs of contemporary LLM workloads.

\begin{figure}[t!]
    \centering
    \includegraphics[width=0.9\linewidth]{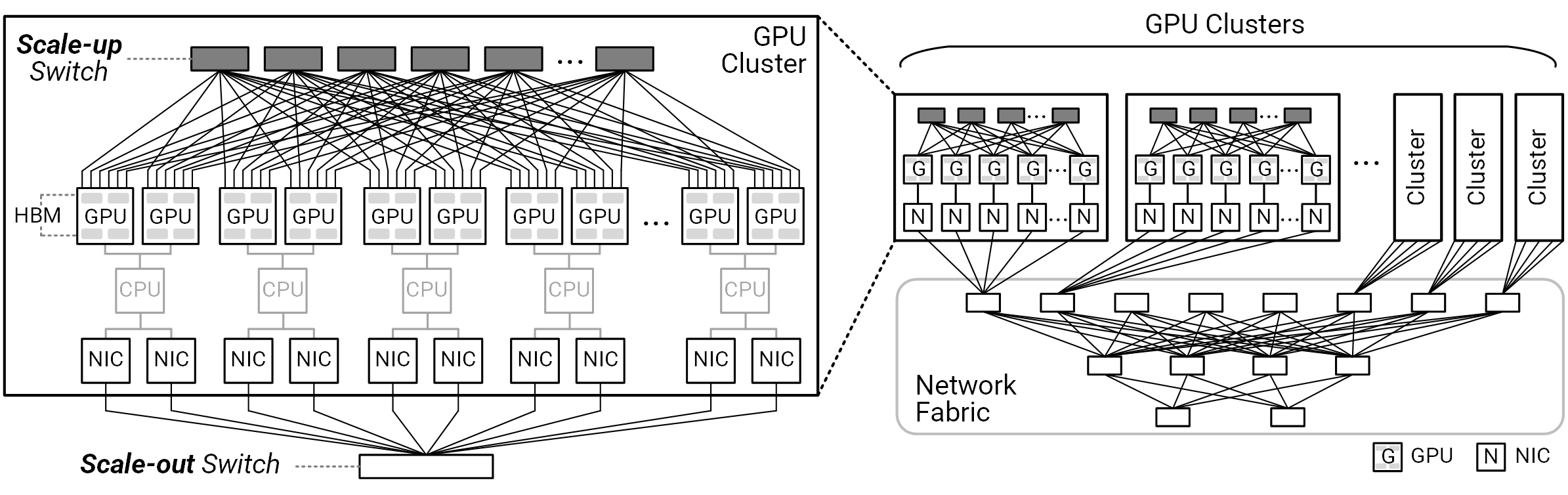}
    \begin{subfigure}{\linewidth}
        \begin{tabularx}{\textwidth}{
            p{\dimexpr.55\linewidth-2\tabcolsep-1.3333\arrayrulewidth}
            p{\dimexpr.49\linewidth-2\tabcolsep-1.3333\arrayrulewidth}
            }
            \caption{Scale-up and scale-out strategy for multi-gpu deployment.} \label{fig:scale_upout} &
            \caption{Interconnected GPU clusters via network fabric.} \label{fig:networked_cluster}
        \end{tabularx}
    \end{subfigure}
    \caption{Scale-up vs. scale-out and networked GPUs.} \label{fig:multigpu}
\end{figure}

\subsection{Scaling Multi-Accelerator Systems: Scale-up and Scale-out}
\label{subsection:3_2}
Addressing inter-GPU communication challenges in multi-GPU training and inference requires sophisticated hardware interconnect and network solutions tailored to varying performance and scalability requirements. Two primary architectural strategies, \textit{scale-up} and \textit{scale-out}, are commonly adopted to handle these different operational scenarios. Generally, scale-up architectures utilize high-speed ``interconnects'' such as NVLink \cite{nvlink1,nvlink2}, NVLink Fusion \cite{fusion,fusion1}, UALink \cite{ualink,ualink1}, and CXL, while scale-out architectures employ high-bandwidth, ``long-distance networks'' like Ethernet \cite{ethernet,ethernet1,ethernet2} or InfiniBand \cite{infiniband,infiniband1,infiniband2}.

\paragraph{Scale-up architecture: High-speed direct interconnect.} Figure \ref{fig:scale_upout} compares scale-up and scale-out GPU interconnect architectures for multi-GPU deployments. The scale-up strategy tightly couples a limited number of GPUs using specialized high-speed, accelerator-centric interconnects. These direct, high-bandwidth connections optimize data transfer efficiency, which is particularly beneficial for workloads involving frequent and intensive data exchanges within closely integrated GPU clusters. Scale-up solutions are advantageous for tasks requiring maximum intra-node communication speed and minimal latency, thereby significantly improving performance for computationally intensive scenarios such as training, real-time inference, and GPU-heavy AI operations.

Prior to the emergence of LLMs and the latest generation of data center architectures, the number of GPUs requiring these closely integrated direct interconnects was limited. However, as models grow more complex and data volumes increase significantly, the number of GPUs needing such connectivity is rising exponentially. Furthermore, many optimization techniques for LLMs now require high-speed, low-latency data exchanges and consistent I/O data sharing. Consequently, there is a growing trend in modern data centers to deploy more GPUs per rack, aiming to enhance computational efficiency, reduce communication overhead, and lower the total cost of ownership (TCO). For example, NVIDIA has introduced a new architecture, featuring compact and liquid-cooled node units designed specifically for high-density GPU deployments. These nodes utilize high-speed NVLink and NVSwitch interconnects to tightly integrate up to 72 GPUs per rack. This type of direct interconnect designs enhances computational throughput, optimizes inter-GPU communication efficiency, and improves thermal management, thereby reducing operational complexity and lowering the TCO.

\paragraph{Scale-out architecture: Long-distance network interface.} In contrast, the scale-out approach is designed for extensive, data center-scale deployments that may involve thousands of GPUs distributed across multiple racks or nodes. This strategy enables broader scalability and more flexible resource management. Scale-out architectures mainly utilize ``long-distance'' network-based fabrics, relying on \textit{network interface cards} (NICs) and RDMA-enabled communication protocols for GPU-to-GPU interactions \cite{rdmagpu,rdmagpu1,rdmagpu2}. As illustrated in Figure~\ref{fig:networked_cluster}, GPUs are organized into clusters that are interconnected via network fabrics, allowing for modular and dynamic system configurations.

While long-distance network fabrics provide excellent scalability, flexibility, and potentially higher aggregate bandwidth, they unfortunately introduce additional overhead. This overhead arises from complex hardware designs, sophisticated network protocols, and software-mediated communications. Specifically, data serialization and deserialization, network protocol processing, and software-level interactions significantly increase communication latency compared to tightly coupled, hardware-based scale-up architectures. Therefore, carefully evaluating these trade-offs between scale-up and scale-out architectures is essential when designing large-scale, distributed AI systems to ensure performance and efficiency objectives are effectively achieved.

On the other hand, CPUs still play a crucial supporting role in GPU-centric AI infrastructures for both scale-up and scale-out systems. While GPUs handle primary computational tasks, CPUs provide system orchestration capabilities, managing GPU coordination, data transfers, and network interfaces. Each GPU or GPU cluster thus integrates one or more CPUs and NICs as fundamental components. In the past, there have been various approaches toward resource disaggregation, similar to those presented in this technical report. However, complete physical resource disaggregation has not been fully realized because CPUs should act as host processors, managing the bus interfaces and memory controllers that interface directly with GPUs or accelerators. Instead, recent industry trends emphasize tighter integration within single nodes or adopt modular scaling methods. An illustrative example of such node-level integration is NVIDIA's latest GPU model (Grace Blackwell), which will be further examined in the subsequent subsection.

\subsection{Large-Scale AI Infrastructure: Hierarchical Data Center Architecture with Blackwell Instances}
\label{subsection:3_3}
\begin{figure}[t!]
    \centering
    \includegraphics[width=\linewidth]{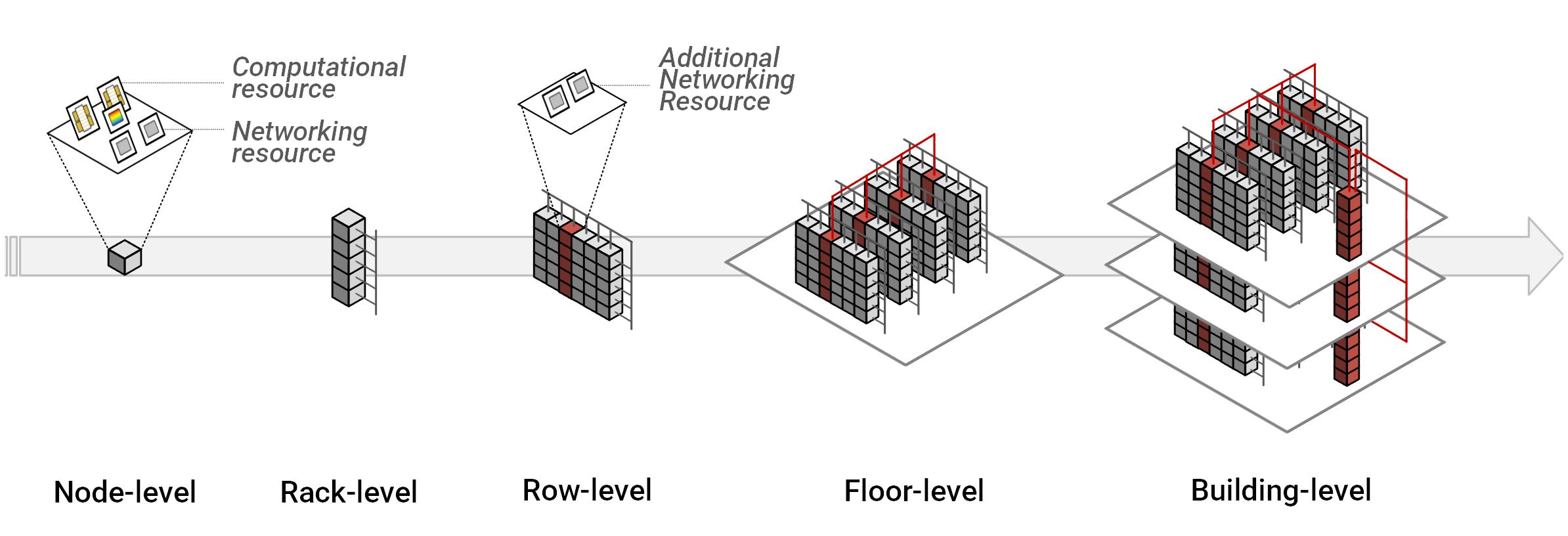}
    \caption{Hierarchical data center architecture.}
    \label{fig:hierarchy}
\end{figure}

Real-world data centers in practice adopt hierarchical architectures to support diverse workloads and varying infrastructure demands. As shown in Figure \ref{fig:hierarchy}, these hierarchical designs integrate computational and networking resources across multiple abstraction levels, which are namely, \emph{nodes}, \emph{racks}, \emph{rows}, \emph{floors}, and entire \emph{buildings} \cite{floor1,floor2,floor3,floor4,floor5}. Specifically, node-level components form the foundational computing units, which are further aggregated into racks to enhance computational density and interconnectivity. Subsequently, racks are organized into rows and then systematically expanded into floor-level structures, forming integrated building-scale deployments.

In the following subsections, we detail each hierarchical level, starting with the fundamental node configurations and gradually progressing toward comprehensive building-level integration. To illustrate practical node configurations, we reference NVIDIA's recent Grace Blackwell architecture \cite{GB200,GB200200,GB300,GB300300,Blackwell1,Blackwell2} as a representative example of contemporary designs. The Blackwell architecture introduces notable hardware enhancements, such as increased \textit{high-bandwidth memory} (HBM \cite{hbm1,hbm2,hbm3,hbm3e}) capacity per GPU and closer CPU-GPU integration, directly addressing key challenges including memory management, inter-GPU communication, and computational coordination. While Blackwell provides specific context, our primary emphasis remains on the broader understanding of general data center architectures, which highlights their respective advantages and limitations.This structured discussion lays the groundwork for exploring advanced designs in the following sections.

\begin{figure}[t!]
    \centering
    \includegraphics[width=\linewidth]{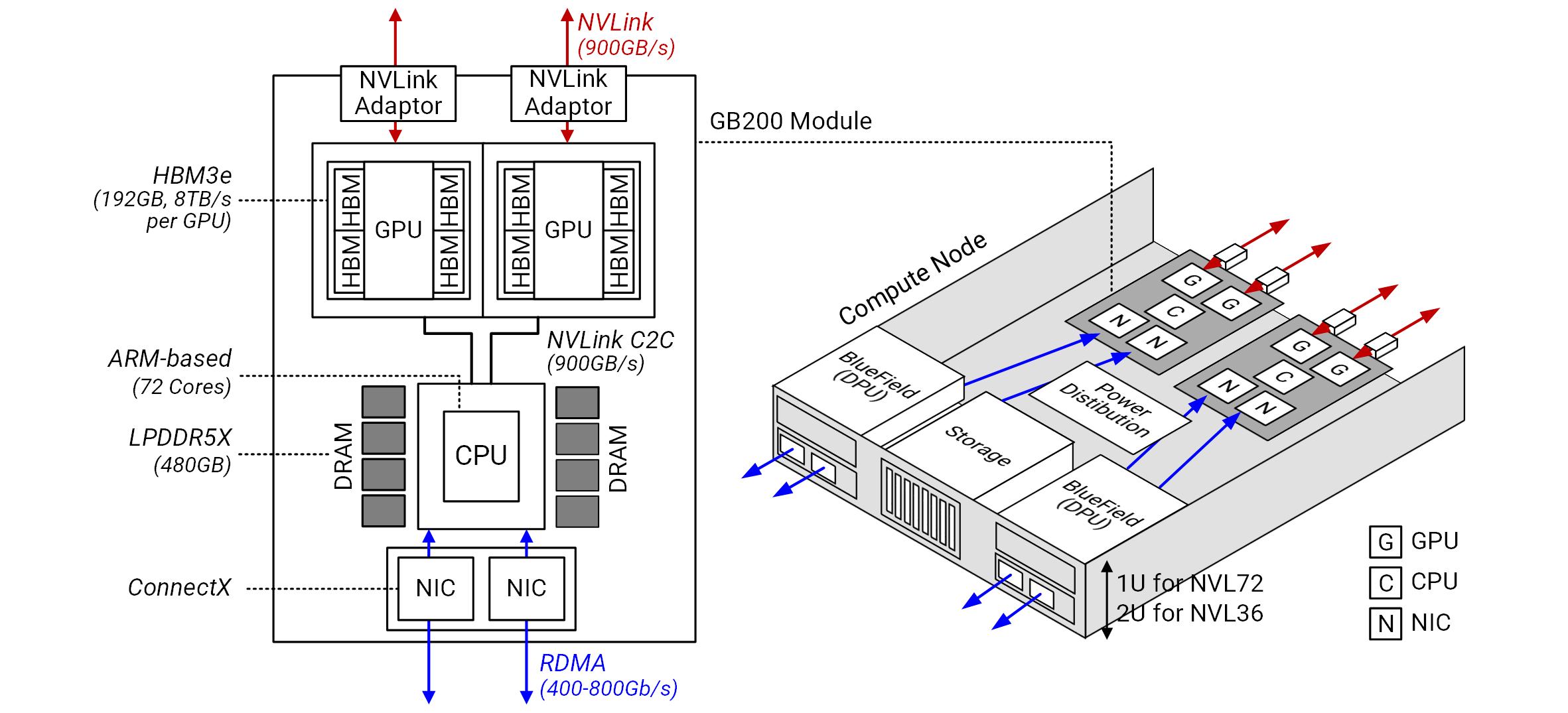}
    \begin{subfigure}{\linewidth}
        \begin{tabularx}{\textwidth}{
            p{\dimexpr.50\linewidth-2\tabcolsep-1.3333\arrayrulewidth}
            p{\dimexpr.50\linewidth-2\tabcolsep-1.3333\arrayrulewidth}
            }
            \caption{Blackwell architecture (GB200 module).} \label{fig:gb200_module} &
            \caption{A compute node setup with two GB200 modules.} \label{fig:gb200_node}
        \end{tabularx}
    \end{subfigure}
    \caption{Node-level configuration of GB200.}
    \label{fig:nv_node}
\end{figure}

\paragraph{Node- and rack-level configuration: Hierarchical integration of CPU-GPU modules.} The fundamental building block of modern AI data centers is the ``compute node'', an integrated computational unit comprising CPUs, GPUs, memory, and network interfaces. Figure \ref{fig:gb200_module} illustrates a representative node setup using the GB200 configuration as an example. Specifically, each GB200 integrates one high-performance ARM-based CPU with 72 cores and two GPUs, tightly coupled via NVLink \emph{chip-to-chip} (C2C) interface \cite{c2c,c2c1}. Each GPU contains approximately 192 GB of HBM3e, delivering bandwidths up to 8 TB/s per GPU \cite{blackwelldata,Blackwell1}, supporting large-scale AI models and inference workloads. In addition, the CPU provides up to 480 GB of LPDDR5X DRAM \cite{blackwelldata,gracecpu}, which maintains low-latency and unified memory communication with GPUs via NVLink C2C, achieving approximately 900 GB/s of bandwidth \cite{c2c,c2c1,wei20239}. This tightly integrated design results in a memory-unified computational domain within each GB200.

Each compute node also integrates advanced NICs to enable high-speed connectivity with external network fabrics. Common examples of the NICs include NVIDIA's BlueField data processing units (DPUs) \cite{dpu} and ConnectX adapters \cite{connectx,connectx1}, which are directly installed within individual nodes. These NICs provide hardware acceleration for networking functions, support high-bandwidth communication (typically 400 to 800 Gb/s per node), and enable RDMA capabilities for high-throughput data transfers. The integration of such NICs within a node ensures that every computational unit can participate in data center-scale networking, supporting both internal cluster operations and communication with broader network fabrics. Note that, as shown in Figure \ref{fig:gb200_node}, a compute node can contain two GB200 modules (two CPUs and four GPUs), packaged into compact 1U or 2U form factors optimized for dense rack-level deployments \cite{GB200200,Rackdeploy}.

\begin{figure}[b!]
    \centering
    \includegraphics[width=\linewidth, trim=0 10 0 0, clip]{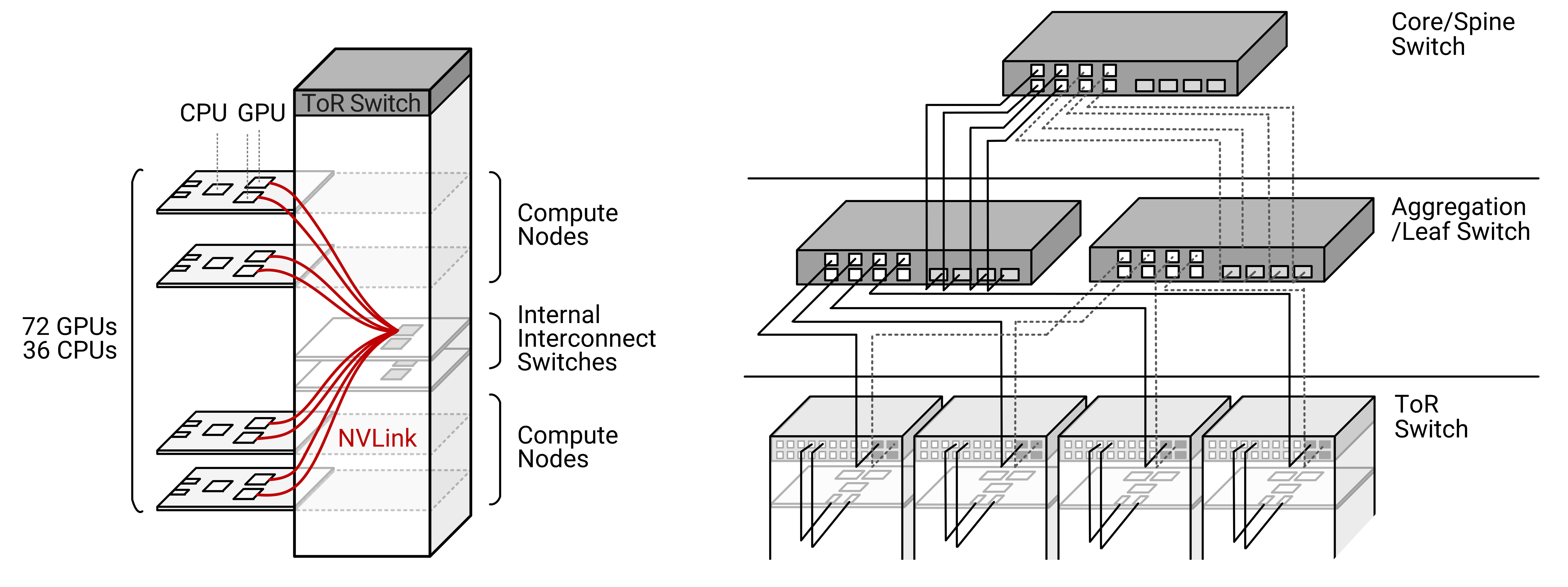}
    \begin{subfigure}{\linewidth}
        \begin{tabularx}{\textwidth}{
            p{\dimexpr.45\linewidth-2\tabcolsep-1.3333\arrayrulewidth}
            p{\dimexpr.53\linewidth-2\tabcolsep-1.3333\arrayrulewidth}
            }
            \caption{Internal connectivity through NVSwitch.} \label{fig:nvl72_nvswitch} &
            \caption{External connectivity through ToR switch.} \label{fig:nvl72_tor_switch}
        \end{tabularx}
    \end{subfigure}
    \caption{Rack-level configuration of NVL72.}
    \label{fig:nv_rack}
\end{figure}

At the rack level, multiple compute nodes aggregate into high-density computational clusters. Figure \ref{fig:nvl72_nvswitch} illustrates a representative rack-level architecture, using the GB200-based configuration as a concrete example to clarify typical design choices. In this example, nodes interconnect via rack-scale internal interconnect fabrics (e.g., NVLink and UALink). A standard rack can accommodate up to 36 GB200 modules, collectively comprising 72 GPUs and 36 CPUs \cite{GB200,GB200200}. Internally, all GPUs within the rack are interconnected using dedicated internal interconnect switches (e.g., NVSwitches \cite{nvswitch,nvswitch1}) carefully distributed across the rack. Such internal interconnects provide high-bandwidth, low-latency communication, ideal for workloads requiring intensive intra-rack data transfers such as large-scale model training and real-time inference tasks. This internal connectivity thus creates an efficient, ``scale-up domain'' within each rack.

Simultaneously, as illustrated in Figure \ref{fig:nvl72_tor_switch}, each node within the rack connects directly to \textit{top-of-rack} (ToR) network switches. The ToR switches aggregate traffic from all internal nodes and manage external connectivity, typically at bandwidths ranging from 400 to 800 Gb/s. ToR switches are positioned at the rack's upper portion, managing node-level network communication, reducing cable complexity, and minimizing latency for external network interactions. By simplifying cable management, this configuration can improve operational efficiency and scalability.

To enable the seamless expansion of data center infrastructure beyond the confines of a single rack, ToR switches are equipped with uplink ports that connect to higher-level, long-distance network infrastructures, such as aggregation switches \cite{aggswitch,aggswitch1} or leaf switches \cite{leafswitch,leafswitch1} within a spine-leaf topology \cite{leafswitch1,sltopology, sltopology1}. Through these hierarchical connections, ToR switches play a pivotal part in scaling networks to encompass multiple racks, rows, floors, and ultimately entire buildings, as will be discussed shortly. Note that ToR switches not only facilitate immediate rack-level communication but also act as the bridge to higher tiers of the data center network. Thus, ToR switches serve a dual role: aggregating and optimizing internal rack-level communication and providing flexible, high-bandwidth connectivity for scalable expansion across larger data center deployments. The combined use of intra-rack interconnects such as NVLink and external network connectivity via ToR switches forms a critical infrastructure in modern AI system design, simultaneously addressing GPU performance and scalability demands, yet also introducing inherent scalability limitations.

\begin{figure}[t!]
    \centering
    \includegraphics[width=0.8\linewidth]{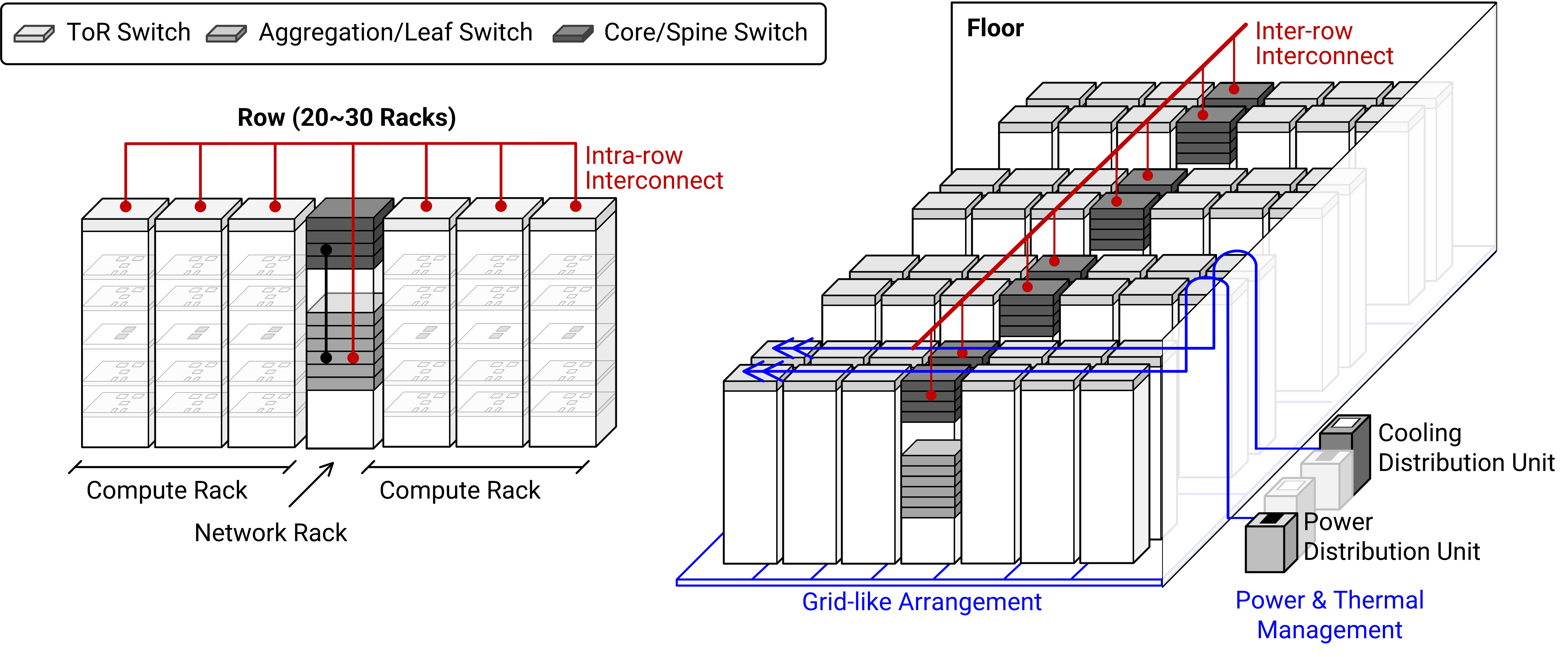}
    \vspace{-5pt}
    \begin{subfigure}{\linewidth}
        \begin{tabularx}{\textwidth}{
            p{\dimexpr.45\linewidth-2\tabcolsep-1.3333\arrayrulewidth}
            p{\dimexpr.50\linewidth-2\tabcolsep-1.3333\arrayrulewidth}
            }
            \caption{Row-level configuration.} \label{fig:row_level} &
            \caption{Floor-level configuration.} \label{fig:floor_level}
        \end{tabularx}
    \end{subfigure}
    \caption{Row and floor-level configuration.}
    \label{fig:row_floor}
\end{figure}

\paragraph{Row and floor-level configurations: Scaling infrastructure.} In modern data center architectures, dedicated network racks, distinct from compute racks, serve as centralized aggregation points within each row. These network racks house aggregation switches or spine-leaf switches that interconnect the ToR switches of multiple compute racks in the same row, forming the backbone of intra-row communication. Figure \ref{fig:row_level} illustrates a typical row-level configuration in modern data centers, showing multiple compute racks connected to these network racks via InfiniBand or Ethernet switches \cite{etherswitch1,etherswitch2}. Each row consists of several compute racks, each containing densely packed compute nodes, and one or more centrally positioned network racks dedicated to switching and aggregation.

As shown in the figure, the network racks located within the row (often at the center or end) are also populated with high-bandwidth switches, such as InfiniBand (Quantum-2) \cite{infiniswitch} or Ethernet (Spectrum-X) \cite{etherswitch}, which aggregate and route traffic between all compute racks in the row. These switches support extensive bandwidth capacities, typically ranging from 200 to 800 Gb/s per port, enabling efficient data exchanges among the compute racks. By centralizing the switching infrastructure within dedicated network racks, cable management is simplified, network latency is minimized, and scalability is enhanced, as additional compute racks can be seamlessly added and interconnected through the aggregation switches.

While such row-level communication structures have assisted in scaling data centers by aggregating existing nodes and racks, recent operational environments handling large models and extensive data, such as LLMs, introduce several structural limitations, particularly related to high-speed inter-GPU synchronization. As the frequent synchronization required among GPUs significantly reduces GPU utilization, strategic infrastructure planning at the row level is becoming increasingly critical for ensuring the stability and performance of large-scale AI workloads. In this technical report, \textit{we emphasize the importance of both inter-row and intra-row communications}, which are currently managed by scale-out architectures. However, \textit{this emphasis also sets the stage for further exploration into potential enhancements through scale-up architectures}.

On the other hand, the transition from row-level to floor-level configuration represents a critical step in scaling data center infrastructure. Specifically, optimizations at the row-level (e.g., efficient intra-row networking, organized cable management, and streamlined thermal strategies) must be cohesively integrated across multiple rows at the floor-level. Typically, floor-level configurations interconnect multiple rows, forming grid-like layouts to optimize spatial efficiency and inter-row connectivity \cite{floor2,rasmussen2007data}. For instance, a single floor generally aggregates several rows, each containing approximately 20 to 30 racks, resulting in coordinated management of hundreds of racks. Figure~\ref{fig:floor_level} illustrates such a floor-level arrangement, emphasizing the coordinated interconnections between rows to facilitate efficient data flow and resource sharing. At this scale, spatial organization, efficient inter-row networking, and comprehensive infrastructure management become crucial. In addition, careful design considerations for scale-up and scale-out domains, based on the characteristics of specific interconnect and network technologies, are essential. Due to the hardware layout characteristics of data centers, frequent inter-row and intra-row communications occur, driven primarily by synchronization requirements for parallel GPU and accelerator processing or memory access data exchanges. Currently, however, most of these row-level communications rely on scale-out domain connections, resulting in relatively high overhead during data transfers. In a subsequent section, we will discuss various strategies to enhance scale-up architectures, enabling low-latency, high-speed intra-row and inter-row communications.

Note that advanced thermal and power management solutions, such as liquid-cooling distribution units \cite{liquid,cooling1} and power distribution units \cite{pdu1,pdu2,pdu3} integrated at the row or rack level, are also important for dissipating heat from densely packed computing nodes. These strategies collectively ensure stable operating conditions, minimize performance degradation, and enhance the overall reliability and efficiency of large-scale AI infrastructure deployments.

\begin{figure}[t!]
    \centering
    \includegraphics[width=0.9\linewidth]{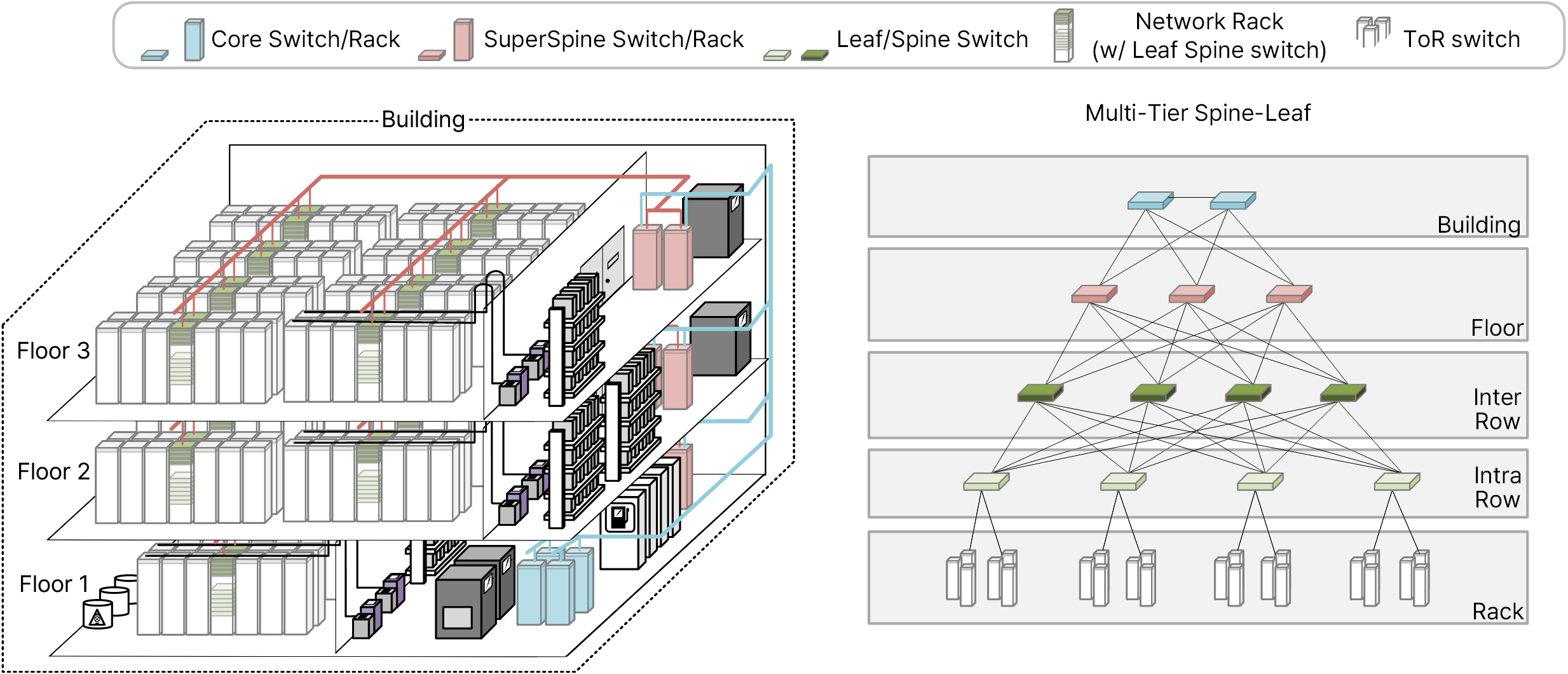}
    \vspace{-5pt}
    \begin{subfigure}{\linewidth}
        \begin{tabularx}{\textwidth}{
            p{\dimexpr.49\linewidth-2\tabcolsep-1.3333\arrayrulewidth}
            p{\dimexpr.59\linewidth-2\tabcolsep-1.3333\arrayrulewidth}
            }
            \caption{Building-level configuration.} \label{fig:building} &
            \caption{Multi-tier spine-leaf data center topology.} \label{fig:overall-topo}
        \end{tabularx}
    \end{subfigure}
    \caption{Overall data center topology and building configuration.}
    \label{fig:overall-datacenter}
\end{figure}

\paragraph{Building-level integration: From AI infrastructure formation to campus-scale.} As illustrated in Figure \ref{fig:building}, building-level integration represents the highest tier of hierarchical data center organization, interconnecting multiple floors, each composed of interconnected rows and racks, into unified, large-scale AI infrastructure. At this scale, managing coherent resource allocation, data movement, and network coordination across thousands to tens of thousands of GPUs distributed over multiple floors poses significant operational complexities. For instance, modern large-scale deployments often connect multiple floors using long-distance network technologies, logically enabling extensive GPU integration across buildings \cite{longreach,longreach1,longreach2}.

However, scaling infrastructure to the building-level introduces unique challenges beyond those encountered at lower hierarchical levels. Specifically, communication between floors increases network latency and congestion, often limiting practical GPU utilization to approximately half of the theoretical peak performance~\cite{floor1,peakperf,lebiednik2016survey}. In typical, building-level integration employs hierarchical network topologies, such as multi-tier spine-leaf or multi-level fat-tree architectures, interconnecting multiple floors to balance communication load, reduce latency, and manage congestion effectively (cf. Figure \ref{fig:overall-topo}). Moreover, handling power distribution, thermal conditions, and fault tolerance across multiple floors presents additional difficulties that further complicate operational efficiency \cite{scalable1,scalable2}. To mitigate these challenges, automated monitoring and centralized resource management systems become indispensable, providing real-time visibility into computational loads, network performance, thermal management, and hardware status.

Unfortunately, despite these sophisticated management strategies, fundamental communication bottlenecks, such as inter-GPU synchronization overhead and extensive data movement across hierarchical layers, remain structurally unavoidable. These persistent communication challenges structurally constrain scalability and efficiency. Addressing these communication constraints therefore requires a new type of interconnect architectures and composable system designs, which will be explored in depth in the following sections.

\begin{figure}[t!]
   \centering
   \includegraphics[width=0.9\linewidth]{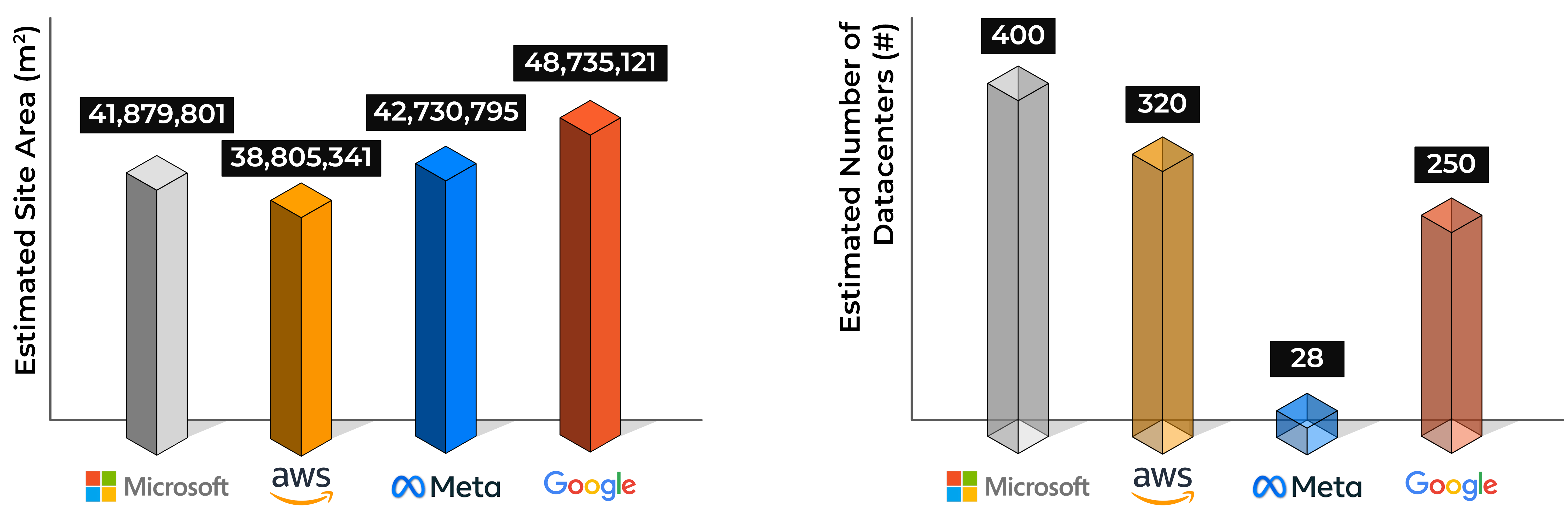}
   \begin{subfigure}{\linewidth}
       \begin{tabularx}{\textwidth}{
           p{\dimexpr.49\linewidth-2\tabcolsep-1.3333\arrayrulewidth}
           p{\dimexpr.49\linewidth-2\tabcolsep-1.3333\arrayrulewidth}
           }
           \caption{{US site area sum per hyperscaler.}} \label{fig:hyper_site} &
           \caption{{Number of hyperscaler's data centers.}} \label{fig:hyper_datacenter}
       \end{tabularx}
   \end{subfigure}
   \caption{Hyperscaler's site area and data center count.}
   \label{fig:hyper}
\end{figure}

Note that multiple building-level structures collectively form campus-scale infrastructures that support large-scale data center deployments. For context, Figures~\ref{fig:hyper_site} and~\ref{fig:hyper_datacenter} illustrate the current scale of data centers operated by major hyperscalers, including Microsoft, Meta, Google, and Amazon, in response to growing demand for AI infrastructure. Figure~\ref{fig:hyper_site} shows the total site area of U.S.-based data centers for each company, including planned facilities projected to be completed by 2027~\cite{baxtel}. Figure~\ref{fig:hyper_datacenter} presents the number of data centers as defined by each hyperscaler~\cite{aws_region,ms_region,ms_az,meta_datacenter,google_region}.

To convey the scale of these deployments, Meta's total site area reaches approximately 42 million~$m^2$, equivalent to about 5,300 standard soccer fields. Microsoft operates nearly 400 data centers worldwide, while AWS and Google manage between 200 and 300 centers. Meta maintains around 30 centers, but each facility is significantly larger in area, resulting in a total site footprint comparable to that of the other hyperscalers. Meta's infrastructure emphasizes large-scale, high-density design, prioritizing capacity and operational efficiency. These differences reflect varied infrastructure scaling strategies, which in turn influence the architectural complexity and efficiency of each hyperscaler's deployment model.

\subsection{Constraints and Challenges in GPU-Integrated AI Infrastructure}
\label{subsection:3_4}

\paragraph{Beyond compute: Why GPU parallelization faces fundamental constraints.} So far, we have discussed hierarchical data center architectures exemplified by the Blackwell configurations, which scale computational resources from nodes to entire buildings, integrating thousands to tens of thousands of GPUs. However, despite such architectural scalability, GPU parallelization encounters intrinsic constraints caused by memory limitations and unavoidable communication overheads. As discussed previously, modern LLM workloads exceed the memory capacities of individual GPUs due to extensive model parameters, large intermediate states generated by attention mechanisms, and sizable activation data. As a result, partitioning computations across multiple GPUs is unavoidable, but each parallelization approach introduces critical performance trade-offs related to inter-GPU synchronization, communication overhead, and operational complexity.

Specifically, model parallelism distributes model parameters across multiple GPUs, addressing memory capacity constraints but introducing frequent and intensive inter-GPU synchronization overhead. Similarly, data parallelism duplicates the entire model across GPUs for parallel batch processing, yet the collective synchronization operations (e.g., All-Reduce) impose significant communication overhead, limiting GPU utilization to approximately 35--40\% of theoretical peak performance \cite{romero2022accelerating,shi2021exploiting,fang2019redsync,xie2022synthesizing,parallel2,dp_util}. Pipeline parallelism segments models sequentially across GPUs to accommodate larger models, but inter-stage data transfers cause pipeline bubbles (GPU idle periods), restricting utilization to about 50\% \cite{gpipe,parallel1,pipefisher,blueconnect,efficientlarge}. Even hybrid approaches, combining multiple parallelization strategies, cannot fully mitigate the inherent communication overhead intrinsic to multi-GPU environments.

Critically, these structural bottlenecks persist despite advances in modern hardware architectures. While the hardware innovations (featuring high-bandwidth NVLink, increased HBM3e memory capacities, and integrated CPU-GPU designs) partly reduce communication latency and improve overall throughput, they cannot eliminate synchronization overhead and complexity in memory management, which are all intrinsic. Thus, \textit{the structural limitations of existing GPU parallelization methods, rooted deeply in communication overhead and synchronization requirements, remain key challenges}. Addressing these intrinsic bottlenecks is essential for the future optimization and architectural evolution of AI infrastructure.

\paragraph{Rethinking resource integration: Limitations of coupled CPU-GPU deployments at scale.} Tightly integrated CPU-GPU modules, as exemplified by the Blackwell architecture, provide performance benefits for specific computational tasks. However, deploying such tightly coupled resources across large-scale data centers presents critical constraints in terms of scalability, flexibility, and operational efficiency. Specifically, these integrated modules enforce rigid resource coupling, restricting the independent scalability of computational, network, and memory resources.

In large-scale AI environments, this tightly coupled approach exacerbates two primary challenges: \textit{significant inter-node communication overhead} and \textit{inflexible memory allocation}. Each CPU-GPU node requires dedicated network connections for inter-node data transfer and synchronization, causing network complexity and communication overhead to escalate linearly with system size. This direct, tightly coupled network topology notably increases latency and synchronization delays among GPUs, degrading performance, for communication-intensive tasks such as LLM training and inference.

In addition, the rigid CPU-to-GPU ratio dictated by integrated modules (e.g., one CPU per two GPUs in GB200/300) often results in underutilized CPUs as the infrastructure scales, leading to inefficient resource use and unnecessary operational costs. Moreover, fixed memory binding within these modules prevents independent memory scaling, forcing proportional increases in memory capacity with node additions. This inflexibility causes either severe memory underutilization or insufficient memory capacity, thus restricting the effective handling of varying AI workloads.

Lastly, tightly coupled CPU-GPU modules inherently complicate maintenance and upgrades. Component-level failures necessitate replacing entire modules, increasing downtime and operational expenses. Furthermore, integrated architectures limit the timely incorporation of technological advancements, as simultaneous CPU-GPU upgrades are typically required, reducing system agility and delaying modernization efforts.

These structural and operational constraints highlight the fundamental limitations of integrated CPU-GPU modules for meeting the dynamic demands of large-scale AI deployments. To address these critical bottlenecks, future data center architectures must adopt modular, independently scalable designs for CPUs, GPUs, memory, and network resources. Such disaggregated approaches will ensure scalability, operational flexibility, and efficient resource utilization at data center scale \cite{indep,indep1,indep2,indep3,indep4}.

\section{Leveraging CXL for Diverse AI Performance Metrics}
\label{section:4}
\begin{figure}[t!]
    \centering
    \includegraphics[width=\linewidth]{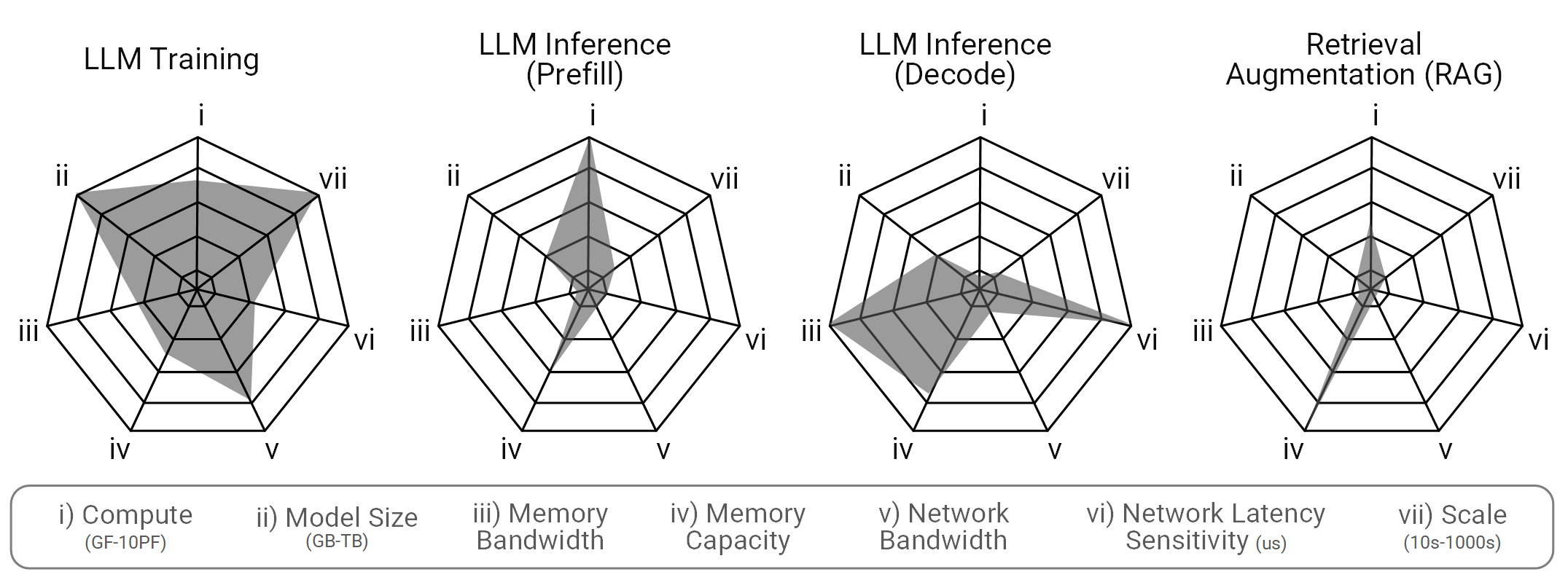}
    \caption{Relative importance of performance metrics across different operational scenarios.}
    \label{fig:metrics}
\end{figure}

Modern AI infrastructures, particularly those supporting large-scale LLM deployments, must concurrently satisfy multiple distinct performance metrics, including i) computational throughput, ii) model size, iii) memory bandwidth, iv) memory capacity, v) network bandwidth, vi) latency sensitivity, and vii) overall system scalability. As shown in Figure \ref{fig:metrics}, these performance dimensions exhibit varying relative importance depending on specific operational scenarios, highlighting the dynamic and workload-specific nature of AI infrastructure demands \cite{kundu2024performance,hu2024characterization,sponner2022ai,jiang2025rago,meta,rdma}.

To illustrate these varying requirements, we analyze distinct workload scenarios, including LLM training \cite{llama3model,touvron2023llama}, the prefill and decode phases of LLM inference \cite{llama3model,mistral2023}, and RAG workloads \cite{ragwork1,ragwork2,ragwork3}, each emphasizing different combinations of these performance metrics. This analysis reveals that no single architectural configuration can optimize all metrics across these diverse scenarios. Addressing these multidimensional and evolving performance demands necessitates modular, composable architectures capable of scaling computational, memory, and network resources.

We advocate that \emph{Compute Express Link} (CXL) technology can address this requirement by enabling resource disaggregation and dynamic composability, thereby providing the flexibility necessary to adapt and efficiently scale resources according to specific AI workload characteristics. In this section, we first discuss the challenges of meeting multidimensional requirements in modern AI infrastructures. We then provide the background by reviewing the evolution of various CXL specifications, emphasizing their architectural developments and their implications for meeting diverse AI infrastructure requirements. Finally, we propose CXL-enabled tray designs and rack architectures tailored to these diverse AI infrastructure scenarios, enabling concurrent optimization of multiple performance metrics across dynamically evolving workloads.

\subsection{Key Challenges in Simultaneously Optimizing Performance Metrics}
\label{subsection:4_1}
To facilitate understanding, we categorize factors influencing the previously mentioned seven performance metrics into four primary groups based on the characteristics of different ML workloads. These groups are computational throughput, memory (capacity and bandwidth), network communication, and latency sensitivity. This subsection first explains why each of those groups is important in running the different workloads and then examines why conventional data center architectures have difficulty optimizing these interconnected dimensions in parallel.

\paragraph{Computational throughput.} For computational throughput, modern LLM training requires extremely large model sizes, involving tens to hundreds of billions of parameters, to achieve adequate expressive power and superior generalization capabilities. Specifically, large models capture complex linguistic relationships and subtle contextual nuances, exhibiting emergent capabilities, such as advanced reasoning and improved context comprehension, which smaller models fail to deliver. For instance, models adopting MoE architectures significantly amplify parameter counts, often into the trillions, by selectively activating specialized expert networks per input token, thus achieving enhanced model capacity and performance. To manage and process these vast parameter sets within practically acceptable timeframes, computational throughput is the matter, and to get high computational throughput, substantial parallelization across thousands of GPUs becomes essential. However, as discussed earlier in Section \ref{subsection:3_4}, deploying large-scale GPU clusters introduces frequent synchronization events, particularly during collective operations, such as gradient aggregations and expert network activations unique to MoE structures. These intensive collective communication patterns escalate network bandwidth demands, creating critical performance bottlenecks. Therefore, practical GPU utilization in realistic training scenarios remains limited, achieving less than half of theoretical peak performance due to synchronization-induced overheads and network congestion \cite{liang2024communication,xiao2020antman,fang2019redsync,efficientlarge,gpipe,pipefisher}.

\paragraph{Memory capacity and bandwidth.} The memory capacity and bandwidth are equally important and represent critical bottlenecks. The exponential growth of model parameters and intermediate activations in modern LLMs, reaching tens to hundreds of billions of parameters, results in memory requirements exceeding hundreds of TBs during training. In practice, model parameters, intermediate activations, optimizer states, gradient buffers, and related metadata must reside concurrently in memory, amplifying overall memory demands \cite{Trillion,memwall1,Trillion1}. These extensive memory requirements considerably surpass the local GPU memory capacity, ranging from tens to a hundred GB in contemporary architectures. Hence, traditional architectures that tightly couple CPUs and GPUs with fixed memory configurations restrict independent scaling of memory resources, limiting their ability to accommodate massive, multi-hundred-TB-scale memory demands. Moreover, workloads leveraging memory-intensive optimizations, such as KV caching and RAG, frequently initiate large-scale memory transactions, exacerbating memory bandwidth constraints. As a result, architectures optimized for computational throughput ironically become inadequate for supporting sustained, high-bandwidth memory traffic, leading to significant inefficiencies and performance degradation.

\paragraph{Network communication.} Efficient network communication is indispensable as sophisticated parallelization strategies distribute workloads across multiple nodes. Conventional tightly integrated architectures predominantly optimize intra-node (and intra-rack) computational capabilities while overlooking extensive inter-node (and inter-rack) communication demands emerging at large scales. As parallelization expands across many nodes, the frequency and volume of inter-node communication grow substantially, often becoming several to tens of times larger than the actual GPU-resident data, surpassing available network bandwidth. For instance, as discussed previously, GPUs collectively manage intermediate activations, optimizer states, and gradient updates totaling hundreds of TBs per iteration in large-scale LLM training scenarios. The required inter-GPU communication arising from frequent synchronization of attention vectors, gradients, and KV caches can escalate to PB-scale data transfers per iteration, exceeding original GPU-resident data sizes and even at rack-scale. As a result, traditional architectures encounter severe network bottlenecks, impeding efficient GPU synchronization and effective scaling across large-scale data center deployments. Note that the immense communication overhead and challenges again underscore the critical importance of scalable, high-bandwidth, low-latency interconnect infrastructure.

\begin{figure}[t!]
   \centering
   \includegraphics[width=\linewidth]{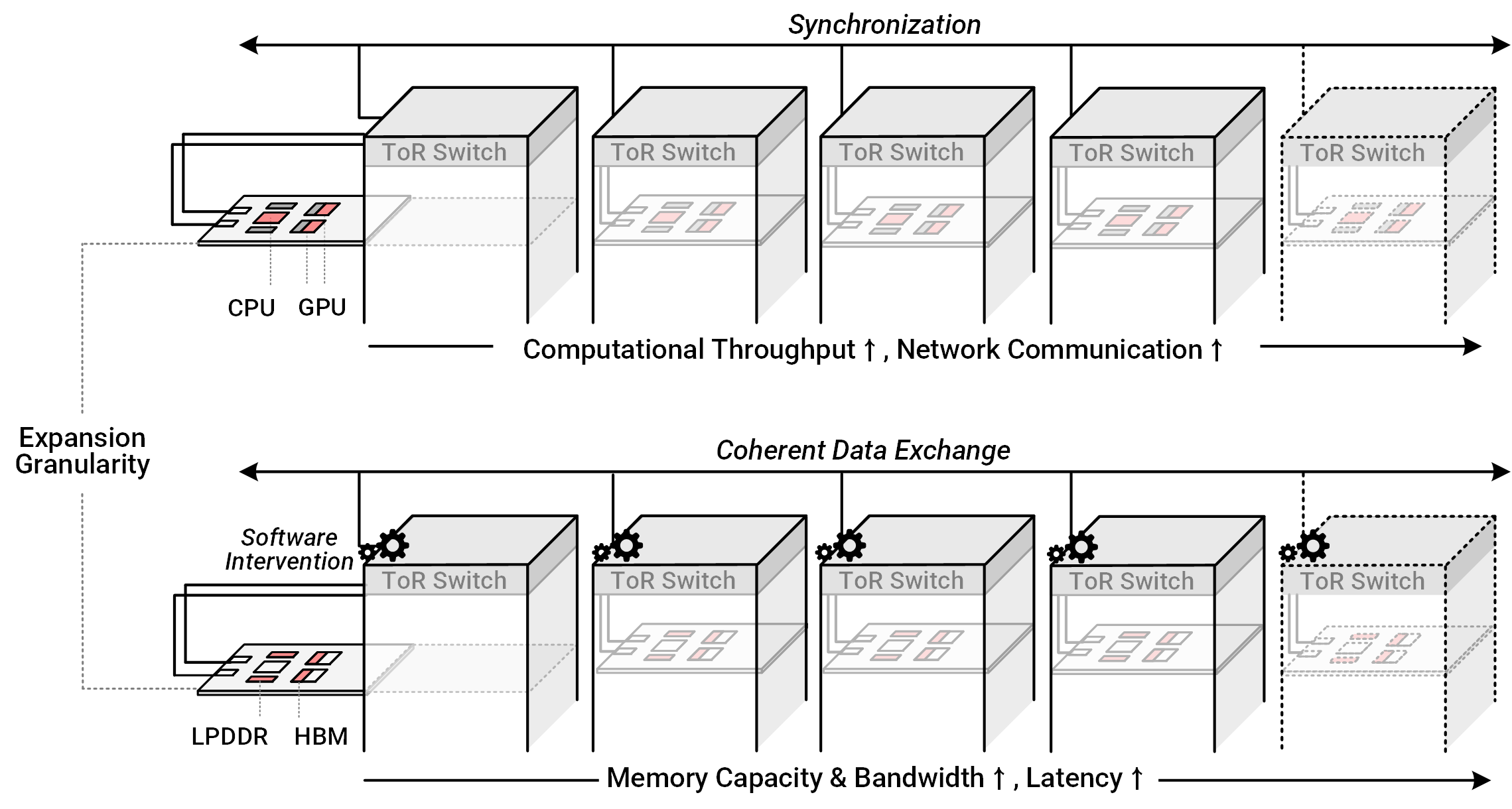}
   \caption{Competing constraints of performance dimensions}
   \label{fig:competing}
\end{figure}

\paragraph{Latency sensitivity.} The latency sensitivity introduces critical challenges, particularly during inference phases such as decode operations in auto-regressive scenarios. Real-time AI workloads require low-latency responses, mandating exchanges of intermediate computational results among GPUs with precise synchronization as soon as possible. However, conventional infrastructures incur significant latency from frequent intermediate data movements across nodes and additional software overhead introduced by traditional communication stacks (e.g., Ethernet or InfiniBand). Specifically, network-based connection technologies involve substantial software overhead due to frequent privilege mode transitions of an operating system (e.g., kernel/user mode switches), redundant memory copy operations, interrupt handling, and protocol processing. These software-induced overheads typically increase latency by tens to hundreds of times compared to hardware-only operable interconnects such as CXL or NVLink, limiting achievable performance and scalability. This inherent latency restricts conventional architectures from meeting real-time responsiveness requirements, thereby constraining their applicability to latency-sensitive AI inferences. \newline

These four performance dimensions (i.e., computational throughput, memory capacity and bandwidth, network communication, and latency sensitivity) involve competing constraints, making simultaneous optimization challenging (cf. Figure \ref{fig:metrics}). Specifically, as shown in Figure \ref{fig:competing}, increasing computational throughput by extensive GPU parallelization intensifies demands on network bandwidth, exacerbating synchronization overhead. Likewise, enhancing memory capacity to accommodate massive parameter sets and activations introduces additional complexity and latency overhead due to frequent coherent data exchanges across nodes. Consequently, conventional architectures characterized by tightly integrated CPU-GPU modules lack the flexibility to scale compute, memory, and networking resources independently. This rigid coupling leads to suboptimal resource utilization and constrained overall system performance.

To address these architectural limitations, we believe that future AI data centers are required to adopt composable architectures enabling modular and independent scaling across compute, memory, and networking domains. CXL emerges as a practical solution, offering advanced features such as resource disaggregation, dynamic composability, and coherent memory pooling. By physically decoupling memory resources from processing units and facilitating direct cache-coherent access, CXL can also reduce latency, minimize synchronization overhead, and enhance overall system efficiency.

\subsection{Background: Evolution of CXL and Composable Architectures}
\label{subsection:4_2}
\begin{figure}[t!]
    \centering
    \includegraphics[width=\linewidth]{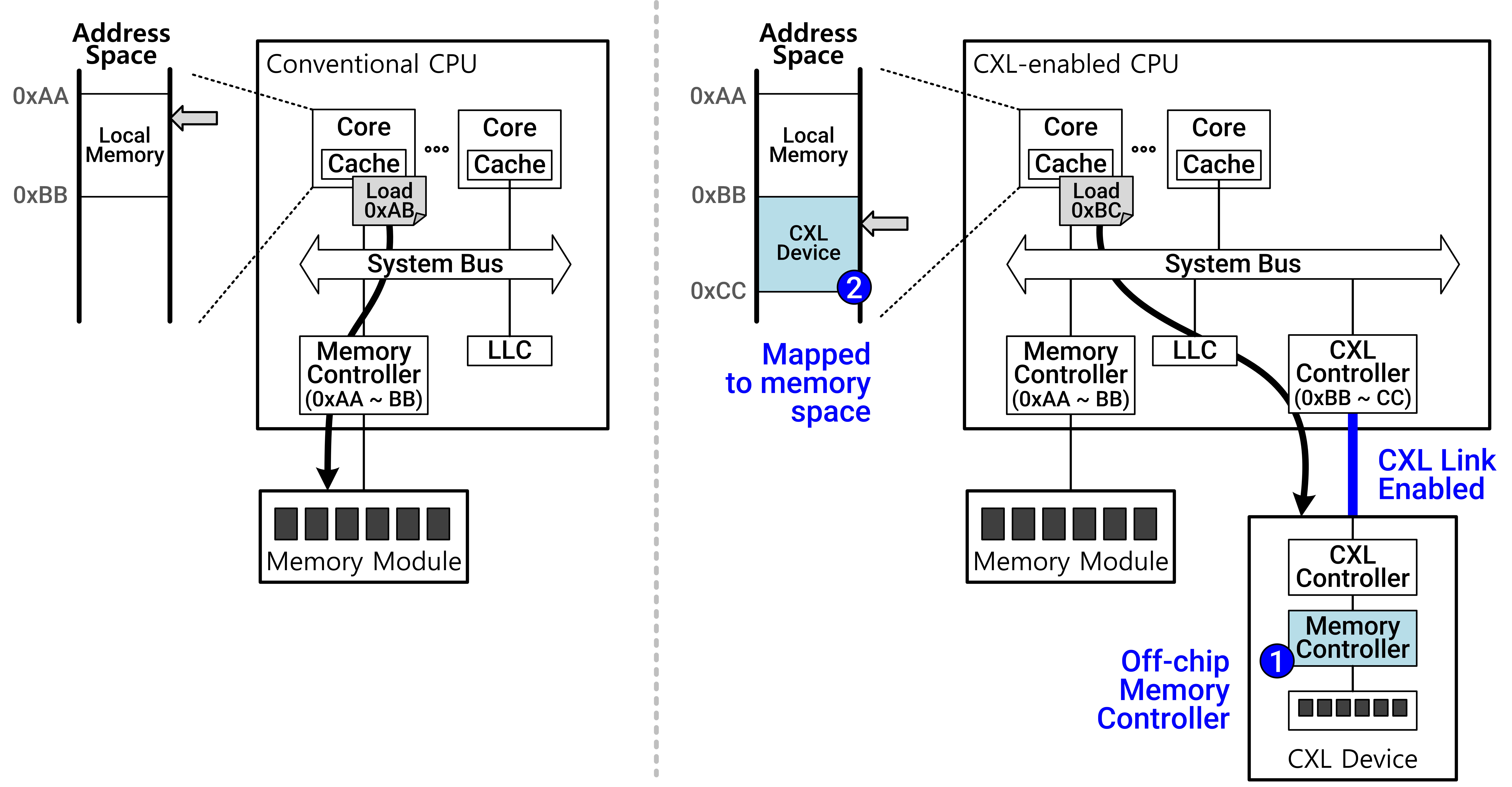}
    \caption{Relocation of memory controller in CXL.}
    \label{fig:mem_disaggr}
\end{figure}

\paragraph{Decoupling memory resources from CPUs: CXL 1.0.} To address the aforementioned scalability, flexibility, and performance issues, data center architectures can leverage various interconnect characteristics offered by CXL to facilitate architectural reconfiguration. In traditional computer architectures, memory controllers are tightly integrated within CPU packages, making memory expansion and physical disaggregation challenging. Physical separation of memory resources from CPUs is essential to accommodate dynamically changing workload demands, improve utilization of imbalanced memory capacities across nodes, and reduce TCO. However, due to architectural constraints, existing data centers have employed RDMA technologies to logically disaggregate memory, allowing multiple computing units to share these resources via software assistance~\cite{rdmadisaggregate,rdmadisaggregate1,rdmadisaggregate2,rdmadisaggregate3}. Although RDMA methods provide architectural flexibility, their inherent software overhead, such as frequent privilege mode transitions, data serialization/deserialization, and redundant memory copy operations, results in significant performance degradation and substantial energy overhead from additional data movement and management.

CXL can directly address these fundamental limitations by physically and logically decoupling memory controllers from CPUs. As depicted in Figure \ref{fig:mem_disaggr}, instead of embedding memory controllers within CPU packages, CXL relocates these controllers onto external memory modules, termed CXL \textit{endpoints}, such as DRAM expansion cards or specialized memory devices. This critical architectural shift allows memory resources to scale independently beyond traditional CPU constraints, enabling true hardware-level memory disaggregation and composability. Unlike the conventional RDMA methods, CXL leverages the existing PCIe-based \emph{physical layer} (PHY), but it delivers a direct, cache-coherent memory interface accessible via standard CPU load/store instructions. This design eliminates software-induced overhead, including context switches and redundant memory copy operations, by establishing direct hardware-mediated memory access paths. In addition, by implementing specialized logical layers (e.g., CXL link layer and transaction layer) atop the established PCIe infrastructure, CXL integrates into existing hardware ecosystems, simplifying architectural adoption without necessitating substantial hardware modifications.

Table~\ref{tab:cxl_version_features} comprehensively compares the key features across CXL versions 1.0, 2.0, and 3.0, emphasizing the progressive enhancements in scalability, connectivity, and advanced functionalities. The initial specification, CXL 1.0 \cite{cxl1.0}, introduced the key concept of memory controller decoupling, laying the foundational framework for composable infrastructure. However, practical scalability under CXL 1.0 remained limited as all types of CXL devices should be located within a node. Each CXL endpoint's external memory controller is constrained by a limited number of memory channels and physical factors, such as signal integrity and attenuation issues \cite{cxl1.0,cxlintro}. Because of those, achievable memory capacities per CXL endpoint device typically remained within the range of 1--2 TB, complicating large-scale memory expansion. Given the limited number of endpoints that each node can accommodate, fully realizing extensive and scalable memory disaggregation required further architectural enhancements, motivating subsequent evolutions starting with CXL 2.0.

\begin{table}[t!]
\centering
\caption{Comparative analysis of different versions of CXL.}
\label{tab:cxl_version_features}
\resizebox{\textwidth}{!}{
\small
\renewcommand{\arraystretch}{1.15}
\setlength{\arrayrulewidth}{0.5pt}
\begin{tabular}{|>{\columncolor{gray!10}\raggedright\arraybackslash}m{6cm}|>{\centering\arraybackslash}m{3cm}|>{\centering\arraybackslash}m{3cm}|>{\centering\arraybackslash}m{3cm}|}
\hline
\rowcolor{gray!40}
\textbf{Feature} & \textbf{CXL 1.0} & \textbf{CXL 2.0} & \textbf{CXL 3.0} \\ \hline
Max Link Rate ($GTs$) & 32 & 32 & 64 \\ \hline
Flit 68-byte (up to 32 $GTs$) & \checkmark & \checkmark & \checkmark \\ \hline
Flit 256-byte (up to 64 $GTs$) & \textbf{--} & \textbf{--} & \checkmark \\ \hline
Memory Controller Decoupling & \checkmark & \checkmark & \checkmark \\ \hline
Memory Expansion & \checkmark & \checkmark & \checkmark \\ \hline
Memory Pooling & \textbf{--} & \checkmark & \checkmark \\ \hline
Memory Sharing & \textbf{--} & \textbf{--} & \checkmark \\ \hline
Switching (Single-level) & \textbf{--} & \checkmark & \checkmark \\ \hline
Switching (Multi-level) & \textbf{--} & \textbf{--} & \checkmark \\ \hline
Hierarchical-based Routing (HBR) & \textbf{--} & \checkmark & \checkmark \\ \hline
Port-based Routing (PBR) & \textbf{--} & \textbf{--} & \checkmark \\ \hline
Hot-plug Support & \textbf{--} & \checkmark & \checkmark \\ \hline
Max Accelerator per Root Port & 1 & 1 & 256 \\ \hline
Max Memory Devices per Root Port & 1 & 256 & 4096 \\ \hline
Back-Invalidation & \textbf{--} & \textbf{--} & \checkmark \\ \hline
Peer-to-Peer Communication & \textbf{--} & \textbf{--} & \checkmark \\ \hline
Release Year & 2019 & 2020 & 2022-23 \\ \hline
\end{tabular}}
\end{table}

\begin{figure}[t!]
    \centering
    \includegraphics[width=\linewidth]{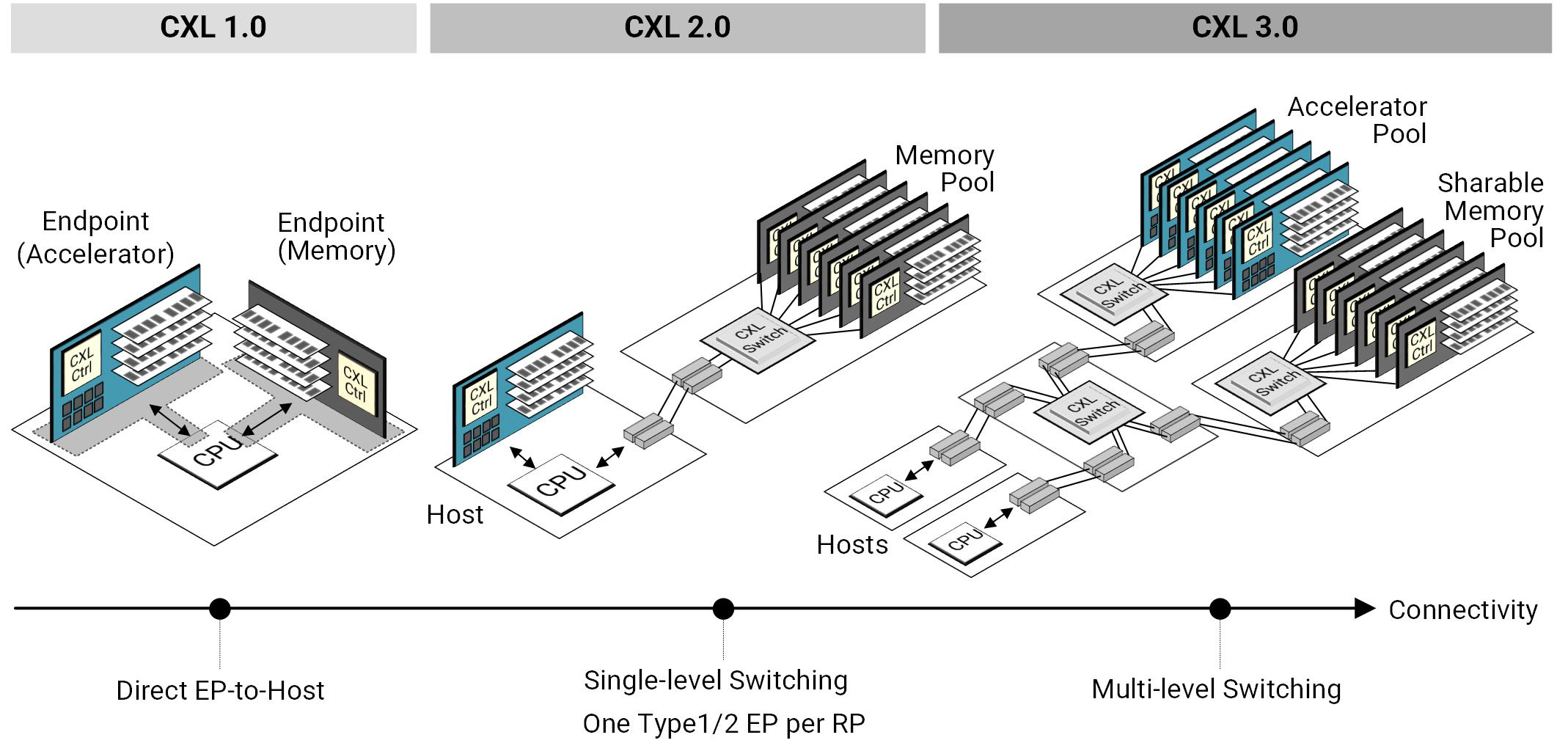}
    \caption{Evolution of CXL: from direct connections to multi-level switching.}
    \label{fig:cxl_history}
\end{figure}

\paragraph{Scalable and composable memory through switch-based architectures: CXL 2.0.} To overcome the memory capacity, endpoint scalability, and rigid connection limitations inherent to CXL 1.0, the CXL 2.0 specification \cite{cxl2.0} introduced an important architectural advancement: dedicated \textit{switch}-based topologies. In contrast to the direct endpoint-to-host connectivity of CXL 1.0, which restricted scalability due to limited memory channels and fixed endpoint configurations, CXL 2.0 enables flexible aggregation and management of multiple memory resources via intermediate CXL switches. As depicted in Figure~\ref{fig:cxl_history}, this introduction of an intermediate switching layer between compute nodes and memory endpoints allows individual hosts to access larger memory pools composed by multiple external endpoints, resolving the connectivity bottlenecks associated with earlier point-to-point CXL implementations.

Internally, CXL 2.0 switches can utilize high-bandwidth crossbar architectures, routing coherent memory transactions among numerous connected endpoints and compute nodes. This hardware-mediated coherent communication reduces latency in accessing large-scale external memory, eliminating substantial software-induced overhead observed in traditional RDMA-based network fabrics. Leveraging PCIe Gen5 technology (32 GT/s per lane), a CXL 2.0 switch can offer configurable multi-port interfaces, each capable of up to 64 GB/s bidirectional bandwidth in standard 16-lane configurations. As a result, a single CXL 2.0 switch can aggregate tens of TBs of memory per node, and because it can connect multiple nodes, it exceeds the scalability constraints inherent to endpoint-centric CXL 1.0 designs. This scalable topology supports modular system expansion and simplifies resource provisioning, decoupling memory expansion from rigid endpoint limitations. Advanced operational features introduced by CXL 2.0, including hot-plug support \cite{cxlintro,cxl2.0}, further enhance system flexibility by allowing dynamic addition or removal of memory endpoints with minimal operational disruption. In addition, host-specific static memory allocation features \cite{cxlintro,pond,breakwall} enable memory resource management, empowering data centers to accommodate evolving workload requirements.

Despite these improvements, CXL 2.0 retained scalability constraints, particularly its inability to support hierarchical multi-level switch configurations. This restriction confined scalability to ``single-layer switch'' architectures, significantly limiting memory pool sizes and the number of devices per root port. Specifically, typical CPU architectures provide only a finite number of root ports, each constrained by fixed link capabilities and stringent bandwidth limits. Therefore, practical deployments of CXL 2.0 in practice support 4 to 16 memory expanders (i.e., Type 3 devices) per CPU root port, well below the theoretical maximum of 256 devices. This limitation becomes even more pronounced for accelerators (i.e., Type 1 and Type 2 devices), which require strict ``one-to-one'' mappings per root port to maintain cache coherence, restricting accelerator scalability and deployment flexibility.

Recognizing these scalability constraints underscored the necessity for further architectural advancements. This motivated subsequent technical enhancements introduced in CXL 3.0, including multi-level switch cascading, advanced routing mechanisms, and comprehensive system-wide memory coherence capabilities.

\paragraph{True composability through multi-level switching and memory sharing: CXL 3.0.} Addressing the scalability and hierarchical limitations inherent to CXL 2.0, the CXL 3.0 specification \cite{cxl3.2} introduces critical architectural enhancements\footnote{In this section, references to CXL 3.0 encompass all subsequent revisions within the CXL 3.x series, including versions such as 3.1, 3.2, and beyond.} that advance true composability and resource sharing in high-performance computing environments. Unlike CXL 2.0, which limited configurations to single-layer switch topologies and constrained the number of endpoint devices, CXL 3.0 explicitly supports multi-level switch topologies, termed \textit{switch cascading}. This hierarchical fabric interconnects multiple CXL switches across several layers, overcoming single-layer limitations and increasing the number of endpoint devices (e.g., memory expanders and accelerators) that a CPU root port can coherently access.

Specifically, CXL 3.0 extends connectivity for memory expanders (Type 3 devices) per CPU root port to as many as 4,096 devices, facilitating extensive memory pools critical for large-scale AI and analytics workloads. Accelerator integration capabilities are similarly enhanced, supporting up to 256 accelerators (Type 1 and Type 2 devices) per root port, exceeding previous single-device constraints. These enhancements enable dynamic and flexible deployment of heterogeneous accelerator clusters within unified composable fabrics, thereby optimizing infrastructure efficiency to effectively accommodate evolving workload demands.

Internally, the multi-level switch fabric in CXL 3.0 introduces a novel \textit{port-based routing} (PBR) mechanism, complementing the \textit{hierarchical-based routing} (HBR) inherited from CXL 2.0. In contrast to HBR, which relies on fixed hierarchical paths and provides only static memory partitioning (exclusive allocations per host without dynamic sharing), PBR dynamically selects optimal routing paths based on real-time port conditions and network congestion. Importantly, PBR supports genuine multi-host memory sharing, enabling multiple hosts to concurrently and coherently access shared memory resources. This capability can improve traffic distribution, reduce latency, and mitigate communication bottlenecks, thereby overcoming  limitations of the earlier static partitioning approaches inherent in CXL 2.0.

In particular, CXL 3.0 introduces robust multi-host memory sharing and comprehensive system-wide cache coherence capabilities, largely enabled by the advanced PBR mechanism. This architectural enhancement enhances computational efficiency, reduces latency, and lowers overhead in large-scale AI workloads. For example, accelerators and compute nodes can directly and coherently share essential data structures, such as embedding tables, KV caches, and intermediate activations, without redundant data transfers or complex software interventions. Furthermore, CXL 3.0 allows accelerator-local high-bandwidth memories to be unified into a single coherent memory pool, broadening their applicability across diverse system configurations. The practical implications and performance advantages of this genuine memory sharing approach are explored in detail in Section~\ref{subsection:5_2}.

Although CXL 3.0 introduces extensive enhancements and maintains backward compatibility with CXL 2.0, its practical adoption requires specific hardware modifications across CPUs, switches, and endpoint devices, to fully leverage the new capabilities. Specifically, while the PBR mechanism primarily operates within CXL switches, corresponding hardware adaptations at endpoint devices are also necessary. Endpoints must handle larger flit sizes (256-byte flits in PBR mode compared to 68-byte flits in HBR mode). In addition, as CXL 3.0 facilitates genuine multi-host memory sharing with comprehensive cache coherence, memory expanders must implement advanced coherence mechanisms, such as back-invalidation, ensuring consistent data visibility and integrity across all shared resources. Endpoints may also support advanced CXL 3.0 features, including direct peer-to-peer communication, enabling accelerators to directly exchange data within a single host domain or access memory resources on other endpoints without host mediation. These features maximize the scalability and flexibility provided by the PBR-enabled fabric architecture, reducing communication latency and overhead.

\subsection{CXL-Enabled Modular Tray and Rack Architecture for AI Data Centers}
\label{subsection:4_3}
\begin{figure}[t!]
    \centering
    \includegraphics[width=0.9\linewidth]{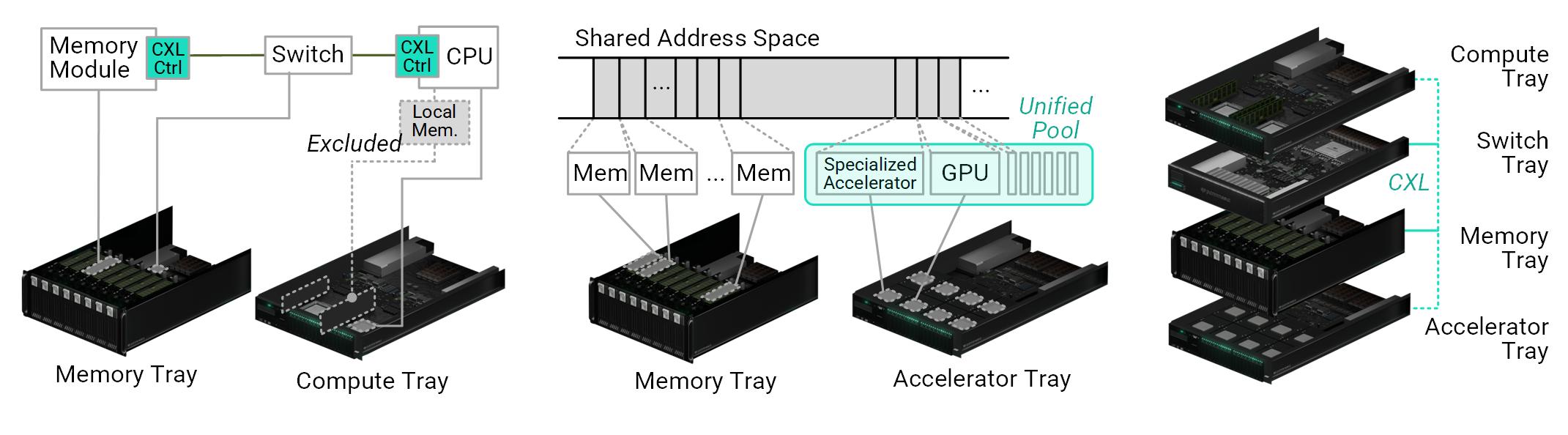}
    \begin{subfigure}{\linewidth}
        \begin{tabularx}{\textwidth}{
            p{\dimexpr.35\linewidth-2\tabcolsep-1.3333\arrayrulewidth}
            p{\dimexpr.35\linewidth-2\tabcolsep-1.3333\arrayrulewidth}
            p{\dimexpr.35\linewidth-2\tabcolsep-1.3333\arrayrulewidth}
            }
            \caption{Disaggregated memory tray.} \label{fig:memtray} &
            \caption{Disaggregated accelerator pool.} \label{fig:acctray} &
            \caption{Connected trays.} \label{fig:trayconn}
        \end{tabularx}
    \end{subfigure}
    \caption{Tray-based disaggregation with CXL cache-coherent sharing.}
    \label{fig:tray}
\end{figure}

Building upon the composability and multi-host memory-sharing capabilities introduced by CXL 3.0, this subsection discusses how these architectural innovations enable a modular, tray-based design for modern AI data centers in practice. Unlike traditional tightly integrated accelerator nodes, \textit{tray-based modular systems supported by CXL} allow independent scaling and dynamic reconfiguration of resources since CXL enables spatially separated placement of endpoints. Each CXL-enabled tray functions as a standardized hardware unit exclusively dedicated to a specific resource type, such as accelerators, CPUs, or memory, thereby enabling precise resource scaling, simplified maintenance, and agile adaptation to evolving AI workload demands.

\paragraph{Modular, tray-based design enabled by CXL technology.} In CXL-based modular architectures, memory resources are fully decoupled from compute and accelerator modules, fundamentally changing resource management strategies in data centers. As illustrated in Figure \ref{fig:memtray}, \textit{memory trays} exclusively integrate DRAM modules aggregated via dedicated CXL controllers or CXL switch(es), forming composable memory pools. These trays are not statically bound to specific compute nodes; instead, they can be flexibly allocated and shared among multiple accelerator or compute trays, enhancing scalability and operational flexibility. For instance, during intensive training phases of large-scale transformer models, additional memory trays can be provisioned to accelerator trays, accommodating increased memory demands without modifying CPU configurations.

Although the high-level concept of modular disaggregation has been previously proposed \cite{indep1,indep2,indep4,modular}, practical realization of true resource disaggregation was infeasible within traditional architectures. Historically, accelerators and memory modules functioned merely as passive peripherals incapable of independently managing memory coherence or direct access. As a result, CPUs hosted memory controllers in a tightly integrated manner and enforced cache consistency, coupling memory and accelerators within fixed server nodes. CXL resolves these constraints by relocating memory controllers externally and establishing standardized cache-coherent communication across all system components. Thus, accelerators can directly access and coherently share disaggregated memory resources without CPU mediation.

On the other hand, within this modular framework, dedicated \textit{accelerator trays} integrate multiple GPUs or specialized accelerators interconnected via high-speed CXL interfaces. This configuration supports efficient cache-coherent communication and direct memory sharing among accelerators. As depicted in Figure \ref{fig:acctray}, employing standardized CXL interfaces enables the integration of various accelerators into a unified pool without coupling CPUs or memory devices. Moreover, the accelerators' local memory can be combined into a shared, coherent memory space, being able to reduce redundant data transfers across the accelerators. Such a design can enhance performance, particularly for frequently accessed intermediate data structures, including KV caches and intermediate activations. In parallel, \textit{compute trays} exclusively host CPUs and, when necessary, network interface cards, deliberately excluding local memory to maintain strict resource disaggregation. Critically, dedicated \textit{CXL switch trays} orchestrate coherent interactions among independently scalable compute, accelerator, and memory trays, facilitating dynamic resource allocation and composability in real time.

By dedicating each tray exclusively to a specific resource type, the full potential of CXL-based resource disaggregation is effectively realized. Accordingly, each resource, compute, accelerator, and memory, can independently scale in alignment with workload demands. This modular architecture significantly enhances operational efficiency and resource utilization, enabling rapid adaptation to diverse and evolving AI workloads, and providing a robust foundation for composable rack-level architectures discussed in subsequent sections.

\begin{figure}[t!]
    \centering
    \includegraphics[width=\linewidth]{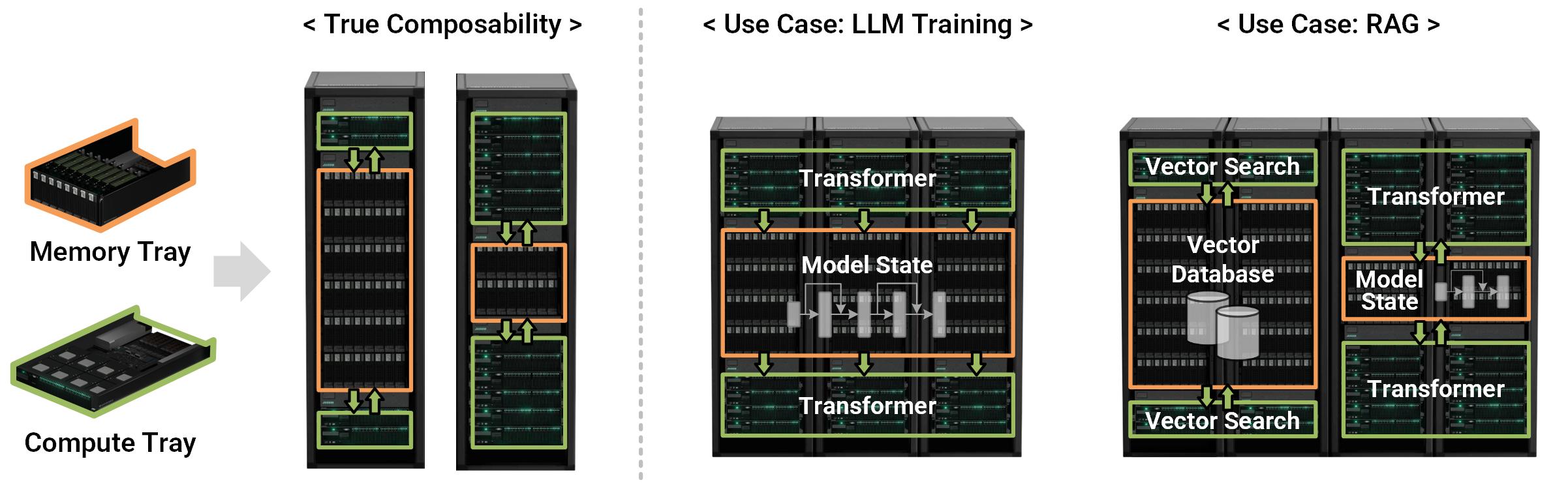}
    \caption{Composable rack architecture with CXL.}
    \label{fig:composable}
\end{figure}

\paragraph{Composable rack architecture with CXL.} At the rack level, modular tray-based architectures enabled by CXL are organized to enhance functional clarity and resource utilization. Each rack comprises dedicated trays for networking, accelerators, CPUs, composable memory expanders, and storage. This design allows for specialized racks interconnected to form composable rows or flexible integration of various tray types within each rack, tailored to specific application needs.

Figure~\ref{fig:trayconn} illustrates one representative configuration of tray-based composability within a modular rack-level architecture enabled by CXL. In this particular layout, accelerator trays and compute trays are interconnected via centrally positioned memory trays managed by high-speed CXL switches. This configuration also provides coherent, low-latency access to pooled memory resources without the direct involvement of CPUs. Accelerator trays can therefore directly share and efficiently access these memory pools, significantly reducing redundant intra-rack data movements. Such a design is especially advantageous for workloads characterized by frequent reuse of intermediate data structures, including transformer inference scenarios and KV caching tasks.

In parallel, Figure~\ref{fig:composable} highlights another illustrative example of composable architectures deployed at rack scale, optimized specifically for diverse AI workloads. Here, accelerator trays and memory trays are fully modularized and decoupled, enabling dynamic scaling and flexible allocation of memory resources. Utilizing high-speed, cache-coherent communication through CXL, accelerators achieve efficient data sharing, thereby improving computational throughput and overall resource utilization. This modular approach particularly benefits workloads with dynamically varying computational and memory demands, such as large-scale LLM training and RAG workloads.

Note that a key advantage of the proposed modular architecture is importantly its optimized intra- and inter-rack networking strategy. Specifically, \textit{this design integrates all racks and nodes within a row into a unified scale-up domain}. As discussed previously, traditional data center designs employing ToR switches inevitably classify inter-rack accelerator communication into the scale-out domain. When high-speed accelerator links such as UALink or NVLink are unavailable, unfortunately, intra-rack communication also depends on long-distance networks. This dependence significantly degrades performance, as each node is treated as an independent scale-out endpoint. Even when UALink or NVLink is available, non-accelerator devices within racks must still communicate through scale-out paths, exacerbating performance bottlenecks, particularly for workloads with intensive GPU-to-GPU interactions.

In contrast, leveraging CXL's support for hierarchical switch cascading, our modular architecture replaces conventional ToR switches by positioning dedicated CXL switch trays centrally as middle-of-rack (MoR) modules. Depending on bandwidth demands, additional CXL switch trays can be composably integrated, enabling flexible scaling and enhanced connectivity within racks and rows. Thus, racks and nodes within a row form an efficient scale-up domain over all interconnect fabrics rather than having long-distance networks. Traditional Ethernet or InfiniBand switches, primarily supporting inter-row communication, can be centrally organized into dedicated network racks. Through this hierarchical approach, network traffic is effectively aggregated and distributed, reducing congestion and latency, and maintaining balanced bandwidth allocation. Consequently, this infrastructure enables coherent, dynamic resource sharing across large-scale accelerator and memory pools, meeting the diverse requirements of modern AI workloads.

On the other hand, from an operational viewpoint, modular tray-based designs simplify maintenance and upgrades as well. Since memory, accelerator, compute, and networking components are physically and logically decoupled, each resource type can be independently replaced, upgraded, or scaled with minimal disruption. This modularity reduces system downtime, accelerates adoption of emerging technologies, and ensures infrastructure agility, enabling rapid adaptation to evolving workload demands while maintaining cost-effectiveness over time.

\begin{table}[t!]
\centering
\caption{Performance comparison between conventional and CXL-enabled architectures.}
\label{tab:conventioanl_cxltraycxl_architecture_comparison}
\resizebox{\textwidth}{!}{
\small
\renewcommand{\arraystretch}{1.15}
\setlength{\arrayrulewidth}{0.5pt}
\begin{tabular}{|>{\columncolor{gray!10}\centering\arraybackslash}m{5.5cm}|>{\centering\arraybackslash}m{6.5cm}|>{\centering\arraybackslash}m{6.5cm}|}
\hline
\rowcolor{gray!40}
\textbf{Essential Performance Metrics of Modern AI Infrastructure} & \textbf{Conventional Architecture} & \textbf{CXL-enabled Tray-based Architecture} \\[0pt] \hline
Scalability & Node-level or rack-level scale-up with limited resource expansion &  Row-level scale-up enabling flexible expansion of computational, interconnect, and memory resources  \\ \hline
Latency & High-latency, protocol overhead and software intervention delays with RDMA ($>$1 µs) & Low-latency, hardware-mediated protocol and cache-coherent (100--250 ns) \\ \hline
Memory Capacity & Low and fixed, tightly coupled CPU and GPUs architecture (192--288 GB per GPU) & Massive and flexible, dynamic composable memory pool ($>$ tens of TBs per node) \\ \hline
Memory Bandwidth & Low efficiency (memory copy for access external memory) & High efficiency (traffic reduction with coherent and pooled memory) \\ \hline
Computational Flexibility & Low flexibility (fixed or coarse-grained resource allocation) & High flexibility (dynamic and fine-grained resource allocation) \\ \hline
\end{tabular}}
\end{table}

\paragraph{How CXL meets diverse performance metrics.} Table~\ref{tab:conventioanl_cxltraycxl_architecture_comparison} analyzes how the CXL-enabled tray-based architecture can address critical performance metrics required by modern AI infrastructures (cf. Section~\ref{subsection:4_1}). Specifically, the proposed composable architecture provides a scalable and unified modular framework that enhances computational flexibility, expands memory capacity, optimizes memory bandwidth, reduces latency, and improves overall system scalability as below:

\begin{itemize}

\item \textbf{Scalability:} CXL's hierarchical switch design enables effective scalability through multi-level cascading of interconnect switches, facilitating incremental and seamless expansion of both computational and memory resources without architectural bottlenecks. Specifically, this scalable architecture coherently aggregates thousands of accelerators and memory endpoints into unified resource pools, addressing critical performance metrics, including computational throughput, model size, and overall system scalability. Thus, CXL-based infrastructures can accommodate rapidly increasing model sizes and complex parallel workloads, eliminating the need for disruptive hardware upgrades and supporting continuous adaptation to evolving AI demands.

\item \textbf{Latency:} CXL can fundamentally address latency constraints by providing direct, cache-coherent communication paths between memory and accelerator trays. Unlike conventional network-based approaches such as RDMA, which incur substantial protocol overhead and software-induced delays, CXL facilitates hardware-mediated memory interactions via direct load/store instructions. Specifically, accelerator and memory trays communicate through dedicated CXL interfaces, reducing round-trip latency. This low-latency communication is beneficial during latency-sensitive operations, including the decode phase of LLM inference. Thanks to the extended scale-up domain, CXL architectures can also optimize both network bandwidth utilization and latency sensitivity metrics, ensuring responsive and efficient AI deployments.

\item \textbf{Memory capacity and bandwidth:} The modular tray-based memory configuration enabled by CXL aggregates numerous memory expansion modules, each providing substantial memory capacity and high-bandwidth interfaces. By coherently pooling these memory modules through dedicated CXL controllers or switches, accelerators can directly access shared memory resources, mitigating performance bottlenecks related to limited memory bandwidth and insufficient capacity. Specifically, such cache-coherent pooled memory structures, implemented via CXL.cache, remove redundant data transfers and memory traffic, greatly benefiting workloads characterized by frequent, high-bandwidth memory accesses, such as RAG and KV caching scenarios.

\item \textbf{Computational flexibility:} CXL-based tray architectures support flexible and fine-grained resource scaling. Specifically, accelerator, memory, and compute trays can each be independently provisioned, upgraded, or reconfigured, enabling precise resource allocation aligned with specific workload requirements. For instance, GPU trays can scale to handle computationally intensive training phases or the inference prefill stage and reconfigure to meet stringent latency constraints during inference decode operations. This dynamic composability can address multiple critical performance metrics, including computational throughput, network bandwidth, and memory capacity, by optimizing resource allocation according to workload-specific demands, thus improving overall system utilization and operational efficiency.

\end{itemize}

By integrating these architectural solutions into a unified framework, CXL-based infrastructures effectively scale and dynamically adapt to evolving AI workload demands, optimizing key performance metrics. Specifically, extending the scale-up domain can eliminate communication overhead among resources, thus significantly enhancing overall performance. In addition, minimizing unnecessary scale-up switches further optimizes total cost of ownership. Moreover, cache-coherent memory sharing enables accelerators to directly access pooled memory without CPU intervention, reducing redundant data transfers and improving computational throughput as well as cost efficiency. Collectively, these capabilities position CXL-enabled modular architectures as a robust and flexible foundation capable of addressing the diverse and increasingly demanding requirements of large-scale AI workloads

\section{Composable CXL Architectures: From Integration Strategies to Empirical Validations}
\label{section:5}
In previous sections, we discussed how composable CXL architectures can address scalability and performance challenges inherent in modern AI infrastructures. Specifically, by supporting dynamic disaggregation and integration of memory and accelerator resources, these composable systems provide greater flexibility and adaptability compared to conventional RDMA-based approaches. However, the structural approach described thus far primarily focuses on modular tray-based configurations enabled by CXL. Practical implementations, such as the detailed composition of individual trays and the specific strategies for connecting GPUs or accelerators within these trays, require additional exploration. In this section, we specifically explore detailed implementation aspects, presenting integration strategies for composable CXL infrastructures. We also validate their effectiveness through empirical evaluations conducted across diverse real-world workloads.

\subsection{Memory and Accelerator Management in Composable CXL Data Centers}
\label{subsection:5_1}
This subsection discusses several considerations including architectural requirements for dedicated memory pools, efficient allocation and interconnection strategies for accelerators, and the development of comprehensive software frameworks necessary to manage memory and accelerator resources coherently.

\paragraph{Dedicated memory pool implementation and management.} Implementing dedicated memory pools involves several architectural considerations. First, it is necessary to determine the hardware architecture and practical methods for memory tray implementation. Second, identifying cost-effective switch placement and defining their roles within the composable infrastructure is essential. Finally, selecting appropriate backend memory media for these dedicated memory pools must be thoroughly evaluated.

\begin{figure}[t!]
   \centering
   \includegraphics[width=\linewidth]{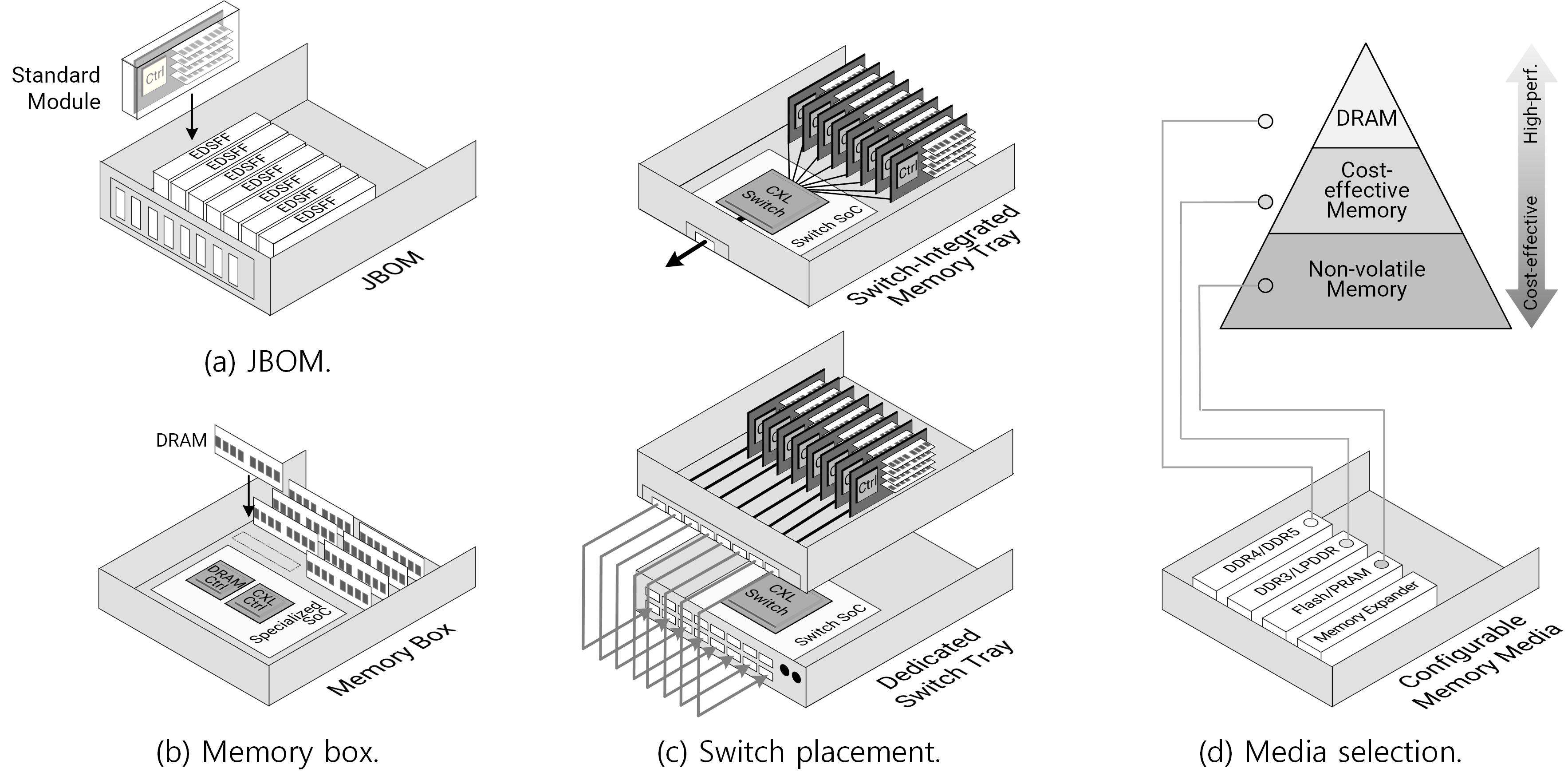}
   \caption{Tray-based disaggregation with CXL cache-coherent sharing.}
   \label{fig:memory_tray}
\end{figure}

Initially, let us consider the hardware architecture and practical implementation methods for configuring memory trays. As illustrated in Figure \ref{fig:memory_tray}a, memory trays can be configured either as \textit{Just a Bunch of Memory} (JBOM) units or as specialized, dedicated memory boxes. In a JBOM configuration, memory expanders utilize standard enterprise and data-center storage form factors, such as EDSFF modules \cite{edsff0,edsff,edsff1,edsff2}, arranged in arrays. Although many memory vendors adopt this standardized approach, it introduces increased costs and complexity due to vendor-specific performance variations and greater operational overhead. Specifically, replacing memory media in JBOM units requires simultaneous replacement of both CXL and memory controllers, despite their longer operational lifespans compared to memory media, thereby increasing the total cost of ownership.

Alternatively, memory trays can adopt dedicated memory boxes integrating specialized system-on-chips (SoCs) equipped with multiple DRAM and CXL controllers. As illustrated in Figure \ref{fig:memory_tray}b, this configuration directly supports high-performance raw or \textit{dual inline memory modules} (DIMM \cite{JACOB2008409, dimm, dimm-2}) by fully decoupling memory controllers from backend memory media. Such decoupling reduces maintenance complexity and operational costs. Moreover, by embedding both CXL and memory controllers within the memory tray, data centers can reuse legacy DIMMs or older DDR memory modules already present, offering additional cost advantages. This design provides significant flexibility, enabling operators to customize tray specifications, including port counts, bandwidth, and memory module types, to align with specific performance targets and workload requirements. Direct control over memory modules also allows operators to optimize memory media selection (e.g., DDR3 \cite{HARRIS2013474,wang2018design,kang20098,ddr3}, DDR4 \cite{ddr4controller, ddr4power, MicronTN4040}, LPDDR \cite{lpddr3, lpddr5, lpddr-samsung}), balancing performance and cost efficiency from the data centers' viewpoint. However, this strategy increases complexity in data-integrity management, coordination of heterogeneous hardware components, and handling sophisticated SoC designs at some extents.

Another important architectural consideration involves determining optimal switch placement for interconnecting memory expanders or SoC-based controllers within memory trays. As illustrated in Figure \ref{fig:memory_tray}c, integrating switches directly within each memory tray addresses compatibility concerns, version discrepancies, and functional variations among different expanders. Specifically, by treating each tray as a self-contained box, external systems can seamlessly abstract internal hardware variations and diverse memory technologies through standardized interfaces or controlled performance guarantees. This design aligns naturally with the previously discussed JBOM and dedicated memory-box architectures; however, similar drawbacks, particularly higher costs and limited flexibility, arise due to tight integration and reduced hardware adaptability. Alternatively, placing switches externally at dedicated switch trays or MoR/ToR positions allows memory trays to function passively, reducing cost, enhancing operational flexibility, and facilitating diverse, adaptive configurations tailored to specific data-center deployment scenarios.

In addition to tray configurations and switch placements, overall efficiency can be improved by diversifying memory media types and organizing them hierarchically within trays (cf. Figure \ref{fig:memory_tray}d). Traditional DRAM modules (e.g., DDR4 or DDR5) are costly and offer limited flexibility for workload-specific tuning. Thus, employing cost-effective or power-efficient memory solutions, such as LPDDR or DDR3, can be beneficial. Moreover, integrating HBM modules as intermediate buffering layers within memory expanders or trays can enhance performance. Specifically, these HBM modules can accommodate variations in expander performance and mitigate latency introduced by integrated or external switches when constructing dedicated memory pools. Depending on bandwidth requirements and positioning within the system, reusing legacy or lower-cost HBM modules, such as older HBM versions or modules with fewer channels, can effectively balance performance requirements and reduce overall system costs. In addition, emerging non-volatile memory technologies, including flash memory or phase-change memory (PRAM), can provide data persistence capabilities where necessary, further optimizing performance and cost profiles.

\paragraph{Accelerator resource management: Topologies and interconnect strategies.} Implementing accelerator composability requires careful consideration of several factors beyond basic interconnectivity. Specifically, topology design should reflect the unique characteristics of different accelerator vendors and consider various accelerator communication patterns and application-specific requirements. Here, we discuss accelerator resource management considerations in the context of LLM workloads exhibiting strong data locality and intensive data exchanges among adjacent accelerators, employing tensor parallelism. For general-purpose AI data centers characterized by random uniform traffic patterns, hybrid interconnect solutions will be addressed in Section \ref{section:6}.

\begin{figure}[t!]
    \centering
    \includegraphics[width=\linewidth]{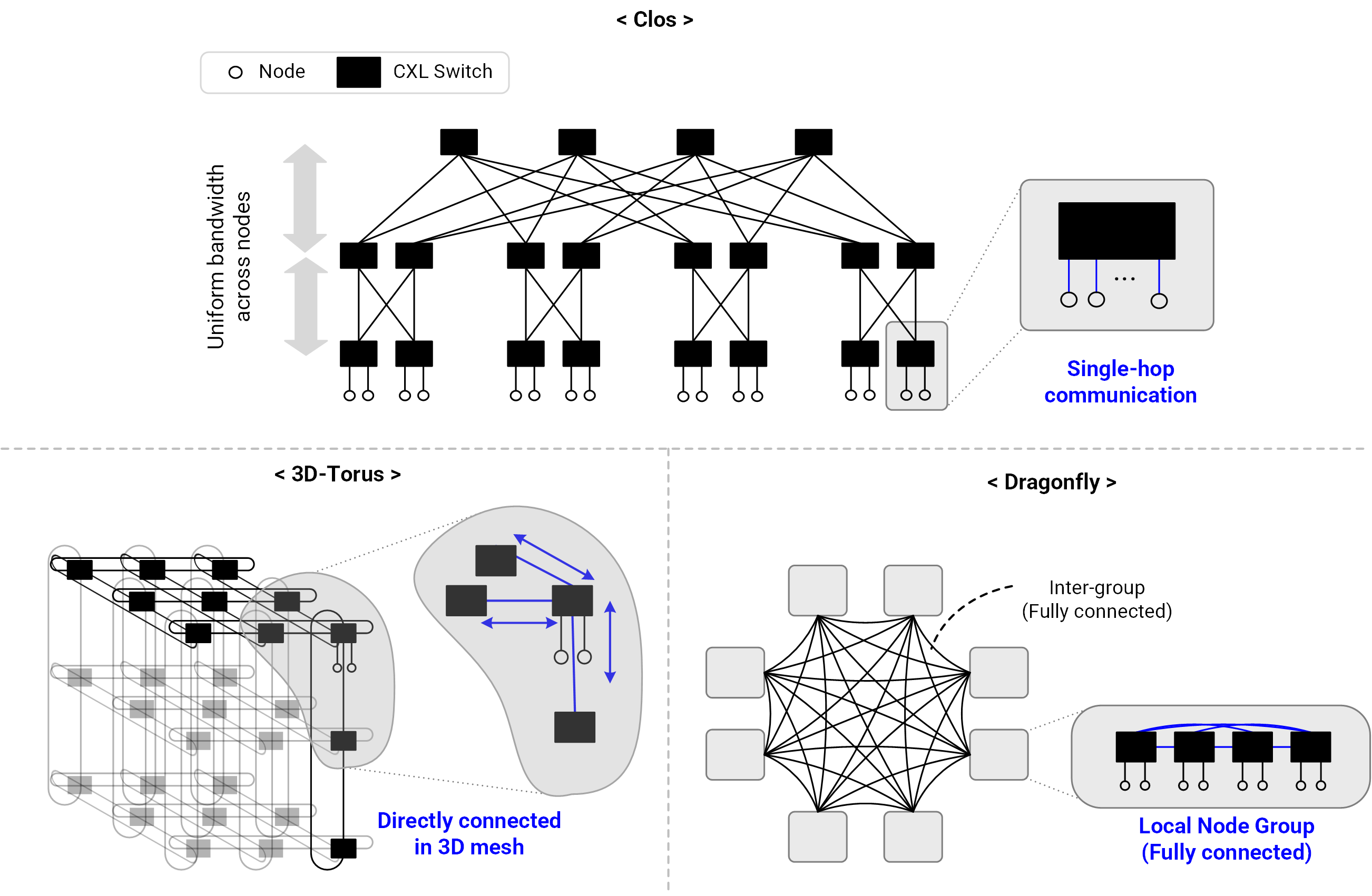}
    \caption{Topology comparison: Clos, 3D Torus, and Dragonfly.}
    \label{fig:topologycompare}
\end{figure}

In contrast to accelerator-optimized interconnect technologies that utilize a single dimensional topology (e.g., UALink and NVLink), CXL supports diverse topological configurations, providing flexibility for accelerator connectivity. Figure \ref{fig:topologycompare} compares the key characteristics of Clos \cite{clos,floor2,clos1}, 3D-Torus \cite{3dtorus,3dtorus1,3dtorus2,3dtorus3}, and DragonFly \cite{dragonfly,dragonfly1,dragonfly2} topologies, which are widely employed interconnect techniques in modern data centers. Specifically, Clos networks deliver uniform bandwidth across nodes through multi-stage switch hierarchies, offering flexibility at the expense of increased complexity and higher implementation costs. In contrast, the 3D-Torus topology directly interconnects nodes in a three-dimensional mesh structure, supporting short-range communications at relatively lower costs. However, it can introduce bottlenecks under heavy long-range communication patterns. The DragonFly topology integrates fully connected local node groups with indirect inter-group links, balancing cost and performance, though certain traffic patterns may degrade overall efficiency.

Considering the communication characteristics of LLM workloads (i.e., intensive data exchanges among nearby accelerators), 3D-Torus or DragonFly topologies might initially appear attractive. However, these approaches become impractical due to the exponential growth in required switch counts as the number of accelerators scales. Thus, maintaining cache coherence among accelerators using a single-hop Clos topology emerges as the most practical solution. This design principle aligns closely with fundamental interconnect strategies utilized by NVLink and UALink (detailed in subsequent sections), enabling local accelerator communication at reasonable hardware cost. Nevertheless, the current CXL specification limits cache-coherent accelerator connections to 256, necessitating alternative topological strategies or the incorporation of additional switches to support larger-scale deployments.

\begin{figure}[t!]
    \centering
    \includegraphics[width=0.9\linewidth]{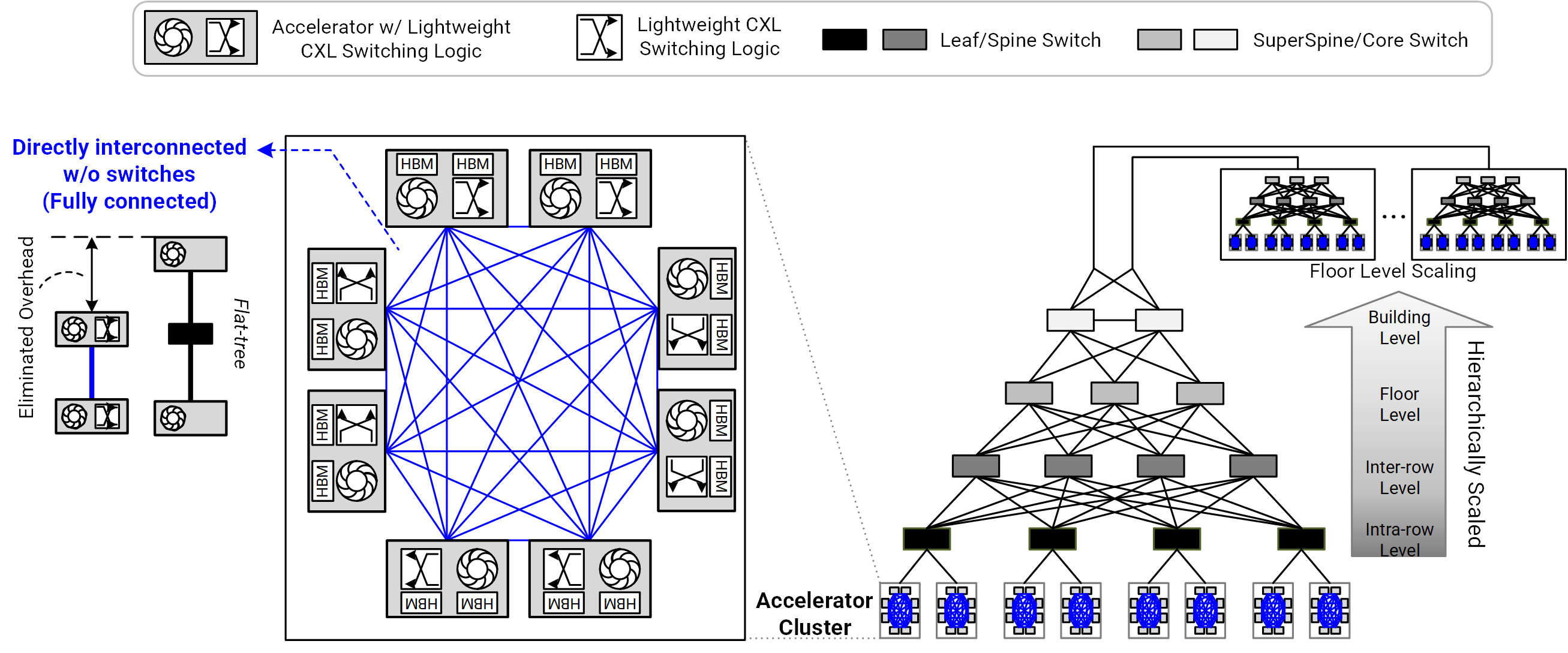}
    \begin{subfigure}{\linewidth}
        \begin{tabularx}{\textwidth}{
            p{\dimexpr.02\linewidth-2\tabcolsep-1.3333\arrayrulewidth}
            p{\dimexpr.48\linewidth-2\tabcolsep-1.3333\arrayrulewidth}
            p{\dimexpr.48\linewidth-2\tabcolsep-1.3333\arrayrulewidth}
            }
            \caption*{} &
            \caption{Directly interconnected accelerator cluster.} \label{fig:integ_acc} &
            \caption{Hierarchically scaled accelerator clusters.} \label{fig:integ_expan}
        \end{tabularx}
    \end{subfigure}
    \caption{Switch placement options.}
    \label{fig:aggressive}
\end{figure}

A more aggressive integration approach is illustrated in Figure \ref{fig:integ_acc}, where accelerators are interconnected using a fully-connected topology without intermediate switches. In this configuration, each accelerator integrates simplified and lightweight internal CXL switching logic, eliminating the overhead associated with external single-hop Clos topologies and additional interconnect hardware within clusters. This fully-connected architecture optimizes data transfers among adjacent accelerators, making it advantageous for LLM workloads exhibiting intensive localized communication patterns. Accelerator clusters structured in this manner can scale hierarchically through external CXL switches across multiple physical levels, such as racks and floors, as depicted by the hierarchical fully-connected topology across multiple physical levels in Figure \ref{fig:integ_expan}. Furthermore, leveraging accelerator-local HBM enhances intra-cluster communication efficiency. While this fully-connected approach can reduce accelerator-to-accelerator data exchange overhead, it demands substantial design revisions and implementation complexity for both accelerators and Type 1 and Type 2 CXL controllers. In addition, inter-cluster communication through external switches may introduce performance imbalances and requires careful management of accelerator-generated bandwidth traffic. Therefore, comprehensive evaluation and meticulous design are important for practical deployment at scale.

\paragraph{Unified management frameworks for composable resources.} The practical deployment of composable CXL infrastructures in AI data centers may require careful consideration of software-related architectural issues as well. Traditional static memory allocation incurs excessive data movements and latency, for frequently accessed data structures such as embedding tables in recommendation models and attention caches in transformer inference. To address these limitations, advanced software frameworks can be employed. These frameworks are required to incorporate predictive memory placement, proactive cache warming strategies, and adaptive eviction policies based on real-time access patterns. For example, dynamically allocating frequently accessed attention caches closer to accelerators can reduce latency and enhance throughput in LLM inference scenarios utilizing tensor parallelism.

Effective accelerator resource management also impacts overall system performance. As discussed previously, conventional GPU-CPU architectures tightly couple computational and memory resources, often leading to inefficient accelerator utilization under varying workloads. In contrast, composable architectures support independent scaling and dynamic resource allocation tailored to evolving workload demands. Achieving this flexibility may necessitate sophisticated software frameworks capable of real-time workload monitoring, predictive resource allocation, priority-driven scheduling, and rapid reconfiguration. Such frameworks are also required precisely coordinate hardware arbitration and minimize inter-accelerator communication latency.

In addition, integrated management of memory and accelerator resources may demand centralized monitoring frameworks beyond traditional static approaches. For example, static resource management techniques are inadequate for handling dynamic resource contention in composable systems. Advanced software solutions can be therefore needed to incorporate real-time telemetry collection, comprehensive performance analytics, including coherent memory efficiency, resource allocation latency, and contention metrics, and automated corrective actions. Future software frameworks could further incorporate reinforcement learning-based orchestration and predictive monitoring optimized explicitly for CXL environments, proactively resolving resource contention and dynamically adjusting allocations. Such advanced capabilities would significantly enhance responsiveness and efficiency for latency-sensitive AI workloads.

\subsection{Practical Case Studies of CXL Infrastructure in AI Workloads}
\label{subsection:5_2}
Building on the theoretical advantages of the previously introduced CXL infrastructure, this subsection provides empirical evidence and practical insights into the advantages of composable CXL infrastructures through real-system prototype evaluations. Specifically, we evaluate representative contemporary workloads from AI and HPC domains, including RAG, graph-based RAG (Graph-RAG), deep learning recommendation models (DLRM), and MPI-based scientific computing applications. Our prototype demonstrates that composable CXL infrastructure mitigates critical performance bottlenecks such as excessive latency, significant data movement overhead, and inefficient memory utilization, which are prevalent in traditional architectures dependent on RDMA-based networking.

\begin{figure}[t!]
    \centering
    \includegraphics[width=0.9\linewidth]{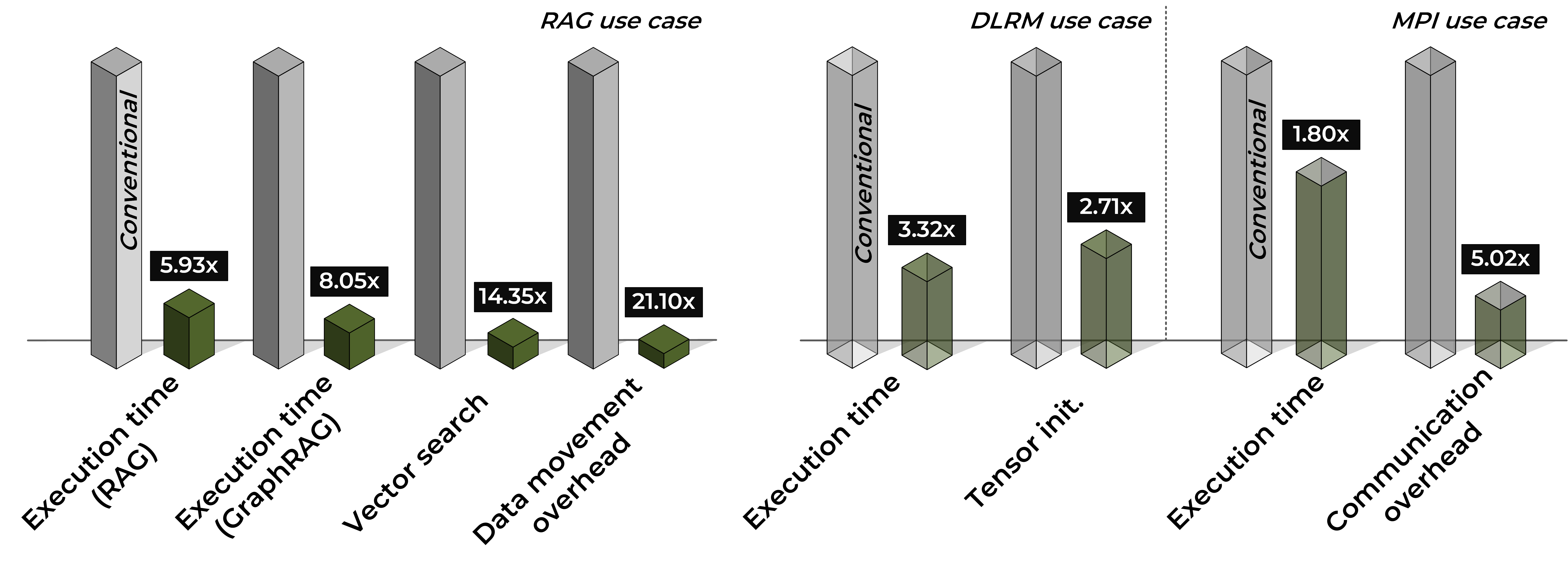}
    \caption{Summary of performance gains: RAG, Graph-RAG, DLRM, and MPI.}
    \label{fig:performance_gains}
\end{figure}

Figure \ref{fig:performance_gains} summarizes the performance improvements achieved by CXL compared to conventional systems across each scenario. Specifically, AI-based search workloads employing RAG and Graph-RAG integrated with LLMs exhibit a 14.35$\times$ reduction in execution time, alongside up to 21.1$\times$ decreases in data movement overhead relative to conventional architectures. Similarly, embedding-intensive DLRM workloads demonstrate approximately 3.32$\times$ faster inference execution and 2.71$\times$ accelerated tensor initialization. Furthermore, MPI-based HPC applications benefit from CXL-enabled direct memory sharing, achieving execution time improvements of approximately 1.8$\times$ and reducing communication overhead by up to 5.02$\times$.

In the subsequent subsections, we detail these scenarios, illustrating how the proposed CXL prototype addresses the performance challenges for each workload. Building upon these empirical insights, we also outline critical architectural implications, providing several guidelines for integrating CXL into future AI data center designs.

\begin{figure}[b!]
    \centering
    \includegraphics[width=\linewidth]{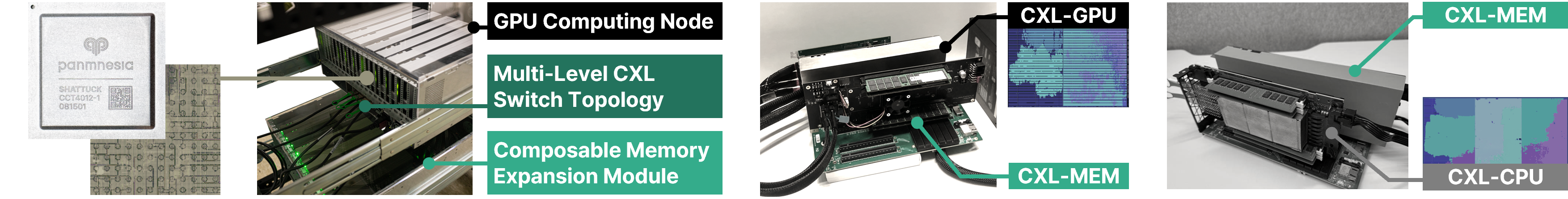}
    \begin{subfigure}{\linewidth}
        \begin{tabularx}{\textwidth}{
            p{\dimexpr.15\linewidth-2\tabcolsep-1.3333\arrayrulewidth}
            p{\dimexpr.32\linewidth-2\tabcolsep-1.3333\arrayrulewidth}
            p{\dimexpr.27\linewidth-2\tabcolsep-1.3333\arrayrulewidth}
            p{\dimexpr.27\linewidth-2\tabcolsep-1.3333\arrayrulewidth}
            }
            \caption{CXL IPs.} \label{fig:infra_ctrlr} &
            \caption{Real-system Prototype.} \label{fig:infra_prototype} &
            \caption{CXL GPU.} \label{fig:infra_gpu} &
            \caption{CXL CPU.} \label{fig:infra_cpu}
        \end{tabularx}
    \end{subfigure}
    \caption{A experimental end-to-end infrastructure compliant with the CXL 3.0 specification.}
    \label{fig:exp_infra}
\end{figure}

\paragraph{Experimental infrastructure.} To evaluate composable CXL-based architectures, we developed a unified experimental infrastructure employing a real-system prototype compliant with the CXL 3.0 specification. Figures \ref{fig:infra_ctrlr} and \ref{fig:infra_prototype} illustrate our silicon-proven CXL controller IPs and a practical setup, respectively. This setup consists of GPU computing nodes interconnected with composable memory expansion modules through a hierarchical CXL switch topology. The memory expanders and switches leverage standard CXL hardware stacks, while the GPU and CPU nodes integrate customized CXL hardware directly within their root ports and endpoint complexes. Due to the current absence of commercially available GPUs and CPUs supporting CXL 3.0, we utilized open-source Vortex GPU~\cite{opensourcegpu0,opensourcegpu00,opensourcegpu,opensourcegpu1} and RISC-V CPU~\cite{opensourcerisc,opensourcerisc1} microarchitectures, which we modified to incorporate essential CXL functionalities of our controller IPs. Prototype implementations of these modified GPUs and CPUs are depicted in Figures \ref{fig:infra_gpu} and \ref{fig:infra_cpu}, respectively.

Memory modules are organized into coherent, dynamically composable pools, presented to GPU computing nodes as distinct non-uniform memory access (NUMA \cite{lameter2013numa,denoyelle2018modeling,majo2013mis}) domains. This composable design enables GPU nodes to directly access shared memory resources, bypassing traditional CPU-mediated memory management or RDMA-based communication protocols. Although our evaluations utilized lightweight, open-source CPU and GPU implementations, our silicon-proven CXL controller and hardware stack IPs can be integrated with various third-party accelerators, NPUs, GPUs, and memory expanders. Specifically, these IPs can be modified to accommodate diverse cache and system-bus interfaces, facilitating straightforward integration into existing hardware platforms.

\begin{figure}[t!]
    \centering
    \includegraphics[width=\linewidth]{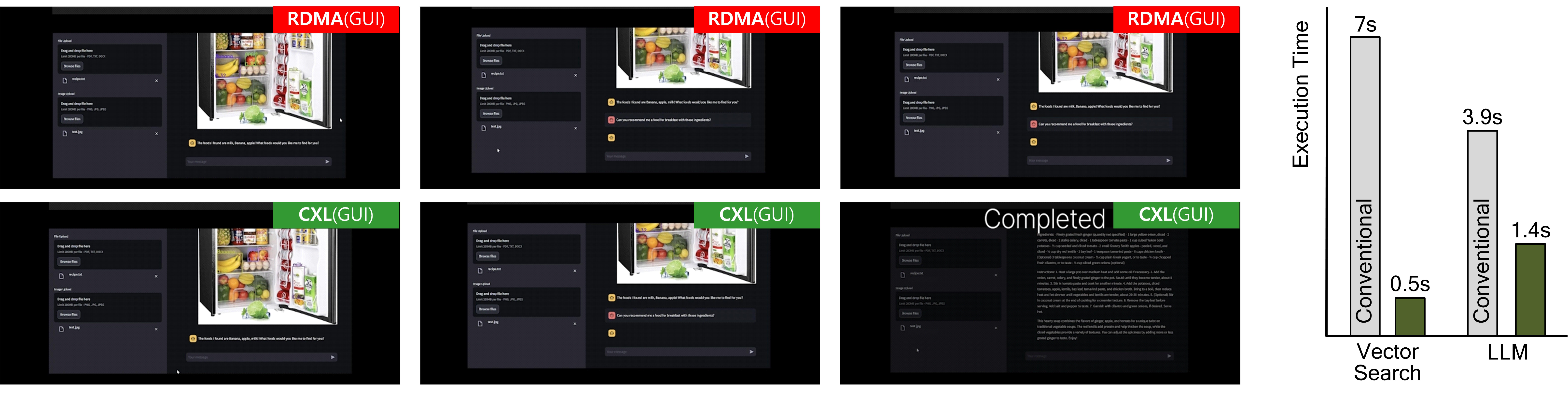}
    \begin{subfigure}{\linewidth}
        \begin{tabularx}{\textwidth}{
            p{\dimexpr.27\linewidth-2\tabcolsep-1.3333\arrayrulewidth}
            p{\dimexpr.27\linewidth-2\tabcolsep-1.3333\arrayrulewidth}
            p{\dimexpr.27\linewidth-2\tabcolsep-1.3333\arrayrulewidth}
            p{\dimexpr.22\linewidth-2\tabcolsep-1.3333\arrayrulewidth}
            }
            \caption{Beginning ({$t_0$}).} \label{fig:rag_t0} &
            \caption{Middle ({$t_1$}).} \label{fig:rag_t1} &
            \caption{End ({$t_2$}).} \label{fig:rag_t2} &
            \caption{Execution time.} \label{fig:rag_graph}
        \end{tabularx}
    \end{subfigure}
    \caption{RAG use case: recipe recommendation (\textit{Demo Video: \href{https://www.youtube.com/watch?v=lrhd4fu0KTc}{[Link]}}).}
    \label{fig:rag_usecase}
\end{figure}

\paragraph{RAG use case: Accelerating interactive retrieval and inference workloads.} Interactive retrieval tasks involving vector matching and real-time inference present significant challenges for traditional infrastructures due to high latency and intensive memory demands. To demonstrate the practical advantages of composable CXL infrastructures, we evaluated a user-friendly RAG scenario integrated with a contemporary LLM. As depicted in Figure~\ref{fig:rag_usecase}, this scenario represents a recipe recommendation system where users upload images of available food ingredients (typical refrigerator items) and specify preferred meal categories, such as breakfast or dinner. Embedding vectors for user-uploaded images are generated using a pre-trained visual-language model (e.g., CLIP \cite{clip,clip1,clip2}), ensuring accurate semantic representations. Subsequently, these embeddings are matched against pre-existing recipe embeddings stored in the system. Compared to conventional infrastructures utilizing RDMA-based interconnects, the composable CXL architecture demonstrates notable performance improvements in vector retrieval. These enhancements arise from reduced memory-access latency and decreased software overhead, which occur in RDMA-based systems.

Following vector retrieval, the retrieved embeddings directly served as inputs for the LLM-based inference phase, generating contextually relevant recipe recommendations. While the conventional RDMA-based systems typically exhibit retrieval and inference latencies ranging from several hundred milliseconds to seconds, our composable CXL infrastructure achieved lower latencies, enabling responses within tens of milliseconds. As illustrated in Figure~\ref{fig:rag_graph}, quantitative evaluations demonstrated that the composable CXL infrastructure completed the vector search and LLM in 0.5s and 1.4s, respectively, which are 14$\times$ and 2.78$\times$ faster than the baseline system. Such latency reductions are critical for user-facing applications, such as recommendation systems, where rapid, interactive responses enhance user experience and satisfaction.

\begin{figure}[b!]
    \centering
    \includegraphics[width=0.9\linewidth]{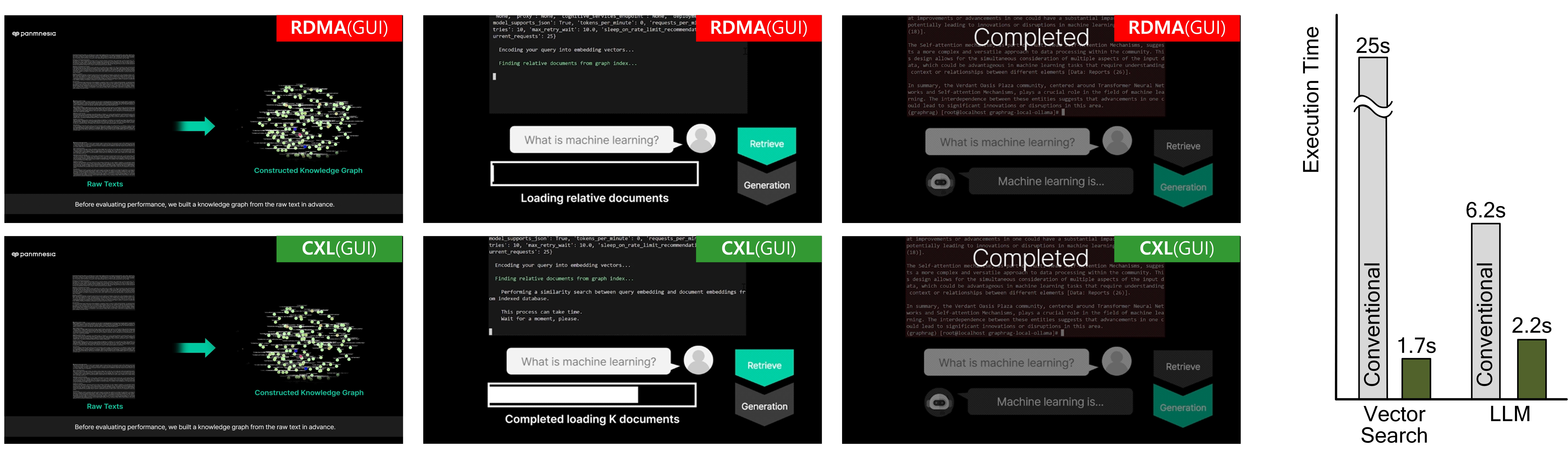}
    \begin{subfigure}{\linewidth}
        \begin{tabularx}{\textwidth}{
            p{\dimexpr.27\linewidth-2\tabcolsep-1.3333\arrayrulewidth}
            p{\dimexpr.27\linewidth-2\tabcolsep-1.3333\arrayrulewidth}
            p{\dimexpr.27\linewidth-2\tabcolsep-1.3333\arrayrulewidth}
            p{\dimexpr.21\linewidth-2\tabcolsep-1.3333\arrayrulewidth}
            }
            \caption{Beginning ({$t_0$}).} \label{fig:graphrag_t0} &
            \caption{Middle ({$t_1$}).} \label{fig:graphrag_t1} &
            \caption{End ({$t_2$}).} \label{fig:graphrag_t2} &
            \caption{Execution time.} \label{fig:graphrag_graph}
        \end{tabularx}
    \end{subfigure}
    \caption{Graph-RAG use case: knowledge graph and query retrieval.}
    \label{fig:graphrag_usecase}
\end{figure}

\paragraph{Graph-RAG use case: Accelerating knowledge graph-based retrieval and inference.} Graph-based RAG workloads, which integrate structured knowledge retrieval with inference, often experience substantial performance degradation due to high latency when accessing external memory resources. To address this challenge, we evaluated a composable CXL infrastructure using a Graph-RAG application scenario that integrates structured knowledge retrieval with LLM inference \cite{mavromatis2024gnn,grag,grag1,jang2023cxl}. As depicted in Figure~\ref{fig:graphrag_usecase}, the evaluation involved two primary operational phases: knowledge graph construction, followed by query-driven retrieval and inference.

Initially, raw textual data sources were processed using standard graph embedding techniques (e.g., RDF embeddings \cite{rdf,rdf1,rdf2,rdf3} or graph neural networks \cite{gnn,gnn1,gnn2}) to construct structured knowledge graphs optimized for efficient semantic retrieval. Subsequently, user queries were transformed into embedding vectors and matched against the structured knowledge graph using approximate nearest neighbor search methods such as HNSW \cite{hnsw0,hnsw,hnsw1} or FAISS \cite{faiss,faiss1}, enhancing retrieval speed. The retrieved embeddings then served as contextual inputs for the LLM inference process, generating coherent and contextually accurate responses.

Compared to the composable CXL infrastructure, the conventional baseline system utilized RDMA over InfiniBand, incurring storage and software-induced latencies during vector retrieval. Empirical analysis demonstrated that the composable CXL architecture reduced total workflow execution time by approximately 8.05$\times$ relative to the conventional RDMA-based baseline. In particular, while the conventional system takes tens of seconds, the composable CXL architecture completes the vector search and LLM inference phases in only 1.7s and 2.2s, respectively (cf. Figure~\ref{fig:graphrag_graph}). This latency reduction and execution speed improvement resulted from the cache-coherent memory pools enabled by CXL, eliminating redundant data copying, bypassing software overhead, and providing direct hardware-mediated memory access.

\begin{figure}[t!]
    \centering
    \includegraphics[width=\linewidth]{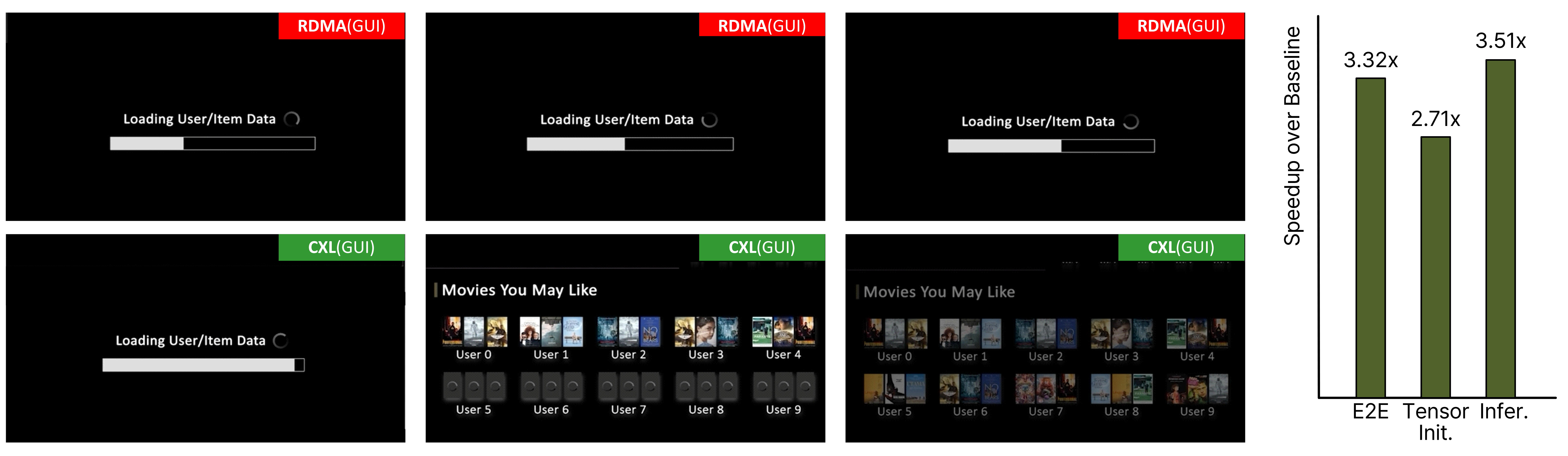}
    \begin{subfigure}{\linewidth}
        \begin{tabularx}{\textwidth}{
            p{\dimexpr.27\linewidth-2\tabcolsep-1.3333\arrayrulewidth}
            p{\dimexpr.27\linewidth-2\tabcolsep-1.3333\arrayrulewidth}
            p{\dimexpr.27\linewidth-2\tabcolsep-1.3333\arrayrulewidth}
            p{\dimexpr.21\linewidth-2\tabcolsep-1.3333\arrayrulewidth}
            }
            \caption{Beginning ({$t_0$}).} \label{fig:dlrm_t0} &
            \caption{Middle ({$t_1$}).} \label{fig:dlrm_t1} &
            \caption{End ({$t_2$}).} \label{fig:dlrm_t2} &
            \caption{Execution time.} \label{fig:dlrm_graph}
        \end{tabularx}
    \end{subfigure}
    \caption{DLRM use case (\textit{Demo Video: \href{https://www.youtube.com/watch?v=_v5zRxum4Ww}{[Link]}}).}
    \label{fig:dlrm_usecase}
\end{figure}

\paragraph{DLRM use case: Accelerating deep learning recommendation workloads.} DLRM workloads \cite{dlrmwork0,dlrmwork,dlrmwork1,dlrmwork2} require efficient embedding lookups, posing challenges related to memory capacity and latency for traditional data center infrastructures. Given these substantial memory and computational demands, it is essential to evaluate composable CXL-based architectures capable of addressing these issues. To this end, we analyzed embedding-intensive tensor initialization and inference phases representative of realistic recommendation scenarios, utilizing the composable infrastructure described previously.

As depicted in Figure~\ref{fig:dlrm_usecase}, the evaluation utilized embedding tables containing hundreds of GBs of parameters, reflecting large-scale production recommendation systems. During tensor initialization, embedding tables were loaded into memory, which is a phase in which the composable CXL infrastructure demonstrated notable performance improvements compared to the conventional RDMA-based baseline. Specifically, the baseline system employed RDMA, commonly deployed in production environments and characterized by higher software-induced communication overhead and latency. In contrast, the composable infrastructure reduced initialization latency and communication overhead via direct hardware-mediated memory access and coherent memory pooling. Following tensor initialization, both infrastructures executed repeated inference computations, simulating realistic operational conditions and verifying sustained performance over multiple inference cycles. Due to accelerated tensor initialization, the composable CXL infrastructure transitioned more rapidly into inference execution, improving overall responsiveness compared to the baseline's prolonged initialization delays.

As shown in Figure~\ref{fig:dlrm_graph}, empirical evaluations demonstrate that the composable CXL infrastructure achieves an overall throughput improvement of approximately 3.32$\times$ compared to the RDMA-based system. As described previously, the composable architecture accelerates tensor initialization and inference phases by 2.71$\times$ and 3.51$\times$, respectively. This performance improvement resulted from CXL's cache-coherent memory pools, enabling accelerator nodes direct hardware-level memory access without network-based communication stack overhead, reducing latency and data transfer overhead. Practically, such performance enhancements translate into improved user experiences for large-scale commercial platforms, including personalized content delivery in e-commerce and streaming services. Consequently, adopting composable, independently scalable CXL infrastructures enhances data center efficiency, meeting the evolving requirements of recommendation workloads.

\paragraph{MPI-based scientific applications: Evaluating memory sharing with CXL.} MPI-based scientific computing applications experience inter-node communication overhead and synchronization latency, limiting performance scalability in traditional network-based architectures \cite{bosilca2011scalability,shipman2006infiniband,zambre2019breaking,huang2023accelerating}. Although our primary focus remains on AI workloads, we evaluated representative MPI-based scientific computing applications to demonstrate the advantages of direct memory sharing enabled by composable CXL infrastructures. MPI-based scientific simulations, such as particle-in-cell (PIC) plasma simulations \cite{pic1,pic2,pic3} and computational fluid dynamics (CFD) simulations \cite{cfd,cfd1}, involve intensive inter-node data exchanges and frequent synchronization of simulation states, closely resembling communication patterns prevalent in distributed AI workloads. Typically, these MPI applications partition computational domains across multiple nodes and regularly synchronize boundary conditions and intermediate simulation data.

\begin{figure}[t!]
    \centering
    \includegraphics[width=0.9\linewidth]{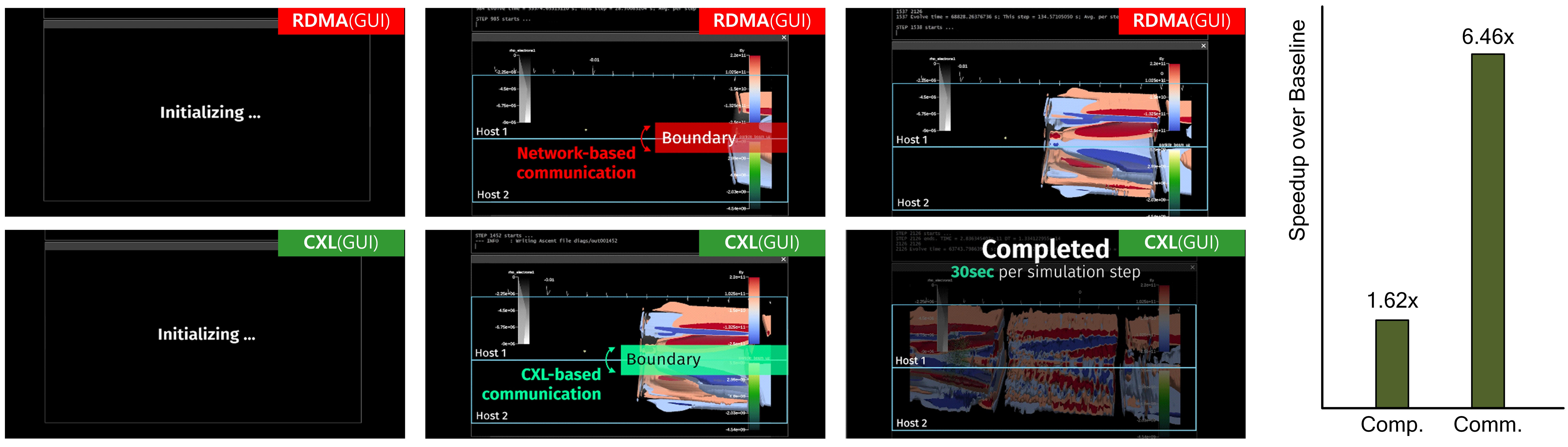}
    \begin{subfigure}{\linewidth}
        \begin{tabularx}{\textwidth}{
            p{\dimexpr.27\linewidth-2\tabcolsep-1.3333\arrayrulewidth}
            p{\dimexpr.27\linewidth-2\tabcolsep-1.3333\arrayrulewidth}
            p{\dimexpr.27\linewidth-2\tabcolsep-1.3333\arrayrulewidth}
            p{\dimexpr.21\linewidth-2\tabcolsep-1.3333\arrayrulewidth}
            }
            \caption{Beginning ({$t_0$}).} \label{fig:plasma_t0} &
            \caption{Middle ({$t_1$}).} \label{fig:plasma_t1} &
            \caption{End ({$t_2$}).} \label{fig:plasma_t2} &
            \caption{Execution time.} \label{fig:plasma_graph}
        \end{tabularx}
    \end{subfigure}
    \caption{MPI use cases: plasma simulation (\textit{Demo Video: \href{https://www.youtube.com/watch?v=zLhYn19EQCM}{[Link]}}).}
    \label{fig:mpi_usecase_plasma}
\end{figure}

To evaluate how composable CXL infrastructure mitigates these communication bottlenecks, we implemented two representative MPI scenarios. Figure~\ref{fig:mpi_usecase_plasma} shows the first scenario utilizing WarpX \cite{pic1}, a PIC framework, simulating interactions among hundreds of millions of charged particles (e.g., electrons and protons) distributed across multiple computational nodes. The conventional RDMA-based infrastructure relying on InfiniBand incurs significant overhead due to system software involvements and data copies across different device/network domains. In contrast, our composable CXL-based setup enabled host CPUs to directly store boundary-related particle data into dynamically composable memory regions shared by CXL.cache. Other nodes accessed this data immediately through direct load operations without invoking traditional software-driven network protocols, significantly reducing communication overhead and latency. Quantitatively, as depicted in Figure~\ref{fig:plasma_graph}, the CXL-based configuration eliminates explicit synchronization overhead inherent to conventional RDMA-based implementations. This elimination results in reductions of computation and communication latencies by 1.62$\times$ and 6.46$\times$, respectively.

\begin{figure}[t!]
    \centering
    \includegraphics[width=0.9\linewidth]{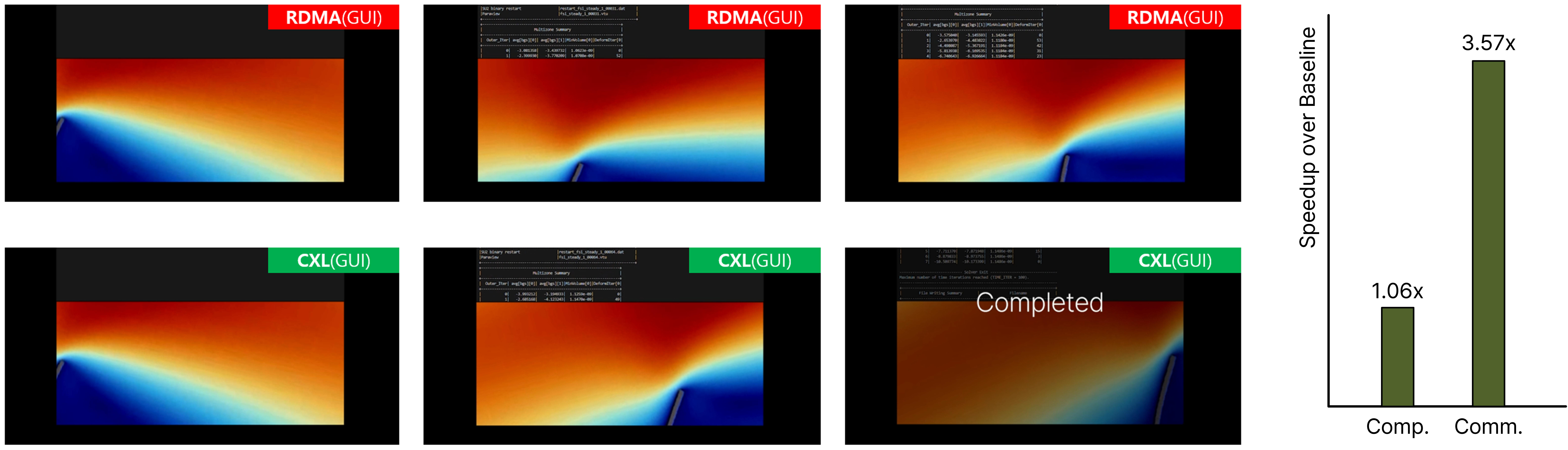}
    \begin{subfigure}{\linewidth}
        \begin{tabularx}{\textwidth}{
            p{\dimexpr.27\linewidth-2\tabcolsep-1.3333\arrayrulewidth}
            p{\dimexpr.27\linewidth-2\tabcolsep-1.3333\arrayrulewidth}
            p{\dimexpr.27\linewidth-2\tabcolsep-1.3333\arrayrulewidth}
            p{\dimexpr.21\linewidth-2\tabcolsep-1.3333\arrayrulewidth}
            }
            \caption{Beginning ({$t_0$}).} \label{fig:cfd_t0} &
            \caption{Middle ({$t_1$}).} \label{fig:cfd_t1} &
            \caption{End ({$t_2$}).} \label{fig:cfd_t2} &
            \caption{Execution time.} \label{fig:cfd_graph}
        \end{tabularx}
    \end{subfigure}
    \caption{MPI use cases: fluid simulation.}
    \label{fig:mpi_usecase_cfd}
\end{figure}

In the second scenario, illustrated in Figure~\ref{fig:mpi_usecase_cfd}, we evaluated a CFD simulation involving intensive synchronization of fluid states across domain partitions. Traditional RDMA-based network communications incurred substantial delays and overhead during synchronization events. By adopting the composable CXL infrastructure, host CPUs directly accessed fluid simulation states stored in composable memory pools, replacing conventional RDMA-based communication with direct shared-memory interactions. In this approach, individual hosts independently performed computations required for CFD simulation. Although certain calculations spanned multiple CPUs, collective communication or explicit synchronization was unnecessary for data aggregation or updates. This is because data consistency and coherence are managed through CXL.cache, enabling CPUs to access uniform memory spaces as if accessing local memory. Consequently, synchronization overhead was significantly reduced, achieving approximately a 1.06$\times$ reduction in computation time and a 3.57$\times$ reduction in communication time compared to the conventional RDMA-based baseline (cf. Figure~\ref{fig:cfd_graph}).

Performance analyses highlight two principal advantages of composable CXL infrastructure. First, the elimination of explicit RDMA-based network operations reduces latency and software overhead through direct, cache-coherent, hardware-mediated memory sharing. Second, resource disaggregation and memory pooling significantly streamline data management, improving scalability and operational efficiency in distributed computing environments. Practically, these enhancements substantially increase scalability and efficiency of large-scale scientific research infrastructures, particularly in domains such as climate modeling, astrophysics, and fusion research, where simulation speed directly impacts research productivity and accuracy. Although scientific MPI workloads fundamentally differ from AI-specific applications, we believe that these evaluations offer valuable insights into potential advantages of employing similar memory-sharing strategies within large-scale distributed AI infrastructures.

\section{Beyond CXL: Optimizing AI Resource Connectivity via Hybrid Link Architectures}
\label{section:6}
While CXL addresses critical memory-capacity expansion and coherent data-sharing challenges, integrating complementary interconnect technologies enables targeted enhancements for specific accelerator-centric workloads requiring efficient intra-accelerator communication, being able to support diverse workload demands and optimizing overall data center efficiency.

\begin{figure}[b!]
    \centering
    \includegraphics[width=0.95\linewidth]{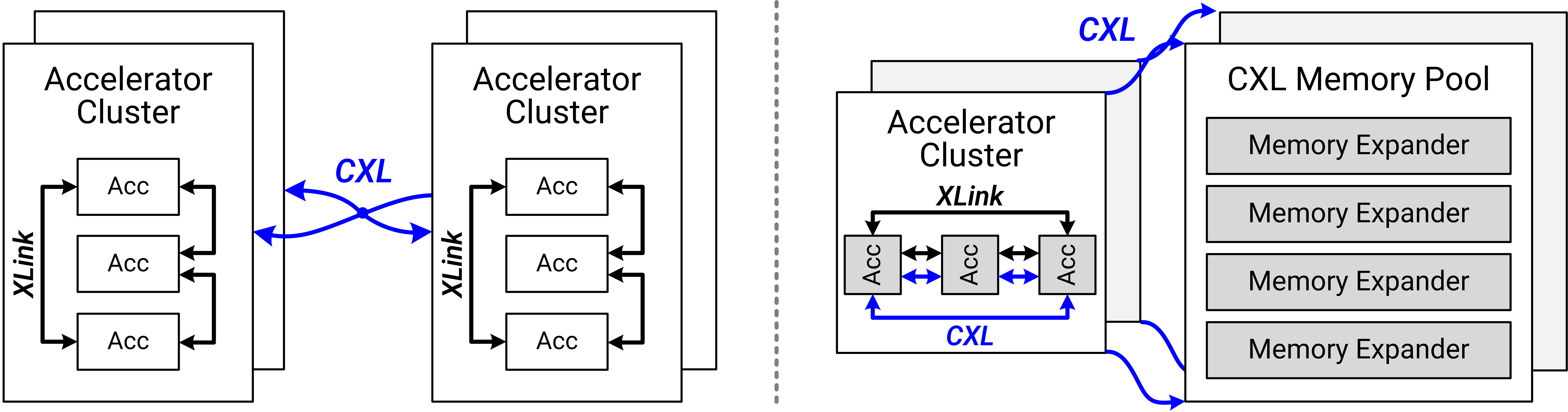}
    \begin{subfigure}{\linewidth}
        \begin{tabularx}{\textwidth}{
            p{\dimexpr.49\linewidth-2\tabcolsep-1.3333\arrayrulewidth}
            p{\dimexpr.49\linewidth-2\tabcolsep-1.3333\arrayrulewidth}
            }
            \caption{Accelerator-centric clusters.} \label{fig:cluster} &
            \caption{Tiered memory architectures.} \label{fig:tiering}
        \end{tabularx}
    \end{subfigure}
    \caption{A high-level viewpoint of hybrid link architectures (CXL-over-XLink).}
    \label{fig:xlink}
\end{figure}

Two prominent accelerator-focused interconnect technologies are \textit{Ultra Accelerator Link} (UALink) and NVIDIA's \textit{NVLink}, collectively termed \textit{Accelerator-Centric Interconnect Link} (XLink) in this technical report. XLink technologies provide direct, point-to-point connections explicitly optimized for accelerator-to-accelerator data exchanges, enhancing performance within tightly integrated accelerator clusters. In contrast to CXL, these XLink technologies do not support protocol-level cache coherence or memory pooling; instead, their focus is efficient, low-latency data transfers among accelerators with a single-hop Clos topology interconnect architecture. While both UALink and NVLink share this common objective, they differ in implementation specifics: UALink employs Ethernet-based communication optimized primarily for large-sized data transfers, whereas NVLink utilizes NVIDIA's proprietary electrical signaling, tailored for small-to-medium-sized data exchanges, such as tensor transfers and gradient synchronization between GPUs.

Integrating CXL and XLink into a unified data center architecture, termed \textit{CXL over XLink}, including \textit{CXL over NVLink} and \textit{CXL over UALink}, leverages their complementary strengths to optimize overall system performance. As depicted in Figures~\ref{fig:cluster} and \ref{fig:tiering}, this integration adopts two architectural proposals: i) ``accelerator-centric clusters,'' optimized specifically for rapid intra-cluster accelerator communication, and ii) ``tiered memory architectures,'' employing disaggregated memory pools to handle large-scale data. XLink typically supports optimized intra-node communication using direct, single-hop Clos topologies, making it effective for bandwidth-sensitive operations such as frequent tensor exchanges and gradient synchronization. However, the single-hop Clos topologies limit scalability, restricting the maximum number of connected accelerators and memory devices. In contrast, CXL enables scalable accelerator interconnections through multi-level switch cascading, facilitating diverse topologies and coherent memory pooling across multiple clusters or data centers. This allows dynamic memory allocation critical for memory-intensive workloads such as KV caching and RAG. Furthermore, CXL supports efficient inter-node data sharing through protocol-level cache coherence and instruction-level memory transactions, minimizing redundant data transfers and improving memory utilization. Composable disaggregation physically separates computational resources from memory pools, enabling independent scaling, simplified maintenance, and flexible hardware upgrades. Compute nodes interconnected via XLink, alongside memory resources managed through CXL, improve operational flexibility, accommodating rapid transitions between compute-intensive training and latency-sensitive inference workloads. In addition, memory resources can be organized hierarchically, optimizing allocation strategies according to specific performance and capacity requirements.

In this section, we first provide an overview of XLink technologies, emphasizing key architectural features and optimizations of UALink and NVLink. We then discuss several hybrid architecture strategies of CXL-over-XLink, highlighting how the complementary strengths of CXL and XLink address diverse and evolving workload requirements in modern AI data centers.

\subsection{Background on Accelerator-Centric Interconnects: UALink and NVLink}
\label{subsection:6_1}
\begin{figure}[t!]
    \centering
    \includegraphics[width=0.9\linewidth]{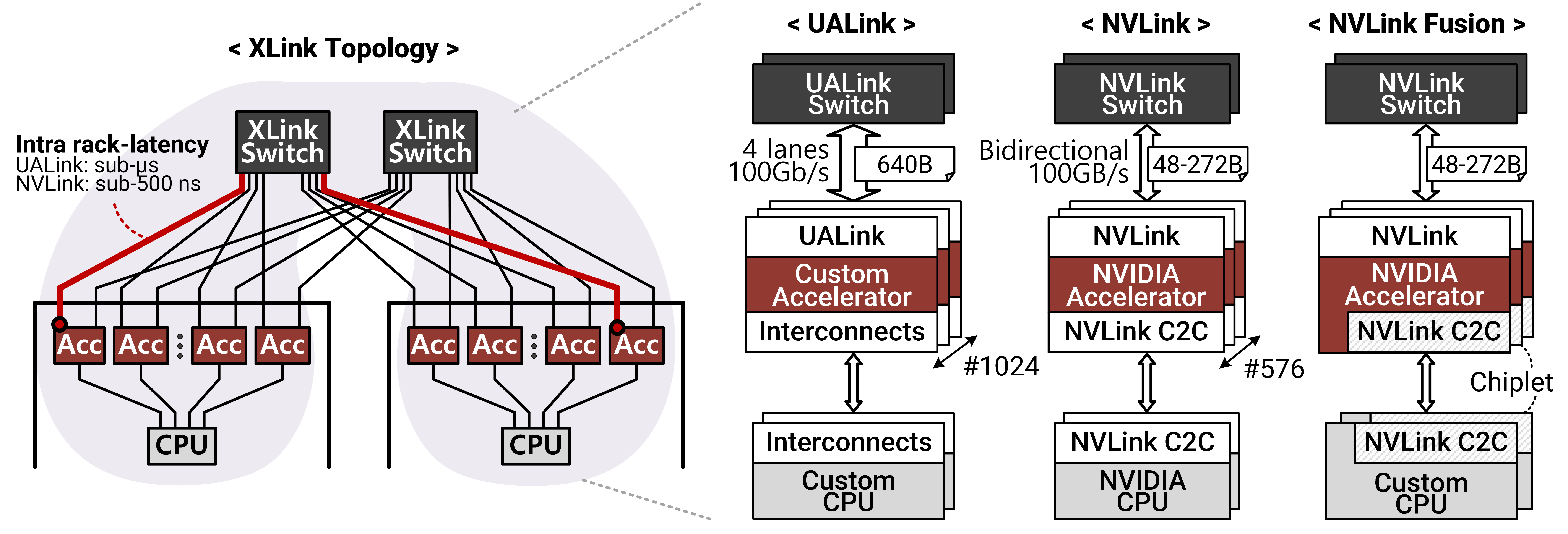}
    \caption{Accelerator-centric interconnects.}
    \label{fig:xlink}
\end{figure}

\paragraph{Ultra Accelerator Link.} UALink is an accelerator-centric interconnect optimized for accelerator-to-accelerator communication within data centers~\cite{ualink,ualink1}. In contrast to CXL, which emphasizes memory disaggregation, coherent memory management, and unified memory spaces, UALink prioritizes direct, high-throughput data transfers between accelerators. UALink is effective for workloads requiring large data transfer and synchronization. Each UALink port typically provides bandwidth up to 100 GB/s via a standard 4-lane configuration.

UALink 1.0~\cite{ualink,buildualink}, introduced in early 2025, shares architectural similarities with GPU-centric interconnects such as NVLink. However, it is designed as an open standard to support diverse accelerators beyond vendor-specific GPUs (e.g., NVIDIA GPUs). UALink utilizes a single-hop Clos switched topology, establishing dedicated, low-latency communication paths among accelerators, theoretically supporting clusters of up to 1,024 accelerators. As illustrated in Figure~\ref{fig:xlink}, this topology simplifies interconnect structure, reducing intra-rack latency to sub-microsecond levels (<1~$\mu$s \cite{ualinkwhitepaper}). By minimizing complexity and enhancing scalability, UALink can address stringent synchronization and high-throughput communication requirements common in densely interconnected accelerator environments.

To maximize throughput, UALink employs 640B data link flits optimized for large data transfers. Although it also supports instruction-level memory access mechanisms, these primarily serve auxiliary management and control functions rather than core high-throughput data operations. Furthermore, UALink explicitly operates as a \textit{non-coherent protocol}; it does not inherently support cache coherence or coherent memory transactions. This design clearly differentiates UALink from CXL, which primarily targets frequent data transactions, coherent memory sharing, cache coherence management, and resource disaggregation.

UALink leverages Ethernet-based topologies tailored for large-scale data transfers and collective communication patterns prevalent in distributed computing environments, such as all-gather operations. Historically, these communication patterns have been extensively used in distributed systems, preceding their adoption in contemporary multi-GPU and accelerator-rich infrastructures~\cite{chan2007collective,algorithm}. To achieve maximum bandwidth and precise data alignment, UALink Ethernet interfaces adopt optimized frame structures, reduced protocol overhead, and hardware-level synchronization mechanisms engineered for accelerator synchronization requirements~\cite{ualink,uascale}.

\paragraph{NVLink and NVLink Fusion.} NVLink is also an interconnect optimized for GPU-to-GPU communication, offering high bandwidth and low latency within GPU-centric data center environments. Introduced before UALink, NVLink enhanced GPU-to-GPU data transfer performance over existing standards (e.g., PCIe). Since its initial release in 2014, NVLink has undergone multiple generational updates, with NVLink 5.0 being the latest version introduced in 2024. NVLink's characteristics are beneficial for deep learning training workloads and HPC applications when deployed with NVIDIA GPUs.

Specifically, NVLink 5.0 provides 50 GB/s of unidirectional bandwidth per link, resulting in 100 GB/s of total bidirectional bandwidth per link~\cite{nvlink5.0}. Deployed primarily through NVIDIA's proprietary NVSwitch crossbar, NVLink efficiently supports configurations ranging from tens of GPUs (e.g., NVLink72~\cite{GB200,GB200200,GB300}) to larger setups utilizing inter-rack network components (e.g., NVLink576 with long-distance network elements). Despite supporting larger-scale deployments, NVLink targets GPU-node or GPU-cluster scales rather than broader rack-level or multi-rack scenarios. Similar to UALink, NVLink employs single-hop Clos (full-mesh) topologies, minimizing latency for critical collective operations such as All-Reduce and All-Gather communications used in transformer-based model training. NVLink 5.0 achieves low-latency communication of less than 500~ns \cite{lutz2020pump}.

In contrast to UALink, NVLink utilizes a smaller 48B$\sim$272B flit\footnote{While the actual flit size in NVLink is 16B (128-bit), each packet is composed of one header flit followed by up to 16 data flits. The minimum transmission unit consists of 2 data flits (32B), and the maximum includes 16 data flits (256B), yielding total packet sizes between 48B and 272B. In this section, the term NVLink flit denotes a complete packet constructed in this format.} \cite{zhang2025nvbleed} optimized for efficient transfer of moderate-sized tensors and gradient data. NVLink supports per-node memory-region unification rather than protocol-level cache coherence, simplifying programming complexity and reducing software overhead at the node-level. Historically, NVLink interoperability has been restricted primarily to NVIDIA products, complicating integration into heterogeneous systems.

\begin{table}[t!]
\centering
\caption{Technical specification comparison of CXL, UALink, and NVLink (Extended).}
\label{tab:cxl_ualink_nvlink_comparison_extended}
\resizebox{\textwidth}{!}{
\small
\renewcommand{\arraystretch}{1.15}
\setlength{\arrayrulewidth}{0.5pt}
\begin{tabular}{|>{\columncolor{gray!10}\raggedright\arraybackslash}m{4.0cm}|>{\raggedright\arraybackslash}m{5.5cm}|>{\raggedright\arraybackslash}m{5.2cm}|>{\raggedright\arraybackslash}m{5.2cm}|}
\hline
\rowcolor{gray!40}
\textbf{Specification} & \textbf{CXL 3.0} & \textbf{UALink 1.0} & \textbf{NVLink 5.0} \\ \hline
Unidirectional Bandwidth (GB/s) & 128 (x16 lanes per link, PCIe 6.0) & 100 (x4 lanes per link) & 50 (x2 lanes per link) \\ \hline
Latency & Hundreds of ns (100--250 ns typical) & Sub-µs ($<$1 µs within rack) & Sub-500 ns ($<$500 ns within rack)\\ \hline
Flit Size & 256B (PBR), 68B (HBR) & 640B & 48B$\sim$272B \\ \hline
Cache Coherency & Yes (hardware-level) & No & No (only support by NVLink C2C) \\ \hline
Memory Pooling & Yes & No (only within UALink connected accelerators)& No (only within NVLink connected GPUs) \\ \hline
Topology & Point-to-point, switched fabric (various topologies) & Point-to-point, switched fabric (only single-hop Clos) & Point-to-point, switched fabric (may only single-hop Clos) \\ \hline
Scalability & Up to 4096 devices & Up to 1024 accelerators & Up to 576 GPUs \\ \hline
Typical Deployment Scale & Rack or multi-rack scale & Intra-rack clusters & GPU-node or GPU-cluster scale \\ \hline
Use-case / Primary Workload & Memory disaggregation, coherent memory pooling & Accelerator-to-accelerator collective transfers & GPU tensor exchanges, gradient synchronization \\ \hline
Consortium & CXL Consortium & UALink Consortium & NVIDIA \\ \hline
Interoperability & Open industry standard & Ethernet-based openness & Proprietary (partial openness via NVLink Fusion) \\ \hline
Initial Release (Year) & CXL 1.0 (2019) & UALink 0.49 (2024) & NVLink 1.0 (2016) \\ \hline
Current Version (Year) & CXL 3.0 (2022), CXL 3.2 (2024) & UALink 1.0 (2025) & NVLink 5.0 (2024), Fusion (2025) \\ \hline
\end{tabular}}
\end{table}

To address this limitation, NVLink Fusion~\cite{fusion,fusion1} has been recently introduced, improving interoperability by enabling connections to external processors such as CPUs, NPUs~\cite{npu,npu1,npu2,npu3}, and AI-specific processors~\cite{mtia,trainium,inferentia,maia,gaudi}. NVLink Fusion provides two primary components: i) a short-reach C2C interface available as coherent IP for external processors, and ii) a Chiplet-based implementation designed for integration with diverse processing units, further optimizing CPU-to-GPU communication.

NVLink Fusion preserves NVLink's intrinsic advantages of high bandwidth and low latency while improving flexibility and shared-memory efficiency between CPUs and GPUs. However, NVLink Fusion is known to require the inclusion of at least one NVIDIA component within interconnected systems \cite{fusion,fusion1,fusion2}. Thus, it currently does not support resource disaggregation or vendor-neutral composability in broader AI infrastructure deployments.

\paragraph{Comparative summary of CXL and XLink technologies.} Table \ref{tab:cxl_ualink_nvlink_comparison_extended} summarizes the characteristics of the three interconnect technologies, CXL, UALink, and NVLink, discussed in this report.

As described previously, CXL separates compute and memory resources physically, supporting coherent memory pools and cache coherence. As illustrated in the table, CXL typically exhibits lower latency than other interconnect technologies, though specific latencies can vary based on actual implementations. Its cache coherence capability enables accelerators to directly service data from local caches, significantly reducing external data transfers and optimizing performance for data with high locality. CXL connections support up to 256 accelerators (Type 1 or Type 2 devices), while memory-type device connections can scale up to 4,096 endpoints within a single interconnect network. In addition, CXL provides practical PBR routing and switch cascading, greatly enhancing scalability, cache coherence management, and flexible memory allocation. Consequently, CXL is well-suited for composable environments requiring frequent synchronization, intensive memory utilization, and flexible resource configuration.

In contrast, XLink technologies emphasize fast and direct accelerator-to-accelerator data transfers, with a primary focus on high-bandwidth connectivity. While per-transfer latency is generally higher than CXL, XLink technologies can achieve greater aggregate bandwidth by employing larger flit sizes or device-specific optimizations for GPUs or accelerators. Specifically, UALink utilizes Ethernet-based networks, transferring large-scale data optimized for frequent inter-accelerator communication and collective communication patterns. NVLink, however, is tailored to GPU-centric workloads, efficiently exchanging small to medium-sized tensor data or gradients through optimized bandwidth and latency via smaller flit sizes. Both UALink and NVLink assume collective communications through data copying and distributed processing rather than data sharing; thus, neither supports hardware-level cache coherence.

Therefore, combining efficient accelerator communication capabilities of XLink technologies with CXL’s flexible memory pooling and cache coherence into a hybrid data center architecture provides a complementary and robust solution. This integration enables accelerators interconnected via UALink or NVLink to leverage CXL-managed memory resources and coherence protocols. Such hybrid architectures can expand the scale-up domain beyond traditional data-parallel approaches, enhancing resource utilization and overall system efficiency.

\subsection{Integrated Accelerator-Centric, CXL-over-XLink Supercluster Architecture}
\label{subsection:6_2}
To accommodate diverse demands of large-scale AI workloads, scaling beyond a single accelerator cluster necessitates efficient inter-cluster communication. Here, the term ``cluster'' refers to a rack-scale, multi-accelerator system as introduced in the previous data center architecture.  Unlike point-to-point intra-cluster networks, inter-cluster connections require more scalable and flexible topologies supporting extensive resource sharing and composability across broader data center infrastructures. CXL can fulfill this requirement through hierarchical, multi-level switching structures, enabling coherent memory pooling among distributed accelerator clusters.

In this subsection, we define a CXL-over-XLink-based \textit{supercluster}, a scalable and hierarchical architecture optimized for accelerator-intensive tasks. A supercluster consists of multiple accelerator clusters interconnected through CXL fabrics. Within each individual accelerator cluster, NVLink or UALink serves as the primary intra-cluster interconnect, providing direct, high-bandwidth, and low-latency communication among accelerators. The detailed configurations of these clusters are described below.

\begin{figure}[t!]
    \centering
    \includegraphics[width=1\linewidth]{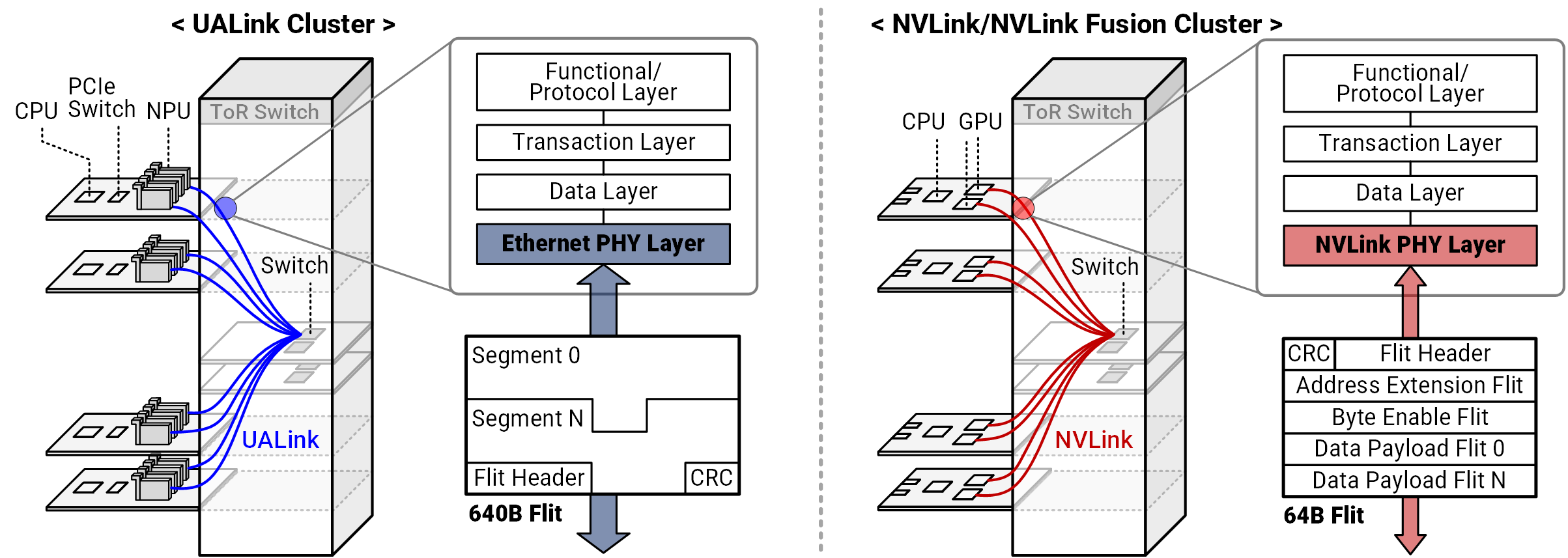}
    \caption{Accelerator-centric intra-cluster design.}
    \label{fig:xlinkcluster}
\end{figure}

\paragraph{Accelerator-centric intra-cluster design with UALink and NVLink} Within the CXL-over-XLink-based supercluster architecture, NVLink and UALink serve as intra-cluster interconnect technologies, enabling efficient construction of accelerator clusters. As previously discussed, these interconnect technologies share fundamental design principles, employing single-hop Clos switching topologies optimized for relatively small-scale accelerator clusters, focusing primarily on intra-accelerator communications (cf. Figure \ref{fig:xlinkcluster}). Specifically, NVLink supports accelerator clusters composed of multiple GPUs or combined GPU-CPU-memory nodes. NVLink integrates up to 72 GPUs interconnected through multiple NVSwitches, while CPUs within each node are interconnected to GPUs via NVLink C2C interfaces. Similarly, UALink explicitly targets accelerator connectivity within clusters or racks, directly attaching CPU modules to accelerators according to application-specific needs. Accelerators in UALink-based clusters communicate exclusively through UALink switches, theoretically supporting single-hop Clos topologies of up to 1,024 accelerators. This scalability benefits smaller logic-sized, AI-specific accelerators such as NPUs. However, for larger logic-sized accelerators like GPUs, which restrict the number of accelerators per node (e.g., two GPUs per node in GB200/300), the practical deployment scale within a rack closely resembles NVLink configurations (i.e., around 72 accelerators). With UALink, CPUs connect to accelerators through PCIe switches; however, alternative short-reach interconnect solutions such as UCIe~\cite{ucie2.0} may also be employed similarly to NVLink C2C.

Accelerator clusters configured in this manner are likely to consist exclusively of one of the two XLink technologies per cluster, rather than mixing hardware components supported by different interconnects within a single cluster. This limitation arises from fundamental technological differences and interoperability constraints between NVLink and UALink. Specifically, each interconnect employs distinct physical layers and data formats. NVLink uses NVIDIA's proprietary high-speed PHY interfaces with relatively small 48B$\sim$272B flits, whereas UALink adopts Ethernet-based PHY interfaces with significantly larger 640B flits. Thus, we believe that these differences in flit formats, protocol operations, and underlying PHY layers severely limit the feasibility of integrating NVLink and UALink hardware within the same cluster. Therefore, from a strategic interoperability perspective, NVLink requires at least one NVIDIA component (e.g., NVIDIA GPUs), restricting integration with fully third-party accelerator configurations.

Therefore, accelerator clusters employing NVLink and NVSwitches within a CXL-over-XLink supercluster predominantly consist of NVIDIA GPUs, complemented by specialized accelerators optimized for computational tasks not efficiently handled by GPUs. For example, data centers deploying NVIDIA GPUs may integrate accelerators tailored for branch-intensive computations (e.g., tree-based models and conditional logic), workloads with irregular control flows, sparse and irregular memory access patterns (such as graph processing or sparse matrix operations), or latency-critical real-time tasks. Such tasks align poorly with GPU architectures optimized for highly parallel computations. Integrating these heterogeneous accelerators within NVLink-based clusters thus enables data centers to accommodate diverse application demands while maintaining optimized intra-cluster communication performance.

In contrast, UALink-based clusters mainly comprise non-NVIDIA accelerators, such as AMD GPUs or AI-specific processors including Meta's MTIA~\cite{mtia}, Amazon's Trainium~\cite{trainium} and Inferentia~\cite{inferentia}, Microsoft's Maia~\cite{maia}, and Intel's Gaudi~\cite{gaudi}. UALink's open, vendor-neutral architecture facilitates diverse accelerator configurations, supporting high-performance intra-cluster communication without dependence on proprietary interfaces. Strategically aligning deployment choices with the interoperability characteristics and architectural strengths of each interconnect ensures optimized intra-cluster performance, enhanced computational throughput, and improved resource efficiency across heterogeneous accelerator environments.

\begin{figure}[t!]
    \centering
    \includegraphics[width=1\linewidth]{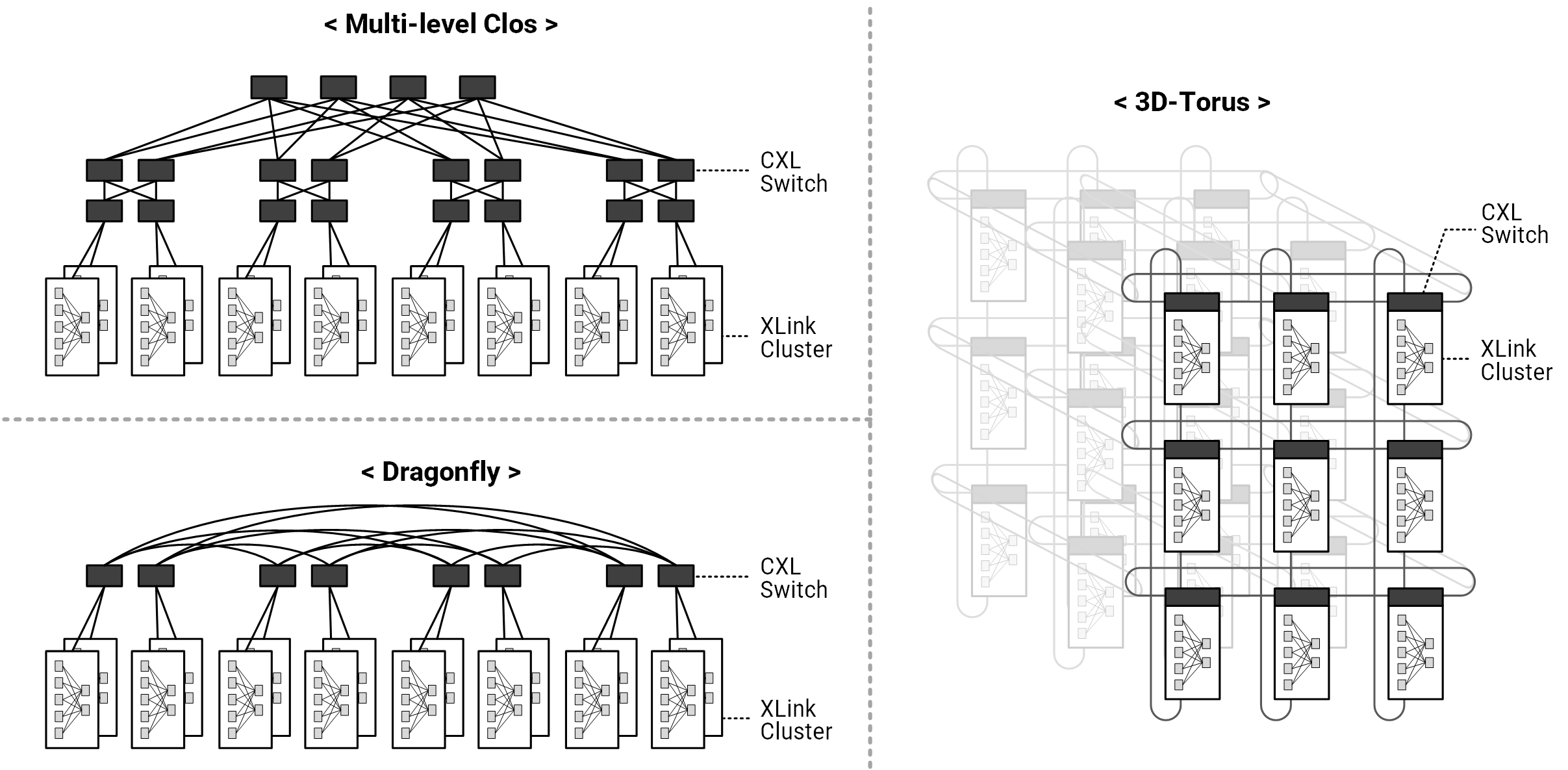}
    \caption{Exemplary CXL-over-XLink supercluster configurations.}
    \label{fig:hybridtopology}
\end{figure}

\paragraph{Scalable inter-cluster communication leveraging CXL.} In a CXL-over-XLink-based supercluster, multiple accelerator clusters interconnected by XLink are integrated into a unified hierarchical architecture through a scalable CXL fabric, forming a large-scale multi-accelerator system. This hybrid interconnect fabric strategy, as presented in this technical report, can reduce latency and data-transfer overhead, accommodating diverse workload characteristics prevalent in contemporary AI data centers. Specifically, while XLink facilitates rapid intra-cluster data exchanges among accelerators, CXL enables scalable and coherent inter-cluster memory sharing using flexible, hierarchical, switch-based fabric architectures. Unlike the single-hop Clos topologies employed by XLink, CXL supports greater scalability, enabling multiple accelerator clusters to dynamically aggregate distributed memory resources into unified composable pools. This pooled accelerator-local memory reduces the dependency on external storage resources, such as off-chip memory or SSDs, maximizing performance gains and enhancing flexibility and utilization of limited accelerator resources. In addition, CXL resolves interoperability limitations between UALink and NVLink clusters by abstracting each cluster as an independent entity, mediating inter-cluster interactions, thus facilitating seamless coexistence and efficient interaction among heterogeneous accelerator clusters within a unified large-scale architecture.

Figure~\ref{fig:hybridtopology} illustrates exemplary fabric architectures formed by hierarchical CXL switches interconnecting multiple UALink-based and NVLink-based clusters into a unified supercluster. Since CXL supports PBR routing and switch cascading it can be used to implement various topologies such as multi-level Clos, 3D-Torus, and DragonFly, satisfying diverse data center requirements. Moreover, CXL enables devices or clusters to be freely added or removed via hot-plugging, even during large-scale AI data center operations. Consequently, CXL-over-XLink-based supercluster configurations can flexibly adapt to specific workload characteristics, integrating diverse accelerators, memory devices, and computing resources into unified scale-up domains.

\begin{figure}[t!]
    \centering
    \includegraphics[width=0.95\linewidth]{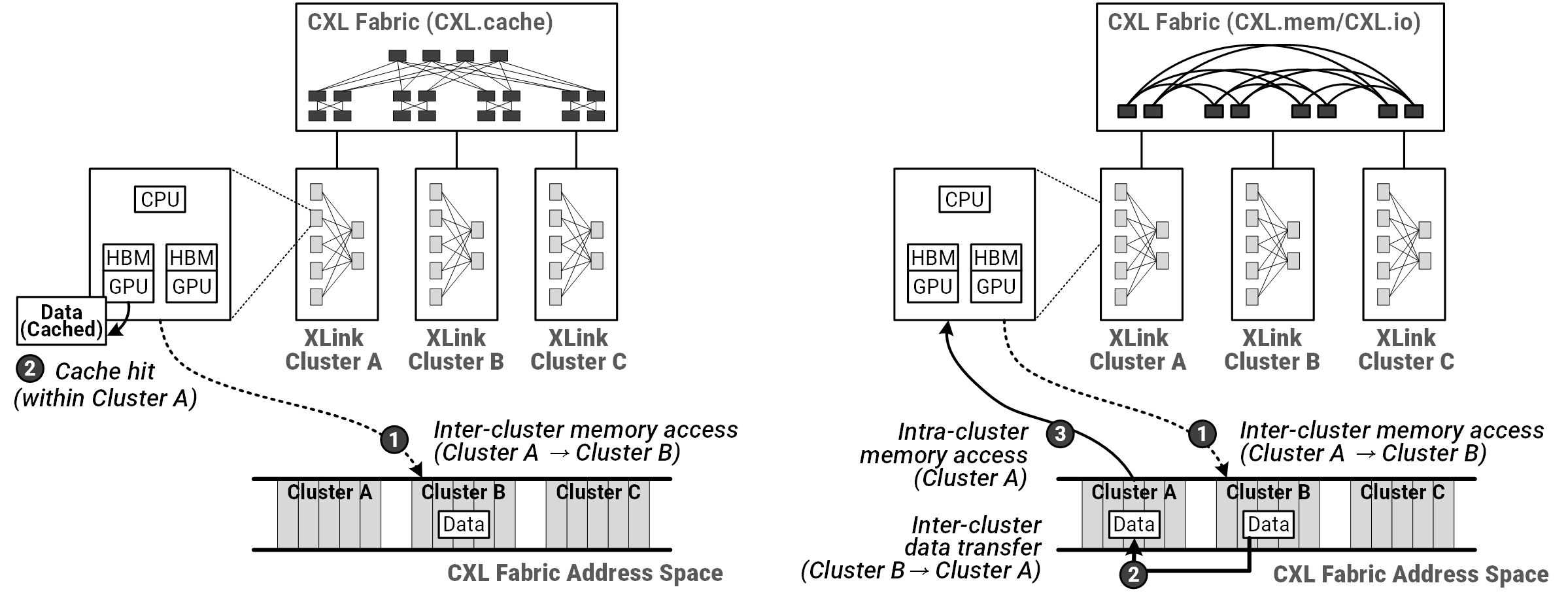}
    \begin{subfigure}{\linewidth}
        \begin{tabularx}{\textwidth}{
            p{\dimexpr.56\linewidth-2\tabcolsep-1.3333\arrayrulewidth}
            p{\dimexpr.42\linewidth-2\tabcolsep-1.3333\arrayrulewidth}
            }
            \caption{Protocol-level cache coherence.} \label{fig:hybridcxl1} &
            \caption{Instruction-level data transfer.} \label{fig:hybridcxl2}
        \end{tabularx}
    \end{subfigure}
    \caption{Cache coherence and data movement management in CXL-over-XLink.}
    \label{fig:hybridcxl}
\end{figure}

Another major advantage of the CXL-over-XLink architecture is its support for inter-cluster protocol-level cache coherence. Leveraging the cache-related sub-protocol (CXL.cache), accelerators within each cluster can directly and coherently access memory resources of other accelerators as well as remote memory resources located in external clusters at instruction-level granularity without software intervention. This approach aggregates local memories of multiple accelerators into a unified memory address space, enabling data to be directly fetched from on-chip accelerator caches. As a result, this minimizes latency and overhead typically associated with conventional inter-node memory access, while also maximizing performance by directly serving localized or shared data from accelerators’ own caches. Furthermore, CXL provides dedicated sub-protocol interfaces for memory access (CXL.mem) and high-speed data transactions (CXL.io), being able to support both small-sized instruction-level data transfers and bulk data transfers. These protocols can also be configured to enable direct device-to-device data management without CPU intervention. As illustrated in Figure \ref{fig:hybridcxl}, this hybrid interconnect architecture significantly reduces redundant data movements among accelerators and facilitates computational acceleration across distributed resources, thus maintaining high performance even in memory-intensive workloads.

Finally, CXL-over-XLink can further optimize data movements within superclusters more aggressively. Specifically, CXL.cache-based coherence enables novel paradigms for collective operations (e.g., broadcast, scatter/gather, and all-reduce), which traditionally incur substantial overhead due to explicit synchronization and redundant data copying. By providing coherent memory access among distributed accelerators, data movements are implicitly managed at the hardware level, enabling accelerators to treat distributed resources as unified memory pools. This fundamentally eliminates overhead associated with explicit synchronization and redundant data copying. This approach not only enhances performance but also simplifies AI model development and management. For example, when programming accelerator kernels, developers can focus exclusively on computational tasks without explicitly managing synchronization or data movement codes. Such software kernels concentrate on parallel computations, while CXL’s protocol-level cache coherence policies transparently handle inter-cluster high-speed memory data transfers without any software intervention. Data accesses, particularly those exhibiting locality across diverse workloads, efficiently utilize accelerator-internal caches, maximizing performance and computational efficiency.

\paragraph{Optimizing hardware and software for integrated XLink and CXL architectures.} Delineating the roles of XLink and CXL within a supercluster architecture provides structural benefits. However, integrating these distinct interconnect technologies presents practical implementation challenges. For example, protocol conversion and data transitions between XLink-based intra-cluster domains and CXL-based inter-cluster domains introduce additional latency due to physical and logical transformations required by differing communication protocols. Such overhead can degrade performance, impacting latency-sensitive workloads. In addition, higher-density accelerator deployments may increase cooling demands, while coherent memory transactions raise concerns about interconnect reliability and error management.

\begin{figure}[t!]
    \centering
    \includegraphics[width=0.9\linewidth]{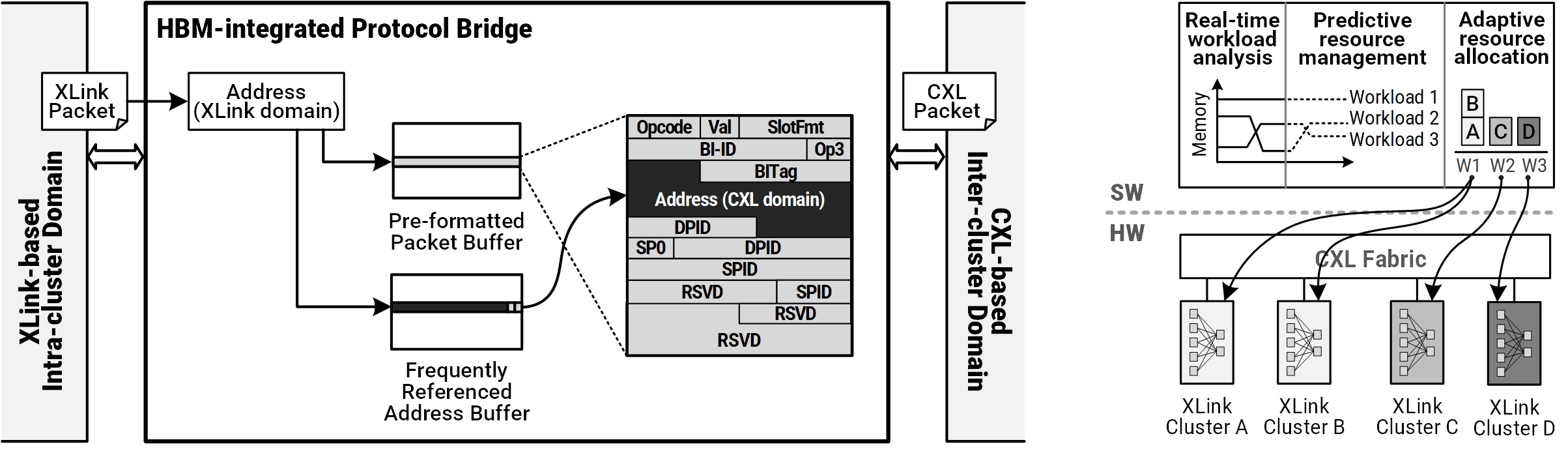}
    \begin{subfigure}{\linewidth}
        \begin{tabularx}{\textwidth}{
            p{\dimexpr.64\linewidth-2\tabcolsep-1.3333\arrayrulewidth}
            p{\dimexpr.07\linewidth-2\tabcolsep-1.3333\arrayrulewidth}
            p{\dimexpr.25\linewidth-2\tabcolsep-1.3333\arrayrulewidth}
            }
            \caption{Specialized bridging interfaces with integrated HBM.} \label{fig:bridge1} &
            \caption*{} &
            \caption{Orchestration software.} \label{fig:bridge2}
        \end{tabularx}
    \end{subfigure}
    \caption{Optimizing hardware and software for integrated XLink and CXL architectures.}
    \label{fig:bridge}
\end{figure}

To address these integration challenges, targeted hardware optimizations are required, including specialized system-on-chip bridging interfaces explicitly designed for rapid data-format conversions and streamlined interconnect protocols that minimize latency and reduce handshaking overhead, as shown in Figure \ref{fig:bridge1}. Incorporating HBM within bridging interfaces further mitigates performance penalties arising from inter-domain protocol conversions. In this example, frequently accessed memory addresses or requests can be cached in HBM, preserving pre-converted formats for immediate reuse and thereby eliminating latency overhead during protocol transitions. Intelligent data-placement strategies can be strategically adopted to minimize unnecessary data movements, reducing associated performance impacts.

Aside from these hardware-level optimization approaches, advanced orchestration and software strategies are equally important for improving system performance. Figure \ref{fig:bridge2} illustrates orchestration software frameworks supporting real-time workload monitoring, predictive resource management, and adaptive resource allocation. These capabilities improve operational efficiency and performance in large-scale integrated supercluster architectures. For example, when a specific cluster frequently accesses external memory, orchestration software can physically move the required data closer to the requesting cluster. This relocation allows accelerators within the cluster to efficiently access and process data locally. In addition, fault-tolerance mechanisms such as data redundancy, replication, and memory checkpointing can be implemented through software. These mechanisms enhance system stability and reliability in large-scale, multi-accelerator environments. Such strategies reduce risks related to component failures, unexpected data corruption, and interconnect disruptions, maintaining stable operation and consistent system performance.

\subsection{Memory Tiers Leveraging XLink and Lightweight CXL Links}
\label{subsection:6_3}
Building upon CXL-over-XLink, we further propose an extended, scalable architecture that integrates a tiered memory hierarchy within supercluster configurations, explicitly designed to address the diverse memory-performance demands of contemporary AI workloads. This structure comprises two distinct memory tiers: i) high-performance local memory managed via XLink and \textit{coherence-centric CXL}, and ii) scalable, composable memory pools enabled through \textit{capacity-oriented CXL}. To effectively establish these memory tiers, we recommend lightweight implementations of CXL specifically tailored for each of these tiers. Finally, we conclude this subsection by introducing data placement and management strategies crucial for strategically leveraging superclusters equipped with these hierarchical memory tiers.

\paragraph{High-performance accelerator-local memory: XLink with coherence support.}
Accelerator clusters within the existing CXL-over-XLink supercluster architecture are interconnected via XLink and utilize high-performance memory technologies, such as HBM or customized high-bandwidth DDR modules. Given that these accelerator structures are pre-defined in each cluster, various HBM modules differing in version or capacity may coexist within a single supercluster. In addition, heterogeneous high-speed memory types may also be intermixed within clusters. As the infrastructure scales, required memory capacities, memory types, and usage patterns differ among clusters, depending on specific workloads and models executed. Since the supercluster architecture already incorporates a CXL fabric for inter-cluster connectivity, this fabric can logically unify distributed high-performance memory resources across clusters into a coherent accelerator-local memory tier.

Within each cluster, accelerator-local memory employs XLink to construct a unified memory address space. Although there can be multiple methods to create this address space, considering the protocol specifications and operational characteristics of XLink, the most fundamental approach is to recognize memory modules of individual accelerators as statically partitioned blocks, forming a unified, linear address space. For instance, UALink can statically partition individual accelerator memory spaces into a unified NUMA-like memory domain across multiple accelerators. Similarly, NVLink constructs a unified address space among multiple devices using virtualization techniques. However, memory unified in this manner does not permit sharing beyond each statically partitioned memory region, requiring software or firmware intervention to explicitly copy data. In addition, due to the absence of protocol-level cache coherence, direct data sharing is infeasible. Consequently, memory accesses targeting regions not locally owned by an accelerator necessitate data transfers via XLink, introducing latency overhead. Such overhead becomes particularly pronounced when accelerators access memory regions across distinct clusters, significantly degrading performance.

To address these performance degradation and latency issues, coherence-centric CXL can be employed. Under the assumption of using cluster-to-cluster CXL connectivity within the proposed CXL-over-XLink architecture without protocol modifications, clusters can designate specific portions of their memory address space and expose them to the inter-cluster CXL fabric in an overlapping, cache-coherent manner, which enable cache coherence selectively for certain applications and datasets. In other words, cache-coherent data sharing becomes possible for designated regions within accelerator-local memory tiers, while other regions continue to be managed through data copying and movement via XLink within a unified address space. Such a partial coherence approach leveraging inter-cluster CXL fabric enhances data locality and performance for specific AI workloads. Moreover, improved data locality confines data accesses within clusters, allowing frequently accessed data to be automatically handled within accelerator on-chip caches, thereby fundamentally eliminating unnecessary data transfers.

For workloads exhibiting high data-sharing and cache coherence requirements, a more advanced and coherence-oriented lightweight application of CXL within clusters can be adopted. In this scenario, dedicated CXL controller logic is integrated into each accelerator, either closely positioned near or directly embedded within the XLink controller. The primary advantage of this configuration is that lightweight, coherence-centric CXL becomes available within clusters alongside XLink, allowing all GPUs or accelerators to perceive a unified memory space and enabling full data sharing. Consequently, explicit collective operations or data movements can be entirely eliminated, resulting in significant performance improvements. Although this design may increase SoC implementation complexity, cost, and the possibility of redundant data transfers, these issues can be effectively mitigated by optimizing the CXL protocol, removing unnecessary protocol features, and focusing explicitly on cache coherence. Such an integrated XLink-CXL controller could be implemented in various ways. However, to eliminate redundancy, bulk data transfers are primarily handled through XLink, while accelerator controllers implement only optimized CXL.cache subprotocols dedicated exclusively to coherence traffic. Despite necessitating detailed controller design and protocol management, this approach ultimately delivers performance benefits, including enhanced cache coherence, simplified data management, and improved computational efficiency across the supercluster.

The combination of accelerator-local memory via XLink and coherence-centric CXL interconnects effectively addresses low-latency and cache-coherence demands at the accelerator-node level. However, modern AI workloads frequently exhibit substantial memory-capacity requirements exceeding the aggregate local memory capacity available at the rack level. Consequently, complementary strategies employing scalable, capacity-oriented composable memory pools, as proposed in the subsequent section, are required to accommodate these extensive memory demands.

\begin{figure}[t!]
    \centering
    \includegraphics[width=0.95\linewidth]{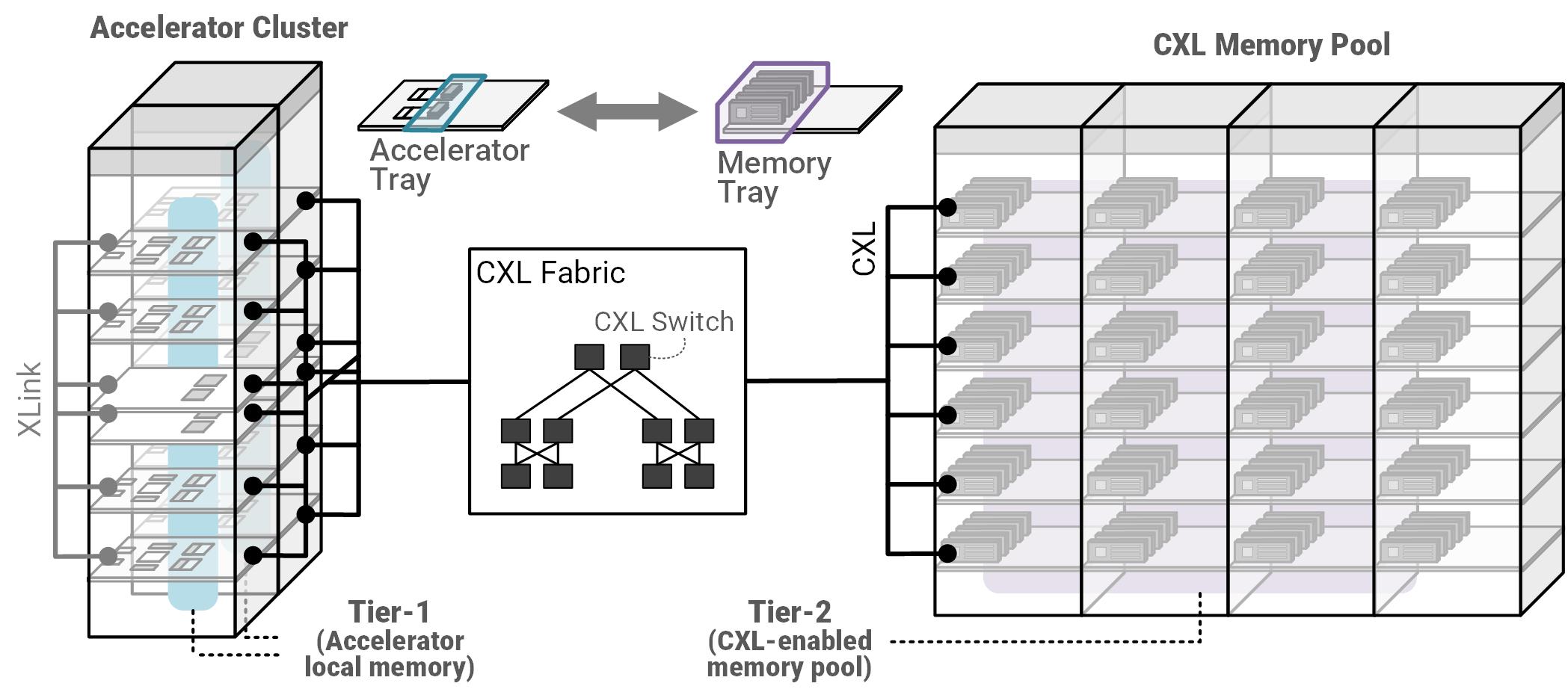}
    \caption{Tiered memory hierarchy configured with CXL memory pool.}
    \label{fig:hybrid_mempool}
\end{figure}

\paragraph{Capacity-oriented composable memory pools: CXL.} Accelerator-local memory serves frequently accessed, performance-critical data. However, modern AI workloads often require significantly larger memory capacities, even at the expense of reduced performance. Representative examples include large-scale embedding tables, caches, and external knowledge bases. To address such challenges, we propose a two-tier composable memory structure integrated within the supercluster architecture, providing flexible memory-capacity expansion.

As illustrated in Figure~\ref{fig:hybrid_mempool}, the proposed two-tier composable memory pools primarily comprise memory trays physically separated from accelerator clusters and interconnected via a dedicated CXL fabric. In a CXL-over-XLink-based supercluster architecture, tier-1 memory already provides accelerators with a unified memory view managed by coherence-centric CXL and XLink controllers. Therefore, accesses to tier-2 memory pools occur only when memory demands surpass the aggregate accelerator memory capacity available at the rack level. These scenarios are analogous to applications such as RAG, where data retrieval traditionally relies on storage systems or distributed file systems with access latencies ranging from milliseconds to tens of seconds. In contrast, the proposed tier-2 composable memory pools  reduce such latency to tens or hundreds of nanoseconds, depending on fabric-switch characteristics. The critical advantage of the proposed approach is capacity scalability, achieved by integrating only memory components within each memory tray, explicitly excluding CPUs or accelerators to maximize memory density and efficiency.

The physical placement of memory trays is related to the number of switch hops and associated latency; thus, these trays can be located anywhere within a CXL-over-XLink-based supercluster, as long as connectivity via CXL fabric is available. Depending on the spatial management requirements of data center designers, memory trays can be flexibly arranged. The tier-2 memory pools' address spaces can be physically configured so that the fabric directly recognizes them, or they can be logically connected through virtual management techniques. Allocating memory trays close to accelerator clusters significantly reduces reliance on slower storage or lower-performance scale-out data access methods. In practice, these external memory resources function as tier-2 capacity-focused memory pools within a hierarchical memory architecture, specifically optimized for memory-capacity expansion.

As in the coherence-centric CXL approach, the proposed capacity-oriented CXL configuration can either utilize the existing CXL-over-XLink fabric or further optimize it to better accommodate large-scale workloads. Given that tier-1 accelerator-local memory already manages cache coherence and handles latency-sensitive data, tier-2 memory pools can be exclusively optimized for memory capacity, simplifying other functionalities and further enhancing cost efficiency. For example, maintaining cache coherence across all memory trays is unnecessary; thus, controllers can be streamlined and efficiency maximized by disabling CXL.cache or CXL.io protocols at switches and endpoints. In particular, when tier-1 memory serves as an exclusive cache, tier-2 composable memory pools can potentially omit CXL.mem altogether, using only the CXL.io protocol for bulk data transfers, similar to traditional storage systems. Regardless of the selected approach, data transfers between accelerator-local memory and composable memory pools persist, making it essential to provide sufficient CXL fabric ports to optimize data transfer performance.

Note that additional optimization strategies beyond protocol simplification are possible for tier-2 capacity-oriented memory pools. As discussed in Section~\ref{subsection:5_1}, memory trays can utilize lower-speed DRAM interfaces instead of high-speed DRAM modules to reduce cost. Alternatively, hybrid memory trays combining high-capacity storage (e.g., flash memory) with modest amounts of HBM can significantly enhance memory capacity while still offering sufficient performance for effective data staging into tier-1 accelerator-local memory. Furthermore, to support the buffering and caching functions of tier-1 accelerator-local memory across extended physical distances, spanning multiple floors or even buildings, the supercluster's memory tiering structure can leverage optical technologies, such as silicon photonics, in place of PCIe PHY for interconnection via CXL.

\paragraph{Discussion on hierarchical data placement and management strategies.}
As discussed previously, integrating accelerator-local memory (XLink-based) with composable memory pools (CXL-based) provides new opportunities for substantial performance and capacity expansion. However, fully leveraging such hierarchical memory structures may require software frameworks capable of intelligent resource management and efficient data placement. Specifically, integrated management methods supporting dynamic allocation of computational tasks across accelerator clusters and optimized memory allocation within composable CXL pools can enhance resource utilization, balance performance, and improve operational efficiency in modern AI data centers.

To fully exploit the hierarchical memory architecture, carefully designed data placement strategies that align with the performance characteristics of each memory tier are critical. Rather than relying on hardware-based management, implementing these placement strategies through software is advantageous. Specifically, software frameworks can precisely evaluate attributes such as data access frequency and latency sensitivity, enabling refined and adaptable placement decisions. For example, latency-critical and frequently accessed data structures, including activation states, embedding vectors, and attention caches, should be placed within accelerator-local memory tiers connected via XLink. In contrast, datasets that are larger or less sensitive to latency are better suited to reside within capacity-oriented composable memory pools facilitated by CXL. Implementing such tier-specific placement requires software-level monitoring of various runtime information, optimizing data placement according to characteristics, thus maximizing resource utilization and overall performance.

Further enhancing the efficiency of sophisticated data-placement strategies may require the addition of advanced software-based orchestration frameworks. For instance, predictive data-migration algorithms that dynamically adjust data placement based on anticipated workload patterns and access-frequency variations can be introduced. Advanced caching policies, such as temperature-aware caching that prioritizes data according to access frequency, as well as intelligent machine-learning-based prefetching, further minimize latency and overhead associated with inter-tier data transfers. However, excessively frequent inter-tier data migrations can introduce performance degradation; thus, comprehensive approaches involving targeted hardware optimizations, efficient protocol-conversion interfaces between XLink and CXL, and carefully designed data-migration policies remain essential.

In summary, large-scale AI applications deployed within multi-accelerator systems integrating CXL and XLink can significantly benefit from hierarchical data management approaches. Specifically, real-time inference workloads utilize accelerator-local memory for rapid data processing and latency-critical operations, while large-scale embedding lookups and external data retrieval efficiently leverage capacity-oriented composable memory pools provided by CXL. Therefore, clearly defining the roles of each memory tier and strategically implementing data-placement policies ensures optimized computational performance and resource utilization, effectively satisfying the diverse operational demands of modern AI data centers.

\section{Conclusion}
\label{subsection:conclusion}
In this technical report, we systematically explored the limitations of traditional GPU-centric architectures in scaling modern AI workloads, highlighting significant performance bottlenecks related to memory capacity, inter-device communication, and resource allocation. To address these challenges, we introduced a composable and modular data center architecture leveraging Compute Express Link (CXL) technology, which disaggregates and dynamically allocates memory, compute, and accelerator resources according to workload-specific demands.

Our empirical evaluations using diverse AI workloads, including retrieval-augmented generation (RAG), Graph-RAG, deep learning recommendation models (DLRM), and MPI-based scientific simulations, demonstrated substantial performance improvements. Specifically, we observed significant reductions in execution latency, communication overhead, and memory management complexity compared to conventional SSD- and RDMA-based infrastructures. These results illustrate how the coherent memory sharing and dynamic composability features of CXL effectively optimize resource utilization and enhance operational flexibility.

In addition, we investigated hybrid architectures integrating dedicated accelerator-centric interconnect technologies (XLink), such as Ultra Accelerator Link (UALink) and NVIDIA's NVLink, alongside CXL. Our analysis revealed the potential ability showing that combining the complementary strengths of these technologies can further enhance scalability, reduce unnecessary long-distance communication, and optimize performance for latency-sensitive intra-accelerator tasks.

Finally, we discussed critical architectural implications of adopting composable CXL infrastructures in real-world data centers. These include dedicated coherent memory pooling, adaptive data placement, accelerator-centric resource management, and sophisticated centralized monitoring frameworks. Future research should focus on addressing deployment challenges at industrial scales, exploring advanced orchestration techniques, and further refining hybrid interconnect strategies. Such efforts are essential to fully realize the potential of composable CXL-based infrastructures in supporting increasingly complex and demanding AI applications.

\phantomsection
\addcontentsline{toc}{section}{References}
\bibliographystyle{IEEEtran}
\bibliography{references}

\phantomsection
\addcontentsline{toc}{section}{Disclaimer}
\setstretch{1.1}
\begin{footnotesize}
\noindent
\subsubsection*{Notices \& Disclaimers}

\noindent Panmnesia, the Panmnesia logo, and other Panmnesia marks are trademarks of Panmnesia, Inc. or its subsidiaries. Other names and brands may be claimed as the property of others.

\noindent All content contained in this document is protected by applicable copyright laws. Any unauthorized use, reproduction, distribution, or transmission of the content is strictly prohibited without prior written consent of Panmnesia.

\noindent All information included herein is provided "AS IS." Panmnesia hereby disclaims all warranties, representations, and guarantees of any kind with respect to the information in this document, including without limitation, warranties of merchantability, non-infringement, accuracy, completeness, timeliness, or fitness for any particular purpose.

\noindent Panmnesia reserves the right to make corrections, modifications, enhancements, improvements, and any other changes to this document, at any time without notice.

\noindent Neither Panmnesia nor any of its affiliates, officers, employees, or representatives shall bear any responsibility or liability whatsoever for any errors, omissions, or consequences arising from the use of or reliance upon any information included herein. Any recipient should conduct their own due diligence before making any decisions based on this information.
\end{footnotesize}

\vfill

\begin{center}
  \small\textbf{© 2025 Panmnesia, Inc. All rights reserved.}
\end{center}

\end{document}